\documentclass[lettersize,journal]{IEEEtran}
\usepackage{amsmath,amsfonts}
\usepackage{algorithmic}
\usepackage{array}
\usepackage{textcomp}
\usepackage{stfloats}
\usepackage{url}
\usepackage{verbatim}
\usepackage{graphicx}
\def\BibTeX{{\rm B\kern-.05em{\sc i\kern-.025em b}\kern-.08em
    T\kern-.1667em\lower.7ex\hbox{E}\kern-.125emX}}
\usepackage{balance}
\usepackage{lipsum}  
\usepackage{textcomp}
\usepackage{tikz}
\usetikzlibrary{positioning}
\usetikzlibrary{arrows}
\usepackage{caption}
\usepackage{subcaption}
\usepackage{xcolor}
\usepackage{wrapfig}

\newcommand{\matrixstyle}[1]{\mathrm{#1}}
\newcommand{\vectorstyle}[1]{\mathbf{#1}}

\newtheorem{theorem}{Theorem}

\newtheorem{definition}{Definition}

\begin{document}

\title{Differential Flatness of Quasi-Static Slider-Pusher Models with Applications in Control}

\author{Sander De Witte, Tom Lefebvre, Thomas Neve, András Retzler and Guillaume Crevecoeur
\thanks{The authors are with the Dynamic Design Lab (D\textsuperscript{2}LAB) of the Department of Electromechanical, Systems and Metal Engineering, Ghent University, B-9052 Ghent, Belgium.}}

\maketitle

\begin{abstract}
    This paper investigates the dynamic properties of planar slider-pusher systems as a motion primitive in manipulation tasks. To that end, we construct a differential kinematic model deriving from the limit surface approach under the quasi-static assumption and with negligible contact friction. The quasi-static model applies to generic slider shapes and circular pusher geometries, enabling a differential kinematic representation of the system. From this model, we analyze differential flatness—a property advantageous for control synthesis and planning—and find that slider-pusher systems with polygon sliders and circular pushers exhibit flatness with the centre of mass as a flat output. Leveraging this property, we propose two control strategies for trajectory tracking: a cascaded quasi-static feedback strategy and a dynamic feedback linearization approach. We validate these strategies through closed-loop simulations incorporating perturbed models and input noise, as well as experimental results using a physical setup with a finger-like pusher and vision-based state detection. The real-world experiments confirm the applicability of the simulation gains, highlighting the potential of the proposed methods for practical manipulation tasks.
\end{abstract}


\section{Introduction}
The ability to manipulate objects by pushing is a resourceful skill for robotic manipulators to master. Expanding the range of manipulation methods with pushing could enable a system to manipulate objects that would otherwise be too large, heavy, or cluttered, to be grasped \cite{yu2016more,chai2022object,stuber2020let}.
Unfortunately, this poses a challenging control problem, due to the the contact dynamics' inherent under-actuation and their hybrid nature resulting from different pushing regimes. It is reasonable to assume that a control method tailored to this task would rely on a dynamic model of the system \cite{hogan2020feedback,raghunathan2022pyrobocop,doshi2020hybrid}. Such a dynamic model can be quite simple but should at least capture the essential principles of movement of the controlled system.
Manipulation by pushing can be formalised by considering slider-pusher systems. Here, the manipulated object is referred to as the slider and the controlled object as the pusher. Accurate modelling of slider-pusher systems is difficult because of the complex contact dynamics involved at (i) the contact point between the slider and the pusher, and, (ii) the contact between the slider and its supporting surface.

The dynamics of slider-pusher systems have been the subject of numerous studies. The earliest references date back almost three decades \cite{mason1986mechanics,goyal1991planar1,goyal1991planar2} with recent literature focusing on data-driven model approaches \cite{ziyan2021planar,yu2016more,bauza2017probabilistic}. Full descriptions of the Newtonian mechanics of the system, including consideration of all interaction forces, are rare and, quite frankly, unnessary. Many authors adopt the quasi-static assumption, which implies that the slider-pusher motions are slow enough for inertial forces to be negligible compared to frictional forces \cite{hogan2020feedback,lynch1992manipulation,zhou2018convex,ghazaei2020quasi}. Under this assumption, a differential kinematic model is obtained, relating the pusher's input velocity to the slider's planar velocities. These earlier studies only consider rectangular sliders.

Even with these simplifications in place, the combined slider-pusher remains a hybrid and under-actuated system \cite{hogan2020feedback}. These mathematical properties necessitate the use of planning and model-based predictive control approaches to manipulate the slider through an actuated pusher, placing significant demands on the control system's computational resources. This reinforces the motivation for employing simple yet principally accurate models.

A second assumption can be made to address these requirements. We propose a quasi-static model that additionally assumes that the friction forces between the slider and pusher are negligible compared to the support forces. The resulting model applies to generic slider shapes and circular pusher geometries.
Starting from this model we explore for which slider geometries the differential kinematic model is differentially flat. Differential flatness is a model property exhibited by certain nonlinear dynamical systems \cite{fliess1995flatness,rigatos2015differential}, and is a useful property for control synthesis. It is particularly advantageous to solve path planning problems \cite{diwold2021trajectory,stoican2015flat,stoical2016obstacle,stoican2017constrained,lefebvre2023differential}. Flatness is also used abundantly in the design and synthesis of asymptotically trajectory tracking control \cite{greeff2018flatness,helling2020flatness,faessler2017differential,aguilar2012trajectory,agrawal2021constructive,greeff2020exploiting,rudolph2021flatness,delaleau1998control,hagenmeyer2003exact}. 
Our analysis suggests that, at least, all slider-pusher systems with polygon sliders and circular pushers are flat.

\begin{figure}[hbt]
    \centering
    \includegraphics[width=\linewidth]{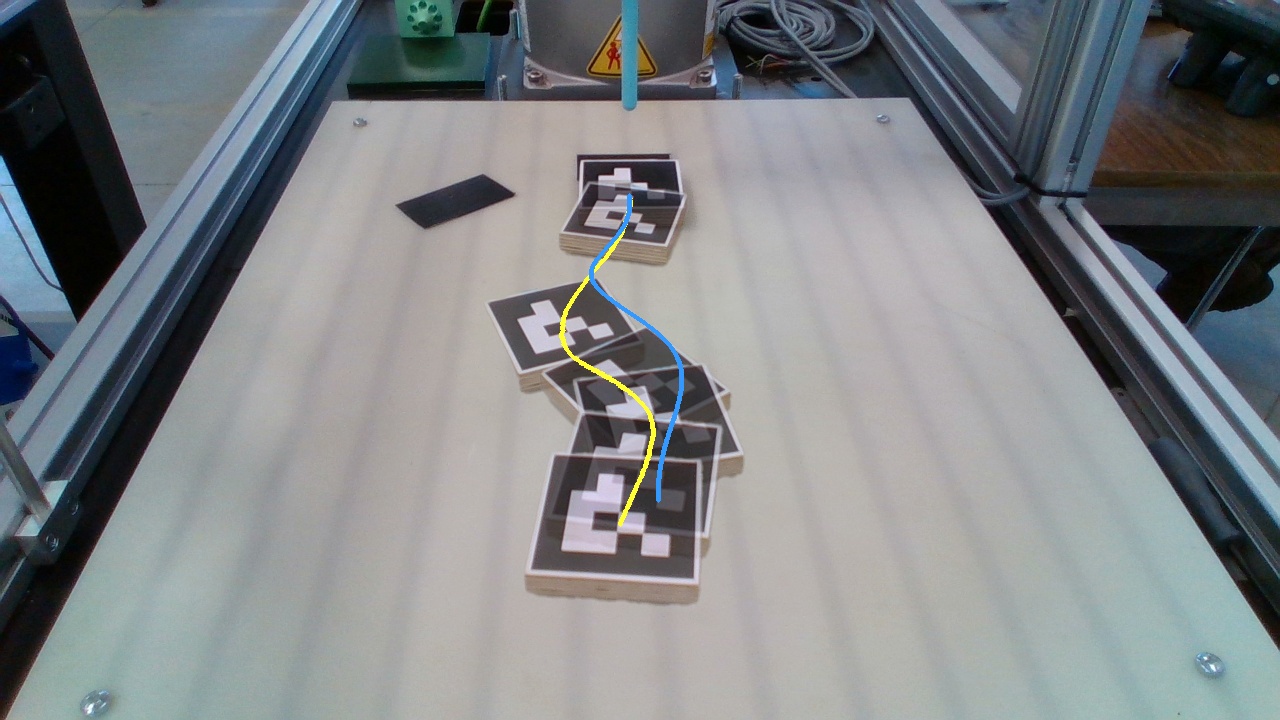}
    \caption{Snapshots of one of the cameras, showing dynamic feedback linearization on the set-up. The reference trajectory is shown in blue, while the executed path is yellow.}
    \label{fig:screengrabs}
\end{figure}

\newpage

Motivated by the system's flatness property, we designed two closed-loop tracking strategies for a rectangular slider. The first ad-hoc control strategy proposes a cascaded (quasi-)static feedback architecture. Additionally, since the system under consideration is flat and thus demonstrably dynamic feedback linearizable \cite{vannieuwstadt1994equivalence,levine2007equivalence,levine2011necessary}, we also design a dynamic feedback linearization approach. Since the system is under-actuated, only position tracking is considered, with both a static and dynamic position signal. In the latter case, it is assumed we have access to every time derivative of the desired position signal.
Next to closed-loop simulation results with perturbed simulation models and input noise, both control approaches are tested on a set-up. The set-up consists of a finger-like end effector, acting as a cylindrical pusher, attached to an industrial robot, with two cameras detecting the slider state. Arbitrary screengrabs taken from one of the two cameras can be seen in Fig. \ref{fig:screengrabs}, showing dynamic feedback linearization in . Real-world results resemble the results achieved in simulation, making it possible to apply the simulation gains directly to the set-up.

\section{Modeling}

\subsection{Kinematics}
\begin{figure}[t]
    \centering
    \includegraphics[width=.6\columnwidth]{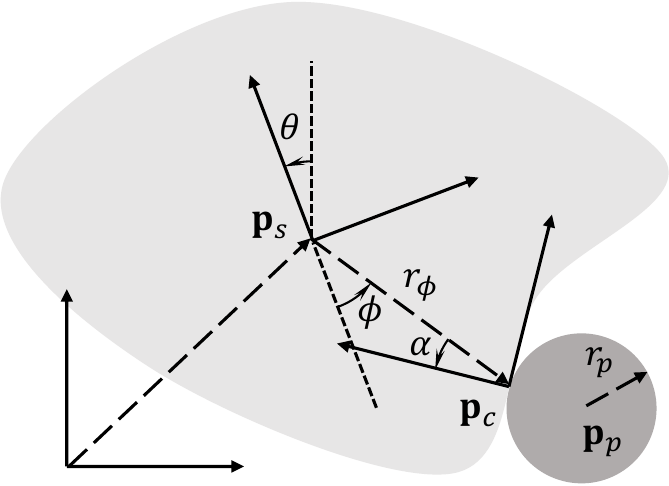}
    \caption{Geometry of the slider-pusher system.}
    \label{fig:overview}
\end{figure}

This section characterizes the kinematics of the slider-pusher system with arbitrary slider geometry. For the sake of simplicity and practical convenience, we only consider spherical pushers and assume that contact is maintained during the entire motion of both slider and pusher. The schematic representation in Fig. \ref{fig:overview} represents the geometry of the problem. 

In the present study, we only consider planar motion to characterise the slider as a planar object. 
We also attach a local frame of reference to the slider for future reference. The origin of this local frame of reference is located at the slider's centre of mass (c.o.m.).

Second, we will consider sliders with arbitrary geometries, meaning that any closed non-intersecting curve can describe the circumference of the slider. To describe the circumference we use polar coordinates in the local frame of reference attached to the slider. The variable $\phi$ quantifies the angle between any point on the slider's circumference and the local vertical axis, starting in the fourth quadrant. The distance between the point and the c.o.m. is determined by a radius function $r(\phi)$ where $r:[0,2\pi]\mapsto \mathbb{R}_{>0}$ and $r(0)=r(2\pi)$. The radius of the pusher is parameterized using the variable $r_p$. 

Since we assume that the slider and pusher maintain contact, the kinematic configuration of the slider is determined completely by the contact point\footnote{In principle we could also parameterize the orientation of the pusher however due to the assumption of frictionless contact and the pusher's spherical geometry its orientation is irrelevant nor is it something we desire to control.}. We can thus parameterize the contact using the angle, $\phi$. 

Conveniently we can gather the kinematic state of the {slider-pusher} system in the vector $\vectorstyle{x}\in \mathbb{R}^4$
\begin{equation}
    \vectorstyle{x} = \begin{pmatrix}
        x & y & \theta & \phi
    \end{pmatrix}^\top
\end{equation}
where $x$ and $y$ denote the Cartesian coordinates of the slider in the global frame of reference, $\theta$ denotes the slider's planar orientation and $\phi$ thus determines the contact point.

Further, we assume that the velocity of the pusher, expressed in the global frame of reference, can be controlled. Thus we can define an input variable, $\vectorstyle{u}\in\mathbb{R}^2$, as 
\begin{equation}
    \vectorstyle{u} = \begin{pmatrix}
        u_x & u_y
    \end{pmatrix}^\top
\end{equation}
where $u_x$ and $u_y$ denote the global velocity of the pusher.

In conclusion, we note that as a result of the variable radius, $r(\phi)$, we can determine a third frame of reference which is rotated over the angle $\alpha$. This final rotation accounts for the change in the circumference normal as a result of the local variation in the radius following a change in the angle $\phi$. It is implied that for given $r(\phi)$ there is a nonlinear but known function that relates the angles $\alpha$ and $\phi$. Here $r'(\phi)$ denotes the derivative of $r(\phi)$. One easily verifies that the angle $\alpha$ can be characterized by the following expression.
\begin{equation}
    \label{eq:alpha1}
    \tan(\alpha) = -\frac{r'(\phi)}{r(\phi)}
\end{equation}
Correspondingly
\begin{equation}
    \begin{aligned}
        \cos(\alpha) &= \frac{r(\phi)}{\sqrt{r(\phi)^2+ r'(\phi)^2}} \\
        \sin(\alpha) &= -\frac{r'(\phi)}{\sqrt{r(\phi)^2+ r'(\phi)^2}}
    \end{aligned}
\end{equation}

\subsection{Quasi-static model}\label{sec:quasi-static-model}

We now seek a relation between the time derivative of the kinematics state vector, $\dot{\vectorstyle{x}}$, the state vector itself, $\vectorstyle{x}$, and the input vector, $\vectorstyle{u}$. The quasi-static model considered in this work is based on two main assumptions. 
\begin{enumerate}
    \item The pusher motions are slow enough that inertial forces are negligible compared to frictional forces.
    \item The friction forces at the contact point are negligible with respect to the friction forces between the slider and the ground.
\end{enumerate}

The first assumption allows us to establish a force-motion model between the contact force exerted by the pusher and the resulting motion of the slider. The second assumption allows us to specify the motion at the contact, establishing a motion-motion model.

\subsubsection{Force-motion model}
The force-motion model for quasi-static pushing can be established through the concept of the limit surface \cite{goyal1991planar1,goyal1991planar2,lynch1992manipulation,ziyan2021planar}. The limit surface is a hypothetical closed and convex hypersurface in the planar wrench space of the slider, $F = (f_x,f_y,\tau_z)$, that encloses the origin and is implicitly defined by the function, $H:\mathbb{R}^3\mapsto\mathbb{R}$. The main idea of the limit surface approach is that any force exerted on the slider that provokes a velocity response but not an acceleration must be on the limit surface. The first assumption implies that inertial forces are negligible, meaning the system operates in a regime where forces directly result in velocities without inducing acceleration.
\begin{equation}
    H(F)=\text{cst.}
\end{equation}

The limit surface approach further states that points on the surface correspond to friction wrenches that are parallel to sliding twist directions of the slider, $V = (\dot{x},\dot{y},\dot{\theta})$. This notion allows us to define a proportional relation between wrenches and twists. Specifically, it follows that
\begin{equation}
    \label{eq:VpropF}
    V \propto \nabla H(F)
\end{equation}

There is no straightforward approach to determine the limit surface. In practice, often the limit surface is assumed to be quadratic
\begin{equation}
    H(F) = F^\top \matrixstyle{L} F 
\end{equation}
where in general we have
\begin{equation}
    \matrixstyle{L}\propto \begin{bmatrix}			\lambda_{xx} & \lambda_{xy} & \lambda_{xz}\\			\lambda_{xy} & \lambda_{yy} & \lambda_{yz}\\
        \lambda_{xy} & \lambda_{yz} & \lambda_{zz} 
    \end{bmatrix}\succ 0 
\end{equation}

Following \cite{hogan2020feedback,zhou2018convex,lynch1992manipulation}, it is hypothesized that the slider's limit surface can be approximated globally by a diagonal ellipsoid, hence
\begin{equation}
    \matrixstyle{L}\propto \begin{bmatrix}			1 &  & \\ & 1 & \\
        & & \tfrac{1}{\beta^2}
    \end{bmatrix}
\end{equation}
The value of $\beta$ may follow from physical considerations \cite{lynch1992manipulation,hogan2020feedback,lefebvre2023differential}, resulting in a geometric definition, or, it can be identified from measurements \cite{zhou2018convex,zhou2019pushing}. It relates to the tendency of the slider to rotate or translate when subject to external influences.

A final restriction on the wrench follows from the observation that we cannot apply a moment at the contact point. It follows that
\begin{equation}
    \label{eq:wrench_con1}
    \tau_z = r_x f_y - r_y f_x
\end{equation}
with $r_x$ and $r_y$ denoting the relative position of the contact to the c.o.m.

Based on these modelling assumptions, it is now possible to map the force exerted by the pusher to the slider's motion. Here it is assumed that the force of the pusher can be controlled. This avoids careful consideration of the motion of the slider itself and therefore the motion of the contact point. To develop a motion-motion model the contact needs to be specified too.

\subsubsection{Motion-motion model}
A motion-motion model can be obtained by further specifying the mechanics at the contact point. Recall that we have assumed frictionless contact implying that we cannot apply a force tangent to the slider circumference. It directly follows that the force is normal to the slider's circumference. This results in another restriction on the wrench.
\begin{equation}
        \label{eq:wrench_con2}
    \tan (\theta+\phi+\alpha) = -\frac{f_x}{f_y}
\end{equation}

To complete the model, we can now express and equate the velocity of the contact point from the perspective of the slider and pusher. Let $\vectorstyle{p}_c^g$ denote the position of the contact point in global coordinates. From the perspective of the slider we have that
\begin{equation}
    \label{eq:pcg1}
    \vectorstyle{p}_c^g = \vectorstyle{p}^g_s - r(\phi) \matrixstyle{R}(\theta+\phi) \vectorstyle{1}_y
\end{equation}
where $\vectorstyle{p}_s^g$ denotes the position of the slider in global coordinates, $R(\cdot)$ represents the standard planar rotation matrix, and $\vectorstyle{1}_y$ represents the unit vector in the second dimension.
        
From the perspective of the pusher, we have that
\begin{equation}
    \label{eq:pcg2}
    \vectorstyle{p}_c^g = \vectorstyle{p}_p^g + r_p \matrixstyle{R}(\theta+\phi+\alpha)\vectorstyle{1}_y
\end{equation}
where $\vectorstyle{p}_p^g$ denotes the position of the pusher in global coordinates.
        
Now we can take the derivative of both expressions (\ref{eq:pcg1}) and (\ref{eq:pcg2}) to obtain an expression for the contact point's motion. From the slider's perspective, we find that
\begin{equation}
    \label{eq:contact1}
    \dot{\vectorstyle{p}}_c^g = \dot{\vectorstyle{p}}_s^g - r'(\phi) \matrixstyle{R}(\theta+\phi)\vectorstyle{1}_y\dot{\phi}+ r(\phi) \matrixstyle{R}(\theta+\phi)\vectorstyle{1}_x(\dot{\phi}+\dot{\theta})
\end{equation}
whereas from the pusher's perspective, we find that
\begin{equation}
    \label{eq:contact2}
    \dot{\vectorstyle{p}}_c^g = \vectorstyle{u} - r_p \matrixstyle{R}(\theta+\phi+\alpha)\vectorstyle{1}_x \left(\dot{\theta}+\dot{\phi}+\dot{\alpha}\right)
\end{equation}
Finally recall that $\alpha$ and $\phi$ are related as dictated by (\ref{eq:alpha1}) and thus so are their time derivatives. 
\begin{equation}
    \label{eq:alpha}
    \begin{aligned}
        \alpha &= -\arctan\left(\frac{r'(\phi)}{r(\phi)}\right) \\
        \dot{\alpha} &= \frac{r’(\phi)^2-r''(\phi)r(\phi)}{r(\phi)^2+r’(\phi)^2}\dot{\phi} = f(\phi) \dot{\phi}
    \end{aligned}
\end{equation}
Note the introduction of the factor $f(\phi)$ for notational convenience.

\subsection{Smooth dynamics}\label{sec:differential-kinematics}
We can now derive a differential kinematic model by solving the equations (\ref{eq:VpropF}), (\ref{eq:wrench_con1}), (\ref{eq:wrench_con2}), (\ref{eq:contact1}) and (\ref{eq:contact2}) for the variables variables $V=(v_x,v_y,\omega_z)$, $F=(f_x,f_y,f_z)$ and $\dot{\phi}$. This model is only valid for sliders with a smooth circumference. An extension to non-smooth sliders is made in the next section. 

Further, it will be beneficial to describe the inputs $\vectorstyle{u}$ in the frame of reference with its vertical axis normal to the slider's circumference instead of the global frame of reference. This frame of reference is characterized by the combined rotation $\theta+\phi+\alpha$. This operation simplifies the analytical expression of the equality constraint imposed on (\ref{eq:contact1}) and (\ref{eq:contact2}). To that end we characterize the local input vector $\tilde{\tilde{\tilde{\vectorstyle{u}}}}$ as
\begin{equation}
    \begin{aligned}
        \vectorstyle{u} &= \matrixstyle{R}(\theta) \tilde{\vectorstyle{u}} \\
        \tilde{\vectorstyle{u}} &= \matrixstyle{R}(\phi) \tilde{\tilde{\vectorstyle{u}}} \\
        \tilde{\tilde{\vectorstyle{u}}} &= \matrixstyle{R}(\alpha) \tilde{\tilde{\tilde{\vectorstyle{u}}}}
    \end{aligned}
    \label{eq:input}
\end{equation}
and its elements
\begin{equation}
\tilde{\tilde{\tilde{\vectorstyle{u}}}} = \begin{pmatrix}
        \tilde{\tilde{\tilde{{u}}}}_x & \tilde{\tilde{\tilde{{u}}}}_y 
    \end{pmatrix}^\top
\end{equation}

As a result of this last alteration, the problem works out to an analytically manageable model.
\begin{equation}
    \label{eq:dk}
    \begin{aligned}
        &\dot{x} = - P(\phi)\sin(\theta+\phi+\alpha)  \tilde{\tilde{\tilde{{u}}}}_y  \\
        &\dot{y} = P(\phi) \cos(\theta+\phi+\alpha)  \tilde{\tilde{\tilde{{u}}}}_y  \\
        &\dot{\theta} = \Theta(\phi)  \tilde{\tilde{\tilde{{u}}}}_y  \\
        &\dot{\phi} = \Phi_x(\phi)  \tilde{\tilde{\tilde{{u}}}}_x + \Phi_y(\phi)  \tilde{\tilde{\tilde{{u}}}}_y   
    \end{aligned}
\end{equation}
where
\begin{equation}
    \begin{aligned}
        P(\phi) &= \frac{\beta^2 r(\phi)^2 + \beta^2 r'(\phi)^2}{\beta^2 r(\phi)^2 + \beta^2 r’(\phi)^2 + r(\phi)^2 r’(\phi)^2} \\
        \Theta(\phi) &= \frac{r(\phi) r’(\phi)\sqrt{r(\phi)^2+ r’(\phi)^2}}{\beta^2 r(\phi)^2 + \beta^2 r’(\phi)^2 + r(\phi)^2 r’(\phi)^2} \\
        \Phi_x(\phi) &= \frac{1}{\sqrt{r(\phi)^2+ r’(\phi)^2}+r_p(1+f(\phi))} \\
        \Phi_y(\phi) &= -\Phi_x(\phi)\frac{ r'(\phi)r(\phi)\left(r(\phi)^2+r_p\sqrt{r(\phi)^2+ r’(\phi)^2}\right)}{\beta^2 r(\phi)^2 +\beta^2 r’(\phi)^2 + r(\phi)^2 r’(\phi)^2}
    \end{aligned}
\end{equation}


Remark that in this analysis we do not take into account possible discrete phenomena (such as geometric locking) that might arise when considering non-convex slider geometries. Hence the model that is presented here only applies globally for convex and only locally for non-convex slider geometries. 

Let us now apply this general model to certain slider geometries by substituting the specific slider circumference's parametrization. Two obvious slider geometries enjoy our initial interest. The (1) rectangular and (2) circular geometries.

\subsubsection{Rectangular slider} The rectangular geometry is characterized by a width $2a$ and a height $2b$. One further verifies that for rectangular geometries $\alpha = - \phi$. As such it follows that $\tilde{\tilde{\tilde{\vectorstyle{u}}}} = \tilde{\vectorstyle{u}}$. The circumference itself is parametrized as follows
\begin{equation}
    \begin{aligned}
        r(\phi) &= b\frac{1}{\cos(\phi)} \\
        r’(\phi) &= b\frac{\tan(\phi)}{\cos(\phi)}
    \end{aligned}
\end{equation}

Substitution of these conditions into the model (\ref{eq:dk}) yields
\begin{equation}
    \begin{aligned}
        \dot{x} &= -\frac{\beta^2}{\beta^2+b^2\tan(\phi)^2} \sin(\theta) \tilde{u}_y \\
        \dot{y} &= \frac{\beta^2}{\beta^2+b^2\tan(\phi)^2} \cos(\theta)  \tilde{u}_y \\
        \dot{\theta} &= \frac{b\tan(\phi)}{\beta^2+b^2\tan(\phi)^2} \tilde{u}_y \\
        \dot{\phi} &= \frac{\cos(\phi)^2}{b} \tilde{u}_x - \frac{\cos(\phi)^2}{b}(b+r_p)\frac{b \tan(\phi) }{\beta^2+b^2\tan(\phi)^2}\tilde{u}_y
    \end{aligned}
\end{equation}

We can compare this model with earlier work where the contact point was described using its distance, $d$, relative to the symmetry axis of the slider \cite{lefebvre2023differential}. It follows that
\begin{equation}
    \begin{aligned}
        d &= b \tan(\phi) \\
        \dot{d} &= b \frac{1}{\cos(\phi)^2} \dot{\phi}
    \end{aligned}
\end{equation}

One then easily verifies that the following alternative model representation can be retrieved, as documented earlier.
\begin{equation}
    \dot{d} = \tilde{u}_x - (b+r_p) \frac{d}{\beta^2 + d^2} \tilde{u}_y 
\end{equation}

In conclusion for uniform pressure distributions we have that 
\begin{equation}
    \beta^2 = \tfrac{1}{3}\sqrt{a^2+b^2}
\end{equation}

\subsubsection{Circular slider} The circular geometry is characterized by a fixed radius, $r_s$. We further have that $\alpha = 0$.
\begin{equation}
    \begin{aligned}
        r(\phi) &= r_s \\
        r'(\phi) &= 0
    \end{aligned}
\end{equation}

Substituting these expressions into model (\ref{eq:dk}) now yields 
\begin{equation}
    \label{eq:ciruclardk}
    \begin{aligned}
        \dot{x} &= - \sin(\phi+\theta) \tilde{\tilde{u}}_y \\
        \dot{y} &=   \cos(\phi+\theta) \tilde{\tilde{u}}_y \\
        \dot{\theta} &= 0 \\
        \dot{\phi} &= \tfrac{1}{r_s+r_p}  \tilde{\tilde{u}}_x
    \end{aligned}
\end{equation}

We know of no earlier studies that describe this model in particular, however, it is easily verified to satisfy physical intuition. Further, note that as a result of the assumed lack of slider-pusher friction the slider's orientation is unaffected by the motion. Without loss of generality for circular sliders we can assume that $\theta = 0$, further simplifying the model. This also means that the distribution of pressures is irrelevant explaining the absence of the parameter $\beta^2$. 

\subsection{Transition and point contact dynamics}
The example of rectangular sliders invites us to extend our model to consider general slider geometries with non-smooth circumferences.

A slider with non-smooth circumference is characterised by a piecewise continuous radius function
\begin{equation}
    r = \{r_i(\phi), \phi \in (\phi_i,\phi_{i+1})\}
\end{equation}
where $r_{i-1}(\phi_{i}) = r_{i}(\phi_{i})$.

At the vertices, $\{\phi=\phi_i\}$, the slider geometry changes discontinuously. This will affect the contact between the slider and the pusher. We can now distinguish between two contact regimes. The first contact regime exists in smooth contact and is described by (\ref{eq:dk}). The other regime is referred to as the \textit{point contact} regime. During the point contact regime, the pusher transitions from one smooth slider face to the next smooth slider face. Again we assume that the contact is maintained throughout. This regime is governed by an alternative set of dynamic equations. We will refer to these alternative dynamics as the transition dynamics.

The point contact regime is characterised by the observation that during the transition, the state variable, $\phi$, is constant. In particular, we have that $\phi=\phi_{i}$ when we transition from the $(i-1)$-th to the $i$-th face. Thus far we had assumed that the position of the pusher was determined uniquely by the location of the slider and the contact point. This only holds for smooth slider circumferences. For sliders with discontinuous circumference and when the pusher is transitioning in between faces, i.e. at the vertex, the pusher's position is no longer uniquely defined by the contact point. During the transition, the value of $\alpha$ is also no longer geometrically linked to the value of $\phi$. We can calculate the value of $\alpha$ just before loosing, $\underline{\alpha}_i$, and just after reestablishing, $\overline{\alpha}_{i}$, smooth contact, with subsequent faces with radii, $r_{i-1}$, and, $r_i$. It follows from (\ref{eq:alpha1}) that
\begin{equation}
        \label{eq:alpha2}
    \begin{aligned}
        \underline{\alpha}_i &=-\arctan \frac{r_{i-1}'(\phi_i)}{r_{i-1}(\phi_i)} \\
        \overline{\alpha}_i &=- \arctan \frac{r_{i}'(\phi_i)}{r_{i}(\phi_i)}
    \end{aligned}
\end{equation}	

\begin{figure}[t]
    \centering
    \includegraphics[width=.6\columnwidth]{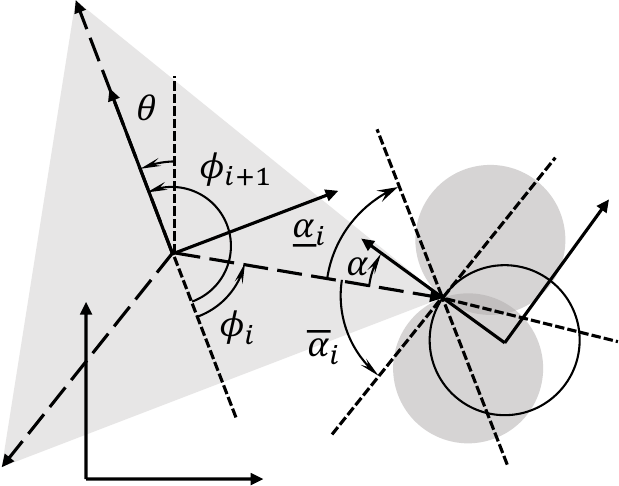}
    \caption{Geometry of the transition and point contact dynamics.}
\label{fig:transition}
\end{figure}

Now consider the diagram sketched in Fig. \ref{fig:transition}. We will discuss the figure based on the assumption that the pusher moves counterclockwise about the slider vertex. This is without loss of generality. In the figure, the grey-coloured pushers represent the location of the pusher just before losing and just after re-establishing contact. The transparent pusher represents the location of the pusher during strict point contact. The transition from one face to the next face is initiated with the angle $\phi(t)$ hitting the value $\phi_i$ and at that same instant $\alpha = \underline{\alpha}_i$. As soon as the point contact regime is initiated we have that $\phi(t) \equiv \phi_i$ so that $\dot{\phi}(t)\equiv 0$. During the point contact regime, the position of the pusher clearly can not be described by $\phi$. Instead, we can use $\alpha$. Equation (\ref{eq:alpha2}) implies that $\alpha$ changes smoothly from $\underline{\alpha}_i$ to $\overline{\alpha}_i$. As soon as $\alpha(t)$ hits $\overline{\alpha}_i$ the point contact changes to a smooth contact regime again.

%

To determine the dynamics during the point contact as a function of $\tilde{u}_x$ and $\tilde{u}_y$ we can repeat the analysis from section \ref{sec:quasi-static-model} keeping $\alpha$ variable, $\phi$ fixed and $r(\phi) = r_i$. This approach then gives rise to the following model
\begin{equation}
    \label{eq:point}
    \begin{aligned}
        \dot{x} &= -\frac{\beta^2}{\beta^2+r_i^2\sin(\alpha)^2}\sin(\theta+\phi_i+\alpha)\tilde{\tilde{\tilde{{u}}}}_y\\
        \dot{y} &= \frac{\beta^2}{\beta^2+r_i^2\sin(\alpha)^2}\cos(\theta+\phi_i+\alpha) \tilde{\tilde{\tilde{{u}}}}_y\\
        \dot{\theta} &= -\frac{r_i}{\beta^2+r_i^2\sin(\alpha)^2}\sin(\alpha) \tilde{\tilde{\tilde{{u}}}}_y \\
        \dot{\alpha} &= \frac{1}{r_p} \tilde{\tilde{\tilde{{u}}}}_x +  \frac{r_i}{r_p} \frac{r_p+r_i\cos(\alpha)}{\beta^2+r_i^2\sin(\alpha)^2} \sin(\alpha)\tilde{\tilde{\tilde{{u}}}}_y
    \end{aligned}
\end{equation}


\subsection{Full dynamics}
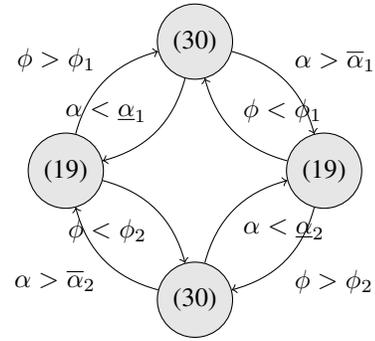
\begin{figure}[t]
    \centering
    \begin{tikzpicture}[
        roundnode/.style={circle,draw=black,fill=black!10,minimum size=10mm},
        scale=0.2]
        \node[roundnode] (1) {(\ref{eq:dk})};
        \node[roundnode] (2) [above right=of 1] {(\ref{eq:point})};
        \node[roundnode] (3) [below right=of 2] {(\ref{eq:dk})};
        \node[roundnode] (4) [below left=of 3] {(\ref{eq:point})};
        \draw[->] (1) edge [bend left=30] node [above left] {$\phi > \phi_1 $} (2);
        \draw[->] (2) edge [bend left=30] node [above right] {$\alpha > \overline{\alpha}_1 $} (3);
        \draw[->] (3) edge [bend left=30] node [below right] {$\phi > \phi_2 $} (4);
        \draw[->] (4) edge [bend left=30] node [below left] {$\alpha > \overline{\alpha}_2 $} (1);

        \draw[->] (2) edge [bend left=30] node [above left] {$\alpha < \underline{\alpha}_1$} (1);
        \draw[->] (3) edge [bend left=30] node [above right] {$\phi < \phi_1 $} (2);
        \draw[->] (4) edge [bend left=30] node [below right] {$\alpha < \underline{\alpha}_2 $} (3);
        \draw[->] (1) edge [bend left=30] node [below left] {$\phi < \phi_2 $} (4);
    \end{tikzpicture}
    \caption{Example of finite state machine describing the dynamics of a $2$-faced slider. The conditions for which a transition between either smooth or point dynamics is triggered by the events denoted above the edges. The nodes refer to the equations of motion corresponding the motion regime that is triggered by the events.}
    \label{fig:fsm}
\end{figure}

Considering the developed transition dynamics to describe the behaviour of the system at the vertex of non-smooth circumferences, we can describe the hybrid dynamics of any non-smooth slider with $n$-faces using a $2n$ bidirectional cyclic finite state machine. An example is given in Fig. \ref{fig:fsm}.

\section{Differential flatness}
In this section, we explore whether there exist deferentially flat quasi-static slider-pusher systems. 

Differential flatness is a system property that allows the representation of dynamically feasible trajectories of underactuated systems in a manner that is no longer subject to differential constraints \cite{rigatos2015differential,welde2021dynamically}. If a system is \textit{flat}, this denotes that all differentially dependent system variables (states and inputs) can be written in terms of a specific set of differentially independent variables and their first and higher derivatives \cite{rigatos2015differential}. Flatness is a resourceful property for both the analysis and controller synthesis of nonlinear dynamical systems. Systems that are known to be flat are various flying robots \cite{faessler2017differential,welde2023role}, gantry cranes \cite{fliess1995flatness}, cars with trailers, etc. but also fully actuated multi-body systems such as robotic manipulators.

The formal definition of differential flatness is given below \cite{fliess1995flatness}. 
\begin{definition}
The system, $\dot{\vectorstyle{x}}=\vectorstyle{f}(\vectorstyle{x},\vectorstyle{u})$, with state $\vectorstyle{x}\in\mathcal{X}\subset\mathbb{R}^{n_x}$ and input $\vectorstyle{u}\in\mathcal{U}\subset\mathbb{R}^{n_u}$, is differentially flat if there exists a variable $\boldsymbol{\zeta}\in\mathcal{Z}\subset\mathbb{R}^{n_u}$, whose components are differentially independent, and operators $\Lambda$, $\Phi$ and $\Psi$ such that the following holds \cite{fliess1995flatness}

\begin{equation}
    \begin{aligned}
        \boldsymbol{\zeta} &= \Lambda(\vectorstyle{x},\vectorstyle{u},\dot{\vectorstyle{u}},\dots)\\
        \vectorstyle{x} &= \Phi(\vectorstyle{\zeta},\dot{\vectorstyle{\zeta}},\ddot{\vectorstyle{\zeta}},\dots)\\
        \vectorstyle{u} &= \Psi(\vectorstyle{\zeta},\dot{\vectorstyle{\zeta}},\ddot{\vectorstyle{\zeta}},\dots)
    \end{aligned}
\end{equation}
\end{definition}



Unfortunately, there exists no straightforward procedure to verify whether a flat output exists, and thus whether the system is differentially flat, given a set of dynamic system equations. General necessary and sufficient conditions have been given but are difficult to apply to examples \cite{levine2011necessary}. Alternatively, studying the symmetries of the system has been shown to aid identification of candidate flat outputs \cite{welde2023role}. Nevertheless, it remains standard practice to verify each system and set of candidate output variables independently. 

\subsection{Centre of mass}\label{sec:flatness-condition} 
We initiate our search selecting the centre of mass as a first potential flat output. By manipulating the equations in (\ref{eq:dk}), we can now try and find expressions for the remaining states, $\theta$ and $\phi$, and the inputs, $\tilde{\tilde{\tilde{u}}}_x$ and $\tilde{\tilde{\tilde{u}}}_y$, as a function of the flat coordinate and its derivatives. 
\begin{equation}
\vectorstyle{\zeta} = \begin{pmatrix}
    x &
    y
\end{pmatrix}^\top
\end{equation}

From the first and second equations, it directly follows that 
\begin{equation}
\label{eq:theta+phi+alpha}
\theta+\phi+\alpha = -\arctan\left(\frac{\dot{x}}{\dot{y}}\right) 
\end{equation}

By differentiating this equation and comparing the result with the other equations we obtain
\begin{equation}
\label{eq:dtheta+dphi+dalpha}
\dot{\theta}+\dot{\phi}+\dot{\alpha} = \frac{\dot{x}\ddot{y}-\ddot{x}\dot{y}}{\dot{x}^2+\dot{y}^2} 
\end{equation}

Now we can also substitute the differential relation in (\ref{eq:alpha}) into the equation above. This operation yields the auxiliary result
\begin{equation}
\label{eq:dtheta1}
\dot{\theta} = \frac{\dot{x}\ddot{y}-\ddot{x}\dot{y}}{\dot{x}^2+\dot{y}^2} - \left(1+f(\phi)\right)\dot{\phi}
\end{equation}

Revisiting the equations in (\ref{eq:dk}), it can also be verified that
\begin{equation}
\label{eq:dtheta2}
\dot{\theta} = \sqrt{\dot{x}^2+\dot{y}^2}\frac{r(\phi) r’(\phi)\sqrt{r(\phi)^2+ r’(\phi)^2}}{\beta^2 r(\phi)^2 + \beta^2 r’(\phi)^2}
\end{equation}

We have arrived at two different equations, (\ref{eq:dtheta1}) and (\ref{eq:dtheta2}), that express $\dot{\theta}$ as a function of the potential flat output, $\vectorstyle{\zeta}$, and the yet expressionless state $\phi$. As such we could try to get rid of $\dot{\theta}$ and find an expression for $\phi$ as a function of the flat output. Unfortunately, the second expression for $\dot{\theta}$ also contains $\dot{\phi}$ introducing a differential dependency which remains to be resolved. However, it is impossible to eliminate $\dot{\phi}$ since this is the only variable that depends on $v$. This implies that we should be able to find an expression for, $\dot{\phi}$, independently to then find an expression for $v$.

The only way to eliminate $\dot{\phi}$ from the equation is by introducing an additional assumption. If the factor preceding $\dot{\phi}$ in equation (\ref{eq:dtheta1}) turns out to equal zero we can follow through with our derivation. This implies that $f(\phi)=-1$. The proposed assumption thus produces a second-order differential equation for the radius function, $r(\phi)$.
\begin{equation}
\label{eq:flat}
r^2+2 (r')^2-r'' r = 0
\end{equation}
This assumption also implies that $\dot{\alpha} = -\dot{\phi}$ and thus $\alpha = -\phi + \phi_0$. 

If this assumption is correct the following equality holds. The right-hand side contains a nonlinear function in $\phi$ that we can invert to find an expression for $\phi$ as a function of the first and second derivatives of the flat coordinate.
\begin{equation}
\begin{aligned}
    \frac{\dot{x}\ddot{y}-\ddot{x}\dot{y}}{\sqrt{\dot{x}^2+\dot{y}^2}^{3}} &= \frac{r(\phi) r’(\phi)\sqrt{r(\phi)^2+ r’(\phi)^2}}{\beta^2 r(\phi)^2 + \beta^2 r’(\phi)^2} = g(\phi) \\
    \phi &= g^{-1}\left(\frac{\dot{x}\ddot{y}-\ddot{x}\dot{y}}{\sqrt{\dot{x}^2+\dot{y}^2}^{3}}\right)
\end{aligned}
\end{equation}

Once we have obtained an expression for $\phi$ we invoke equation (\ref{eq:theta+phi+alpha}) to obtain an expression for $\theta$. 
\begin{equation}
\theta = -\arctan\left(\frac{\dot{x}}{\dot{y}}\right) - \phi +\arctan\left(\frac{r’(\phi)}{r(\phi)}\right)
\end{equation}

Finally, we can take the time derivative of $\theta$ and $\phi$ as a function of the flat coordinates and substitute these expressions back into (\ref{eq:dk}) to obtain expressions for $\tilde{\tilde{\tilde{u}}}_x$ and $\tilde{\tilde{\tilde{u}}}_y$.
\begin{equation}
\begin{aligned}
    \tilde{\tilde{\tilde{u}}}_y &= \frac{\beta^2 r(\phi)^2 + \beta^2 r’(\phi)^2 + r(\phi)^2 r’(\phi)^2}{r(\phi) r’(\phi)\sqrt{r(\phi)^2+ r’(\phi)^2}} \dot{\theta} \\
    \tilde{\tilde{\tilde{u}}}_x &= 					\sqrt{r(\phi)^2+ r’(\phi)^2}\dot{\phi}  + \frac{ r(\phi)^3 r’(\phi)}{\beta^2 r(\phi)^2 +\beta^2 r’(\phi)^2 + r(\phi)^2 r’(\phi)^2}\tilde{\tilde{\tilde{u}}}_y
\end{aligned}
\end{equation}

It appears that the differential equation in (\ref{eq:flat}) determines whether a slider's geometry is flat with respect to the slider's centre of mass. Before we investigate geometries that satisfy this equation, we consider other potential flat coordinates. We restrict our investigation to static points on the slider that may. This is a reasonable restriction from the point of view of potential applications.

\subsection{Other points on the sliders}
To investigate whether there are other points on the slider that are potentially flat, we can consider the arbitrary point, $\vectorstyle{p}'$, expressed in global coordinates
\begin{equation}
\vectorstyle{p}' = \vectorstyle{p} + \matrixstyle{R}(\theta) \vectorstyle{d}
\end{equation}
where $\vectorstyle{p}$ denotes the slider's centre of mass and arbitrary $\vectorstyle{d}\in\mathbb{R}^2$.

Taking the time derivative of the equation above and substituting the expressions in (\ref{eq:dk}), yields the reparametrized model
\begin{equation}
    \begin{aligned}
        \dot{x}' &= \dot{x} - (\sin(\theta)d_x+\cos(\theta)d_y) \dot{\theta} \\
        &= -\left(P(\phi) \sin(\theta+\phi+\alpha)+\Theta(\phi)(\sin(\theta)d_x+\cos(\theta)d_y)\right) \tilde{\tilde{\tilde{u}}}_y \\
        \dot{y}' &= \dot{y} + (\cos(\theta)d_x-\sin(\theta)d_y) \dot{\theta} \\
        &= \left(P(\phi) \cos(\theta+\phi+\alpha)+\Theta(\phi)(\cos(\theta)d_x-\sin(\theta)d_y)\right) \tilde{\tilde{\tilde{u}}}_y 
    \end{aligned}
\end{equation}

An attempt could be made to find values for $d_x$ and $d_y$ so that these equations would simplify. One idea could be to choose $d_x = A\cos(\gamma)$ and $d_y = A\sin(\gamma)$ so that $\sin(\theta)d_x+\cos(\theta)d_y = A \sin(\theta+\gamma)$ and $\cos(\theta)d_x-\sin(\theta)d_y)=A\cos(\theta+\gamma)$. However, since $A$ and $\gamma$ are static, whereas the other angles are dynamic, one verifies that this will not lead to another restriction on the geometry. It only proves that if $\alpha = -\phi + \phi_0$, any static point on the slider qualifies as a flat coordinate.

\begin{figure}[t]
\centering
\includegraphics[width=\columnwidth]{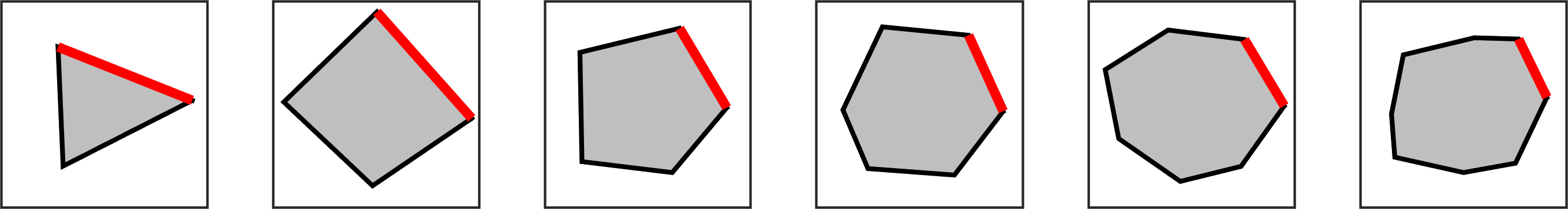}
\caption{Illustration of flat slider geometries.}
\label{fig:sliders}
\end{figure}

\subsection{Flat slider-pusher systems}\label{sec:flat-slider-pusher-systems} 
Now we can characterize any slider-pusher system that is differentially flat by finding a general expression for the explicit solution of (\ref{eq:flat})  The solution of (\ref{eq:flat}) is given by (Appendix \ref{sec:derivation-of-equation-refeqr})
\begin{equation}
\label{eq:r}
r(\phi) = A \frac{1}{\cos{(\phi-B)}}
\end{equation}
which we immediately recognize as parameterizing the polar coordinate expression of the rectangular slider.

Generally, the conditions (\ref{eq:r}) parameterizes all slider-pusher systems with polygon sliders. Some illustrative examples are given in Fig. \ref{fig:sliders}. We can then associate to each vertex an angle, $\phi_i$, and, radius, $r_i$. The entire slider's geometry is then determined by 
\begin{equation}
\label{eq:flatcircumference}
r(\phi) = \left\{r_{i,\phi} = A_i \frac{1}{\cos{(\phi-B_i)}},\phi\in(\phi_i,\phi_{i+1}),i\in{1,N+1}\right\}
\end{equation}
with $N$ the number of faces, $\phi_{N+1}=\phi_1$, $r_{N+1}=r_1$ and
\begin{equation}
\begin{aligned}
    A_i &=  \frac{r_i r_{i+1}}{\sqrt{r_{i}^2 + r_{i+1}^2 -2 r_{i}r_{i+1} \cos{(\Delta \phi_i)}}} \sin{(\Delta \phi_i)}\\
    B_i &= \phi_{i} + \arctan \frac{r_i- r_{i+1}\cos{(\Delta \phi_i)}}{r_{i+1} \sin{(\Delta \phi_i)}}
\end{aligned}
\end{equation}
where $\Delta \phi_i = \phi_{i+1}-\phi_i$.

According to the derivation strategy detailed above, we can derive the following flat expressions

\begin{equation}
\label{eq:invDyn}
\begin{aligned}
    \theta &= -\arctan\left(\frac{\dot{x}}{\dot{y}}\right)+B_i \\
    \phi &= \arctan\left(\frac{\beta^2}{A_i}\frac{\dot{x}\ddot{y}-\ddot{x}\dot{y}}{\sqrt{\dot{x}^2+\dot{y}^2}^{3}}\right)-B_i \\
    \tilde{u}_x &= (A_i+r_p)\frac{\dot{x}\ddot{y}-\ddot{x}\dot{y}}{\dot{x}^2+\dot{y}^2}+\beta^2\frac{\dot{x}\dddot{y} - \dddot{x}\dot{y}}{\sqrt{\dot{x}^2+\dot{y}^2}^{3}} + 3\beta^2 \frac{\left(\ddot{x}\dot{y}-\dot{x}\ddot{y}\right)\left(\dot{x}\ddot{x}+\dot{y}\ddot{y}\right)}{\sqrt{\dot{x}^2+\dot{y}^2}^{5}} \\
    \tilde{u}_y &=  \left(1+\beta^2\frac{\left(\dot{x}\ddot{y}-\ddot{x}\dot{y}\right)^2}{\left(\dot{x}^2+\dot{y}^2\right)^3}\right)\sqrt{\dot{x}^2+\dot{y}^2} 
\end{aligned}
\end{equation}

Let us now again consider the (1) rectangular and (2) circular slider geometries.

\subsubsection{Rectangular slider} 
For rectangular sliders, we limit our discussion to the bottom face. The other faces are simply permutations of this solution. In this case, we have that $B=0$ and $A=b$. If we use the relative distance to parameterize the contact point rather than the contact angle the flat expression becomes
\begin{equation}
\begin{aligned}
d &= \beta^2\frac{\dot{x}\ddot{y}-\ddot{x}\dot{y}}{\sqrt{\dot{x}^2+\dot{y}^2}^3} 
\end{aligned}
\end{equation}

\subsubsection{Circular slider} 
We may also revise the differential kinematics of the circular slider. In the limit of infinite vertices, the regular polygon coincides with the circle. So we would expect the circular geometry to exhibit a flat structure too. However, it appears this does not emerge from the expressions above. Therefore we must recognize that the expressions above only hold for $\phi(t) \in (\phi_i,\phi_{i+1})$. At the vertices, i.e. $\phi(t)=\phi_i$, the slider geometry changes discontinuously. To describe the situation when the pusher transitions from one slider face to the following slider face, we must consider an alternative set of dynamic equations. We will refer to these dynamics as the switching dynamics. The transition from the main dynamics in (\ref{eq:invDyn}) to the switching dynamics will introduce discontinuities in the model. 

Nevertheless, for circular sliders, the discontinuous should vanish again and the switching happens instantaneously. Thus to analyze the situation of the circular slider we may directly consider the differential kinematics in (\ref{eq:ciruclardk}). One easily verifies that 
\begin{equation}
\begin{aligned}
\phi &= -\arctan \left(\frac{\dot{x}}{\dot{y}}\right) \\
\tilde{u}_x &= (r_s+r_p) \frac{\dot{x}\ddot{y}-\ddot{x}\dot{y}}{\dot{x}^2+\dot{y}^2} \\
\tilde{u}_y &= \sqrt{\dot{x}^2+\dot{y}^2}
\end{aligned}
\end{equation}


\section{Tracking}
As mentioned before, flatness is a useful property for control synthesis. It is particularly advantageous to solve path planning problems and is used abundantly in tracking control \cite{diwold2021trajectory,stoican2015flat,stoical2016obstacle,stoican2017constrained,lefebvre2023differential,greeff2018flatness,helling2020flatness,faessler2017differential,aguilar2012trajectory,agrawal2021constructive,greeff2020exploiting,rudolph2021flatness,delaleau1998control,hagenmeyer2003exact}. Flatness-based trajectory optimization applied to slider pusher was investigated by \cite{lefebvre2023differential,dewitte2024reactive,neve2024trajectory}.

This section discusses two closed-loop tracking strategies tailored specifically to flat slider-pusher systems. First, we will discuss an ad-hoc cascaded (quasi-)static feedback strategy. Second, since the system that we will address is flat and therefore provably dynamic feedback linearizable \cite{vannieuwstadt1994equivalence,levine2007equivalence,levine2011necessary}, we also develop a dynamic feedback linearization strategy.

Given the developments in previous sections, we can work with the following rectangular slider model henceforth
\begin{equation}
    \label{eq:model}
    \begin{aligned}
        \dot{x} &= -\frac{\beta^2}{\beta^2+d^2} \sin(\theta) \tilde{u}_y \\
        \dot{y} &= \frac{\beta^2}{\beta^2+d^2} \cos(\theta) \tilde{u}_y \\
        \dot{\theta} &= \frac{d}{\beta^2+d^2} \tilde{u}_y  \\
        \dot{d} &= \tilde{u}_x -(b+r_p) \frac{d}{\beta^2+d^2}\tilde{u}_y  
    \end{aligned}
\end{equation}

Remark that we can also write this model in general state-space form
\begin{equation}
        \label{eq:ss}
    \dot{\vectorstyle{\xi}} = \vectorstyle{f}(\vectorstyle{\xi}) + \vectorstyle{g}(\vectorstyle{\xi})\vectorstyle{\upsilon}
\end{equation}
with state, $\vectorstyle{\xi} = (x,y,\theta,d)$, and, input, $\vectorstyle{\upsilon} = (\tilde{u}_x,\tilde{u}_y)$.

\subsection{Control objective}
The slider-pusher system is under-actuated. Therefore we will only pursue tracking of a desired position signal, given by $x_d$ and $y_d$. The position signal may be static or dynamic. In the latter case, we also assume to have access to every time derivative of the desired position signal. Every control strategy assumes access to a measurement of the slider-pusher state, $\vectorstyle{\xi}$.

\subsection{Cascaded (quasi-)static feedback strategy}
The first control strategy is ad-hoc and based on the following observation. Pushing the slider in the direction of its current heading is straightforward. If the current heading does not align with the direction of the desired position, the slider must be rotated. However, to affect the rotation of the slider we must first change the pusher offset. Remark that we cannot directly establish all of these conditions with only two control inputs. However, we can establish these subgoals one by one, carefully tuning the dominant time scale of every feedback loop responsible for establishing the subgoal. Put differently, we will make use of a cascaded control architecture and use the control model to determine reference signals for the internal feedback loops. 

First, assume we can directly control the velocity. Then we can calculate the required velocities $\dot{x}_r$ and $\dot{y}_r$ by imposing an exponential error decay law with feedforward of the desired velocities, $\dot{x}_d$, and $\dot{y}_d$. This loop determines the outer loop or the first stage in the cascade.
\begin{equation}
    \begin{aligned}
        \dot{x}_r &= \dot{x}_d + K_{p,x} (x_d - x) \\
        \dot{y}_r &= \dot{y}_d + K_{p,y} (y_d - y) 
    \end{aligned}
\end{equation}

Now we must try to realise those required velocities. To that end, we can invert the first two equations of the model, (\ref{eq:model}). This allows us to calculate the required reference normal velocity $\tilde{u}_{y,r}$ and a reference orientation $\theta_r$, and this for given offset $d$. 
\begin{equation}
    \begin{aligned}
        \tilde{u}_{y,r} &= \frac{\beta^2+d^2}{\beta^2} \sqrt{\dot{x}_r^2+\dot{y}_r^2} \\
        \theta_r &= -\arctan \frac{\dot{x}_r}{\dot{y}_r}
    \end{aligned}
\end{equation}
Based on the desired reference orientation, we can calculate the required angular velocity by imposing another exponential error decay law. This loop determines the second stage in the cascade. 
\begin{equation}
    \dot{\theta}_r = K_{p,\theta} (\theta_r - \theta) 
\end{equation}

Finally, we have to realise the required angular velocity. To that end, we can invert the third equation of the model and determine the required pusher offset (the value is chosen that is closest to $d$). 
\begin{equation}
 d_r = \frac{-\tilde{u}_{y,r}\pm\sqrt{\tilde{u}_{y,r}^2-4\dot{\theta}_r^2\beta^2}}{2\dot{\theta}_r}
\end{equation}
Based on the required offset, $d_r$, we can now calculate a required offset velocity $\dot{d}_r$, by imposing a final exponential error decay law. This loop marks the inner loop or the third and final stage of the cascade.
\begin{equation}
    \begin{aligned}
        \dot{d}_r = K_{p,d} (d_r - d) 
    \end{aligned}
\end{equation}

From the final required offset velocity, we can calculate the required value for the tangential velocity $\tilde{u}_{x,r}$. This is done by inverting the fourth model equation.
\begin{equation}
    \tilde{u}_{x,r} = \dot{d}_r + \alpha \frac{d}{\beta^2+d^2} \tilde{u}_{y,r} 
\end{equation}
The required inputs $\tilde{u}_{x,r}$ and $\tilde{u}_{y,r}$ are then applied to the system.

We can further extend this approach by imposing second-order error decay laws instead of a first-order error decay law. We give an example for the $x$ position, the same is done for the other laws\footnote{Note that in the orientation and offset loops, we do not make use of feedforward signals. In principle we could make use of the flat expressions to determine desired signals for these loops too, though given that it is unclear how to reconcile e.g. $\theta_r$, provided $\theta_d$, we only use feedforward in the outer loop.}.
\begin{equation}
    \ddot{x}_r = \ddot{x}_d + K_{d,x} (\dot{x}_d - \dot{x}_r) + K_{p,x} (x_d -x)
\end{equation}
Remark that $\dot{x}_r$ is now part of the feedback law. The variable is calculated by integrating $\ddot{x}_r$. Otherwise, it is used in the same way as in the first-order controller. As a result, the value of $\dot{x}_r$ is now part of the controller's internal dynamics.

The gains, $\{K_{d,i}\}$ and $\{K_{p,i}\}$, can be parametrised by the desired characteristic time constant of the second-order error dynamics. By construction, the dominant time scales should decrease with the cascade stage. Careful tuning will be required to address the different time scales such that all loops interact properly.
\begin{equation}
    \begin{aligned}
        K_{p,i} &= \frac{1}{\tau_i^2} \\
        K_{d,i} &= \frac{2}{\tau_i} 
    \end{aligned}
\end{equation}

\subsection{Dynamic feedback linearization strategy} Our second design strategy follows from the simple fact that any flat system is also dynamic feedback linearizable \cite{levine2007equivalence}. 

\begin{theorem} \textit{System (\ref{eq:ss}) is dynamic feedback linearizable if and only if it is differentially flat. }
\end{theorem}

\begin{definition}[Dynamic feedback linearization \cite{lee2022linearization}]
    \textit{System (\ref{eq:ss}) is dynamic feedback linearizable, if there exist auxiliary states, $\vectorstyle{\gamma}\in\mathcal{R}^{n_\gamma}$; a dynamic feedback, with $\vectorstyle{\nu} \in \mathbb{R}^{n_\upsilon}$
    \begin{equation*}\begin{aligned}
            \dot{\vectorstyle{\gamma}} &= \vectorstyle{a}(\vectorstyle{\xi},\vectorstyle{\gamma}) + \vectorstyle{b}(\vectorstyle{\xi},\vectorstyle{\gamma}) \vectorstyle{\nu} \\
            \vectorstyle{\upsilon} &= \vectorstyle{\alpha}(\vectorstyle{\xi},\vectorstyle{\gamma}) + \vectorstyle{\beta}(\vectorstyle{\xi},\vectorstyle{\gamma}) \vectorstyle{\nu}
        \end{aligned}
    \end{equation*}
    and an extended state transformation, $\vectorstyle{\chi} = \vectorstyle{\eta}(\vectorstyle{\xi},\vectorstyle{\gamma})$ such that $\vectorstyle{\chi}\in\mathbb{R}^{n_\xi + n_\gamma}$ and the extended system 
    \begin{equation*}
        \begin{aligned}
            \dot{\vectorstyle{\xi}} &= \vectorstyle{f}(\vectorstyle{\xi}) + \vectorstyle{g}(\vectorstyle{\xi}) \vectorstyle{\alpha}(\vectorstyle{\xi},\vectorstyle{\gamma}) + \vectorstyle{g}(\vectorstyle{\xi}) \vectorstyle{\beta}(\vectorstyle{\xi},\vectorstyle{\gamma}) \vectorstyle{\nu} \\
            \dot{\vectorstyle{\gamma}} &= \vectorstyle{a}(\vectorstyle{\xi},\vectorstyle{\gamma}) + \vectorstyle{b}(\vectorstyle{\xi},\vectorstyle{\gamma}) \vectorstyle{\nu}
        \end{aligned}
    \end{equation*}
    satisfies, for the transformed extended state, $\vectorstyle{\chi}$
    \begin{equation*}
        \dot{\vectorstyle{\chi}} = \matrixstyle{A} \vectorstyle{\chi} + \matrixstyle{B}\vectorstyle{\nu}
    \end{equation*}
    where the linear system $(\matrixstyle{A},\matrixstyle{B})$, is in Brunovsk\'{y} form.}
\end{definition}

When a system is (dynamic) feedback linearizable, it is straightforward to design a tracking controller, $\vectorstyle{\upsilon}(t,\vectorstyle{\xi}(t),\vectorstyle{\gamma}(t))$, that is asymptotically stable towards the reference, $(\vectorstyle{\xi}_d(t),\vectorstyle{\gamma}_d(t))$, in the extended state space. This is achieved by designing a tracking controller, $\vectorstyle{\nu}(t,\vectorstyle{\chi}(t))$, that is asymptotically stable towards the reference, $\vectorstyle{\chi}_d(t)$, in the transformed extended state space, and then apply the inverse transform to reconstruct the controller in the original state-space. Designing an asymptotically stable tracking controller in the linear state-space is trivial. The principle is demonstrated with linear feedback. 
\begin{equation}
    \begin{aligned}
        \vectorstyle{\chi}(t) &=  \vectorstyle{\eta}(\vectorstyle{\xi}(t),\vectorstyle{\gamma}(t)) \\
        \vectorstyle{\nu}(t,\vectorstyle{\chi}(t)) &= \vectorstyle{\nu}_d(t) + \matrixstyle{K}(\vectorstyle{\chi}_d(t) - \vectorstyle{\chi}(t)) \\
        \vectorstyle{\upsilon}(t,\vectorstyle{\xi}(t),\vectorstyle{\gamma}(t)) &= \vectorstyle{\alpha}(\vectorstyle{\xi}(t),\vectorstyle{\gamma}(t)) + \vectorstyle{\beta}(\vectorstyle{\xi}(t),\vectorstyle{\gamma}(t)) \vectorstyle{\nu}(t,\vectorstyle{\chi}(t)) 
    \end{aligned}
\end{equation}
The gain matrix $\matrixstyle{K}$ can be determined e.g. through LQR design.

These results guarantee the existence of a dynamic feedback linearization strategy, however, it does not specify how to derive it. The difficulty lies in the choice of the auxiliary states, $\vectorstyle{\gamma}$. There is no standard procedure to determine suitable variables.

First note that the transformed extended state, $\vectorstyle{\chi}$, will contain the flat coordinate and higher order derivatives. Then, given that the transformed extended system is in Brunovsk\'{y} form, $\vectorstyle{\nu}$ contains derivatives of one order higher than those in $\vectorstyle{\chi}$. Thirdly, since the flat expressions for $\tilde{u}_x$ and $\tilde{u}_y$ contain derivatives of the flat coordinate of at most degree three, it follows that $\vectorstyle{\chi} = (x,y,\dot{x},\dot{y},\ddot{x},\ddot{y})$ and $\vectorstyle{\nu} = (\dddot{x},\dddot{y})$. Now since there must exist a transformation from $(\vectorstyle{\xi},\vectorstyle{\gamma})$ and $\vectorstyle{\chi}$, and $\vectorstyle{\xi}\in\mathbb{R}^4$ and $\vectorstyle{\chi}\in\mathbb{R}^6$, it follows that $\vectorstyle{\gamma}\in\mathbb{R}^2$. Finally, by definition $\vectorstyle{\gamma}$ needs to be a function of $\vectorstyle{\chi}$. 

Now we must choose. 
\begin{itemize}
    \item Our first choice determines the translational velocity amplitude and the absolute angle of the acceleration vector.
    \begin{equation}
        \begin{aligned}
            \gamma_1 &= \sqrt{\dot{x}^2+\dot{y}^2} \\
            \gamma_2 &= -\arctan \frac{\ddot{x}}{\ddot{y}}
        \end{aligned}
    \end{equation}
    
    Based on this definition and the flat expression for the offset, $d$, the transformation from $(\vectorstyle{\xi},\vectorstyle{\gamma})$ to $\vectorstyle{\chi}$ is given by
    \begin{equation}
        \begin{aligned}
            \dot{x} &= -\gamma_1 s_\theta \\
            \dot{y} &= \gamma_1 c_\theta \\
            \ddot{x} &= -\frac{\gamma_1^2}{\beta^2}\frac{d}{\sin{(\gamma_2-\theta)}} \sin{(\gamma_2)} \\
            \ddot{y} &= \frac{\gamma_1^2}{\beta^2}\frac{d}{ \sin{(\gamma_2-\theta)}} \cos{(\gamma_2)} 
        \end{aligned}
    \end{equation}

    To auxiliary variables' dynamics are governed by
    \begin{equation}
        \begin{aligned}
            \dot{\gamma}_1 &= \frac{1}{\gamma_1} \left(\dot{x}\ddot{x}+\dot{y}\ddot{y}\right) \\
            \dot{\gamma}_2 &= \frac{\dddot{y}\ddot{x}-\dddot{x}\ddot{y}}{\ddot{x}^2+\ddot{y}^2} 
        \end{aligned}
    \end{equation}

    Remark that this choice has a singularity when $\gamma_2 = \theta$ and when $d = 0$ since then $\ddot{x} = \ddot{y} = 0$ causing $\dot{\gamma}_2\rightarrow\pm\infty$.

    \item The singularity of the former choice is a direct result of the fact that when $d=0$, the acceleration angle is ill-defined. To accommodate this issue we replace the second auxiliary variable with acceleration amplitude in the local $y$-axis.
    \begin{equation}
        \begin{aligned}
            \gamma_1 &= \sqrt{\dot{x}^2+\dot{y}^2} \\
            \gamma_2 &= \ddot{y}\cos(\theta) - \ddot{x}\sin(\theta)
        \end{aligned}
    \end{equation}

    The transformation from $(\vectorstyle{\xi},\vectorstyle{\gamma})$ to $\vectorstyle{\chi}$ is now given by
    \begin{equation}
        \begin{aligned}
            \dot{x} &= -\gamma_1 s_\theta \\
            \dot{y} &= \gamma_1 c_\theta \\
            \ddot{x} &= -\gamma_2 \sin(\theta) - \frac{\gamma_1^2}{\beta^2} d \cos(\theta) \\
            \ddot{y} &= \gamma_2 \cos(\theta) - \frac{\gamma_1^2}{\beta^2} d \sin(\theta) 
        \end{aligned}
    \end{equation}

    To auxiliary variables' dynamics are now governed by the following equations
    \begin{equation}
        \begin{aligned}
            \dot{\gamma}_1 &= \gamma_2 \\
            \dot{\gamma}_2 &= \dddot{y}\cos(\theta) - \dddot{x}\sin(\theta) - \ddot{y}\sin(\theta)\dot{\theta} - \ddot{x}\cos(\theta)\dot{\theta}
        \end{aligned}
    \end{equation}

    The benefit of this choice is that it is singularity-free except when $\sqrt{\dot{x}^2+\dot{y}^2} = \gamma_1 = 0$ since then $\dot{\theta}\rightarrow \pm\infty$. However, this singularity is easier to avoid than the singularity at $d=0$.
\end{itemize}


For either choice, the control strategy is completed by designing an asymptotically stabilizing controller in the transformed extended state space. Specifically, we define the tracking controller as
\begin{equation}
    \begin{aligned}
        \nu_x = \dddot{x} & = \dddot{x}_d + K_{2} (\ddot{x}_d-\ddot{x}) + K_{1} (\dot{X}_d-\dot{x}) + K_{0} (x_d - x) \\
        \nu_y = \dddot{y} &=  \dddot{y}_d + K_{2} (\ddot{y}_d-\ddot{y}) + K_{1} (\dot{y}_d-\dot{y}) + K_{0} (y_d - y) 
    \end{aligned}
\end{equation}
The gains, $\matrixstyle{K}$ are chosen such that the closed-loop system is asymptotically stable, meaning that the matrix \(\matrixstyle{A}+\matrixstyle{B}\matrixstyle{K}\) is Hurwitz. In our formulation, the matrices are given by
\begin{equation}
    \begin{aligned}
        \matrixstyle{A} &= \begin{bmatrix}
            0 & 1 & 0 \\
            0 & 0 & 1 \\
            0 & 0 & 0
        \end{bmatrix}, \qquad
        \matrixstyle{B}^\top = \begin{bmatrix}
            0 & 0 & 1
        \end{bmatrix}, \\
        \matrixstyle{K} &= \begin{bmatrix}
            K_0 & K_1 & K_2
        \end{bmatrix}.
    \end{aligned}
\end{equation}
Tuning \(\matrixstyle{K}\) is more straightforward than for the cascaded feedback strategy, as there is no need to carefully coordinate the dominant time scales of different loops.

The gains can be set using straightforward control methods, such as the Linear Quadratic Regulator (LQR) approach. In LQR, the optimal state feedback gain matrix is determined by minimizing a quadratic cost function
\begin{equation}
    J = \int_0^\infty \left( \vectorstyle{x}^\top \matrixstyle{Q} \vectorstyle{x} + \vectorstyle{u}^\top \matrixstyle{R} \vectorstyle{u} \right) dt,
\end{equation}
subject to the system dynamics. Here, $\matrixstyle{Q}$ is the state weighting matrix, which penalizes deviations in the system states, and $\matrixstyle{R}$ is the control weighting matrix, which penalizes excessive control effort.

The optimal gain matrix is then given by
\begin{equation}
    \matrixstyle{K} = \matrixstyle{R}^{-1}\matrixstyle{B}^\top\matrixstyle{P},
\end{equation}
where \(\matrixstyle{P}\) is the unique positive-definite solution to the continuous-time Algebraic Riccati Equation
\begin{equation}
    \matrixstyle{Q} + \matrixstyle{A}^\top\matrixstyle{P} + \matrixstyle{P}\matrixstyle{A} - \matrixstyle{P}\matrixstyle{B}\matrixstyle{R}^{-1}\matrixstyle{B}^\top\matrixstyle{P} = 0.
\end{equation}

\section{Experiments}
We start this section by verifying the quasi-static model found for arbitrary slider geometry and polygon slider geometry with face transition. Next, we apply the two control strategies to a rectangular slider both in simulation and on a physical set-up.

\subsection{Model verification}
\subsubsection{Arbitrary geometry} 
\begin{figure}
    \begin{subfigure}[b]{0.32\linewidth}
        \centering
        \includegraphics[width=.99\textwidth]{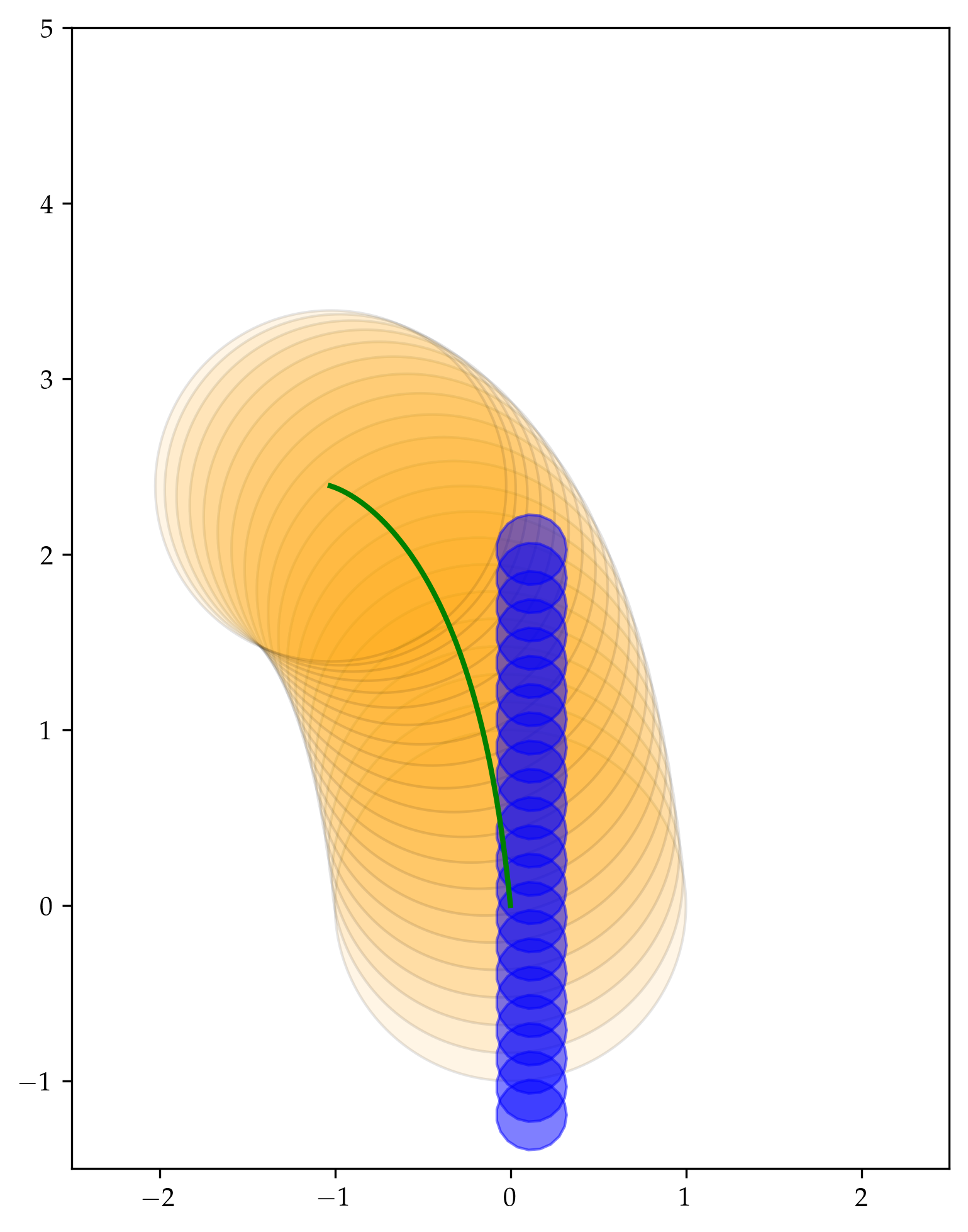}
        \caption{}
    \end{subfigure}
    \begin{subfigure}[b]{0.32\linewidth}
        \centering
        \includegraphics[width=.99\textwidth]{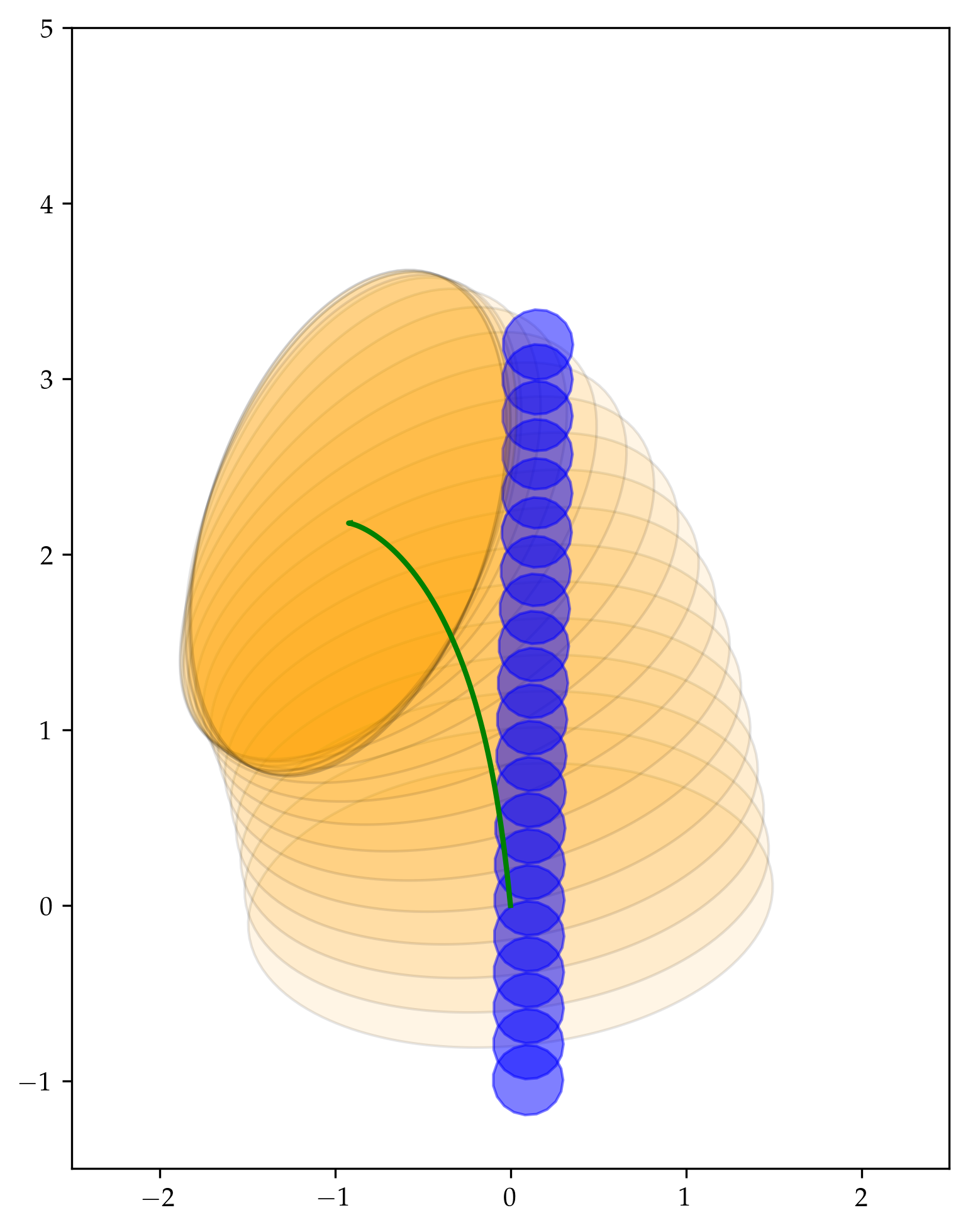}
        \caption{}
    \end{subfigure}
    \begin{subfigure}[b]{0.32\linewidth}
        \centering
        \includegraphics[width=.99\textwidth]{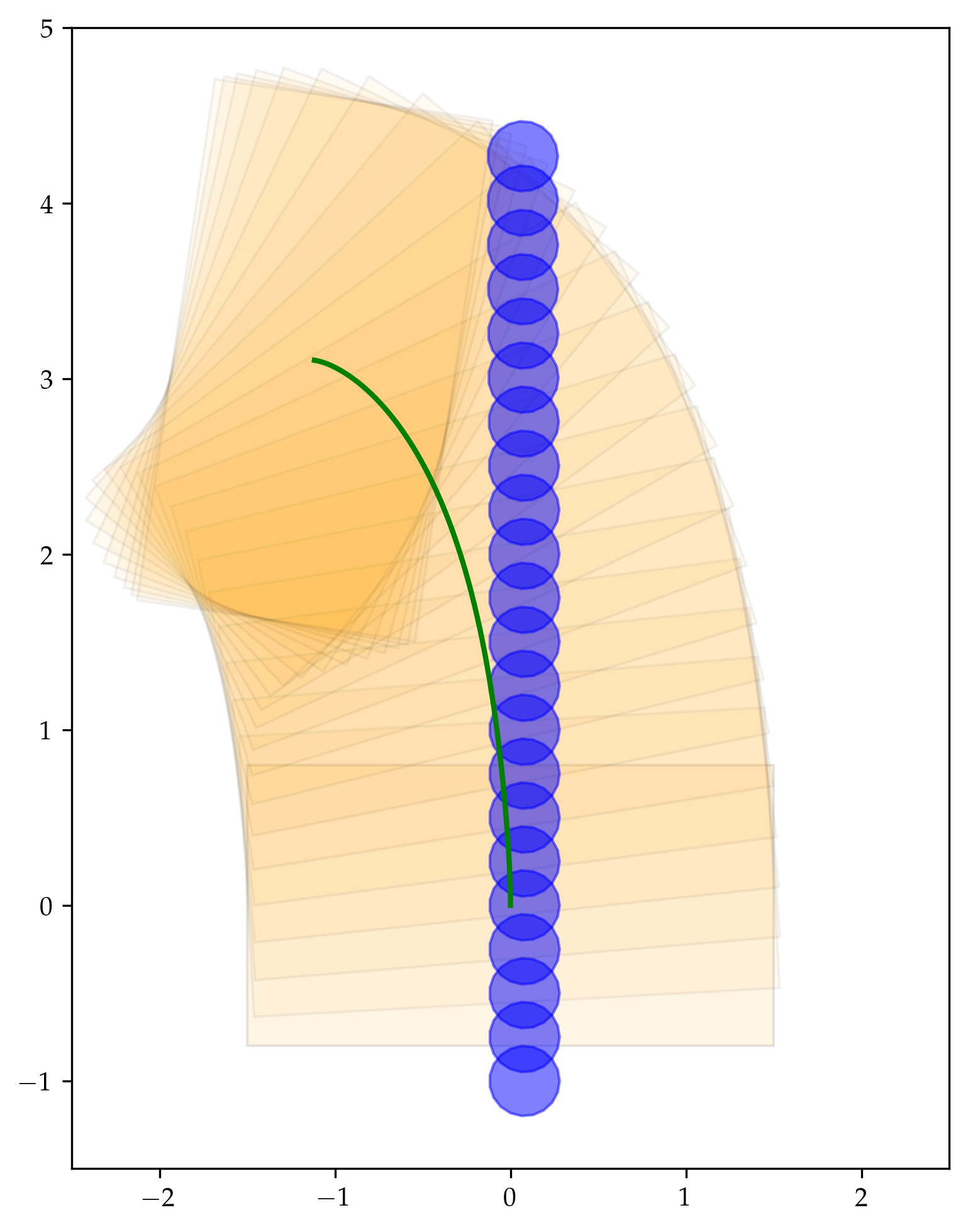}
        \caption{}
    \end{subfigure}
    \caption{A forward push of three slider shapes, using the full dynamics.}
    \label{fig:push_arbitrary}
\end{figure}

We verify the model by applying a forward push to three shapes, a circular, an elliptical and a rectangular slider, as shown in Fig. \ref{fig:push_arbitrary}. A global input speed is defined, with $u_y$ non-zero and $u_x$ zero. The local input vector can be found each time step using equation (\ref{eq:input}).

The radius, $r(\phi)$, for the rectangular and circular slider are as described previously, while we have the following formula for the elliptical slider:
\begin{equation}
    r(\phi) = \frac{ab}{\sqrt{(b\cos{\phi})^2+(a\sin{\phi})^2}}
\end{equation}

\begin{figure*}[th]
    \begin{subfigure}[b]{0.32\linewidth}
        \centering
        \includegraphics[width=.99\textwidth]{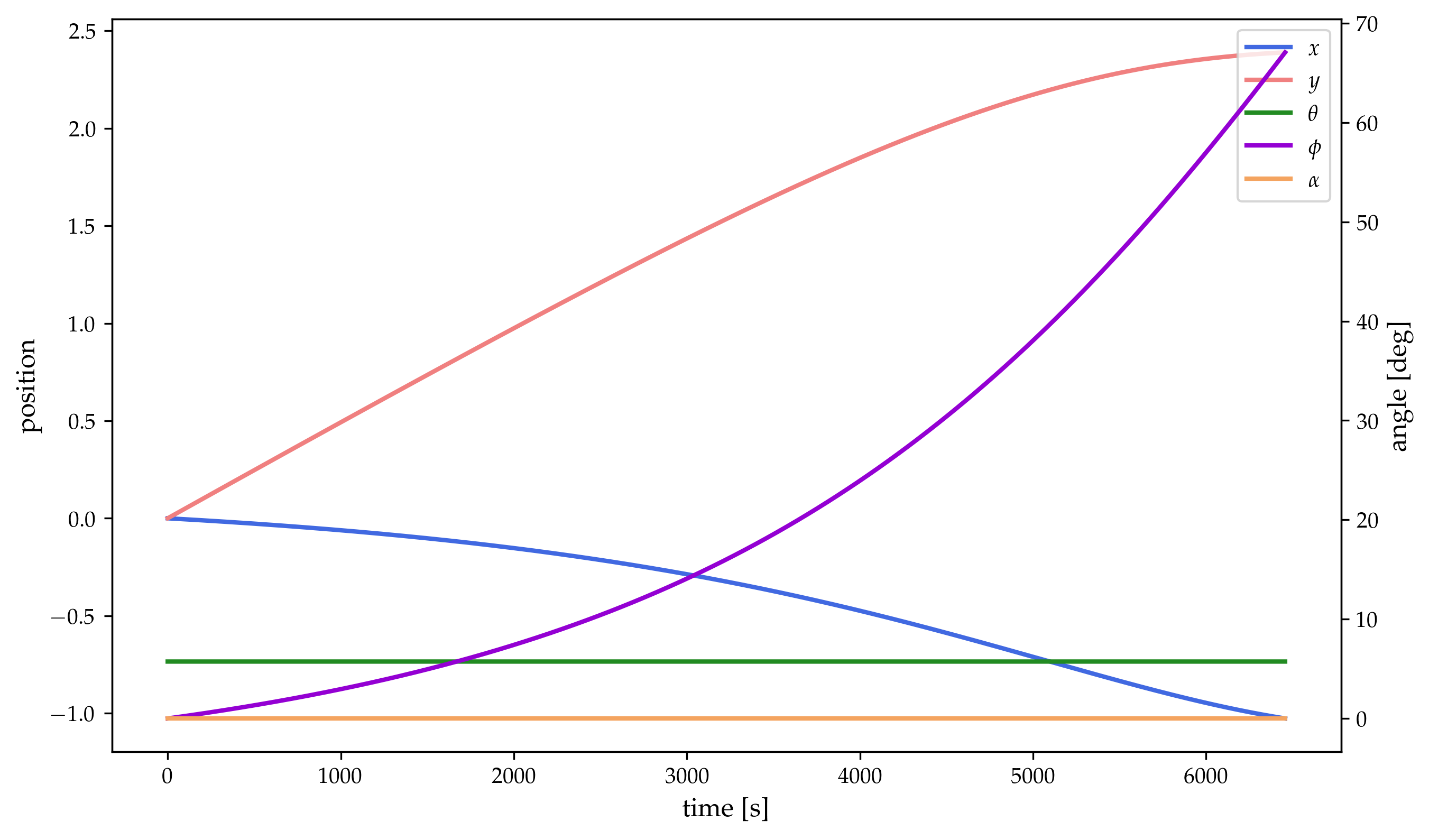}
        \caption{Circular slider}
    \end{subfigure}
    \begin{subfigure}[b]{0.32\linewidth}
        \centering
        \includegraphics[width=.99\textwidth]{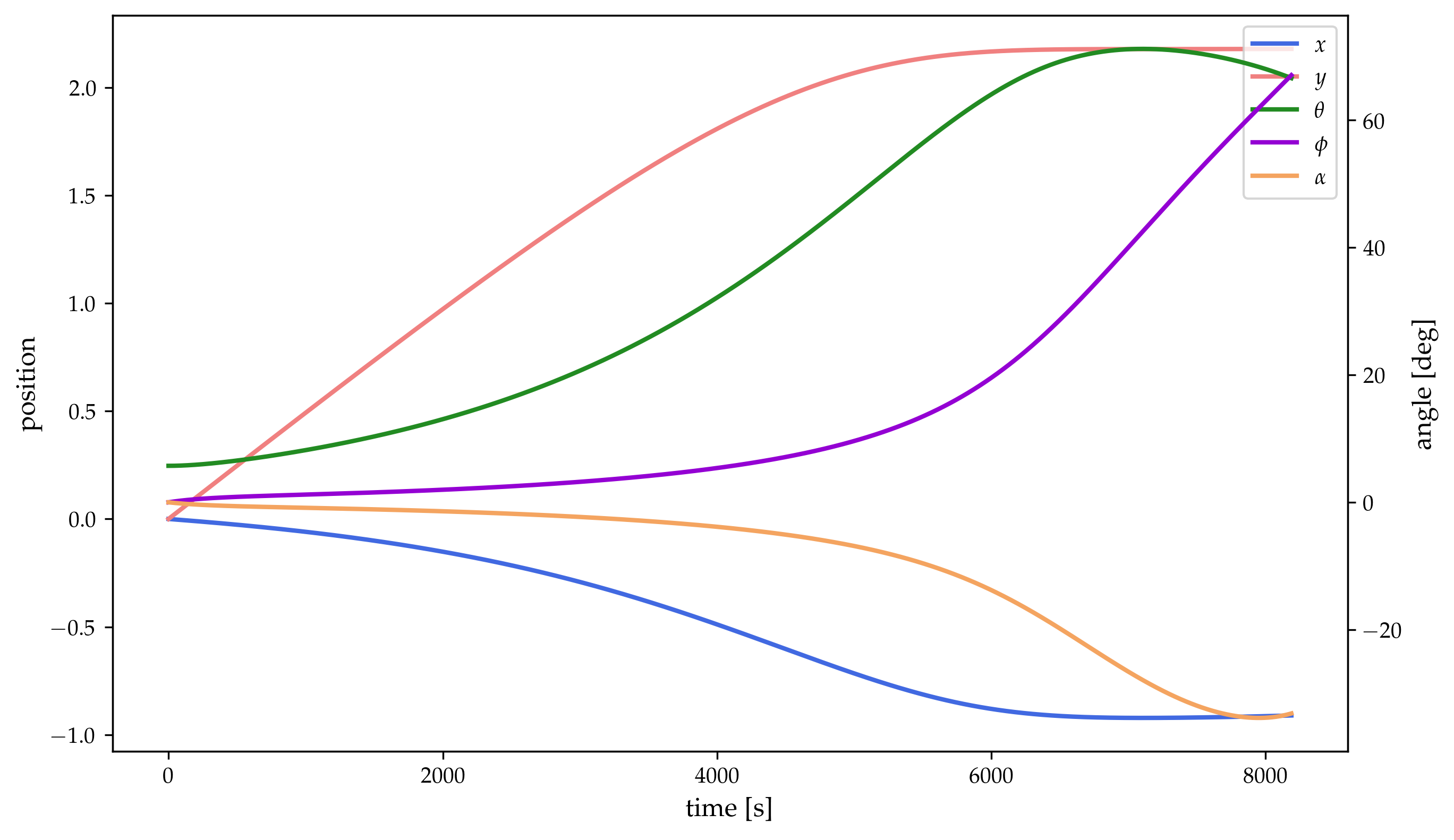}
        \caption{Elliptical slider}
    \end{subfigure}
    \begin{subfigure}[b]{0.32\linewidth}
        \centering
        \includegraphics[width=.99\textwidth]{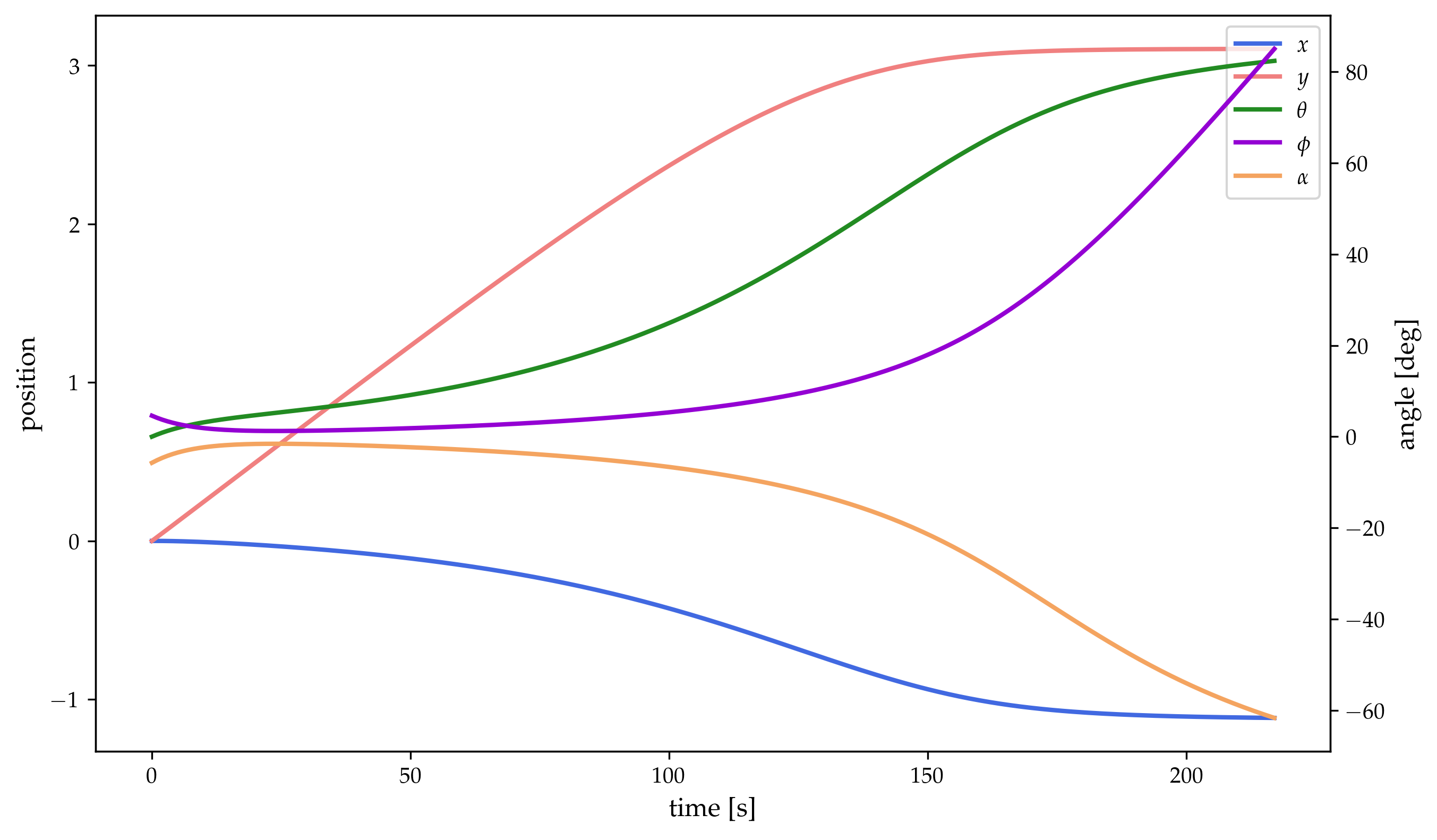}
        \caption{Rectangular slider }
    \end{subfigure}
    \caption{State variables for a forward push of three slider shapes, using the full dynamics.}
    \label{fig:push_arbitrary_states}
\end{figure*}

The simulations are performed by choosing a $\beta$ that makes sense visually. The true $\beta$ can be identified from measurements or may follow from physical considerations, but this is beyond the scope of this paper. These experiments show that the model follows intuition. Additionally, the results of the rectangular slider correspond with the simplified dynamics found earlier, which can easily be seen in \ref{fig:push_arbitrary_states}.

\subsubsection{Polygon geometry} 
\begin{figure}
    \centering
    \includegraphics[width=0.49\linewidth]{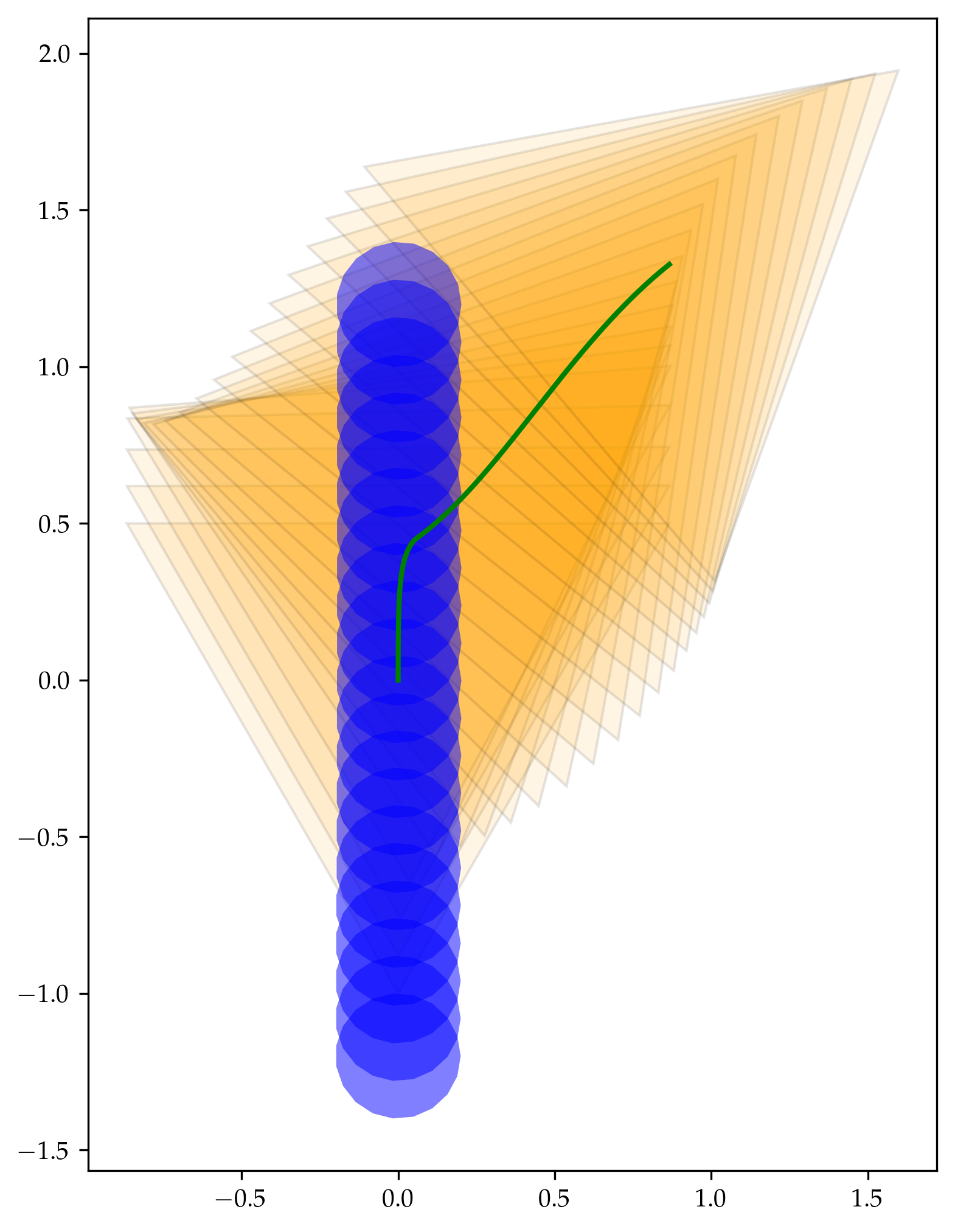}
    \caption{A forward push of a triangular slider, starting from a corner, showing transition and point contact dynamics.}
    \label{fig:push_poly}
\end{figure}

A forward push of a triangular slider is shown in Fig. \ref{fig:push_poly}, showing the point contact dynamics. The simulation is stopped once $\alpha(t) = \overline{\alpha}_{i}$. From that moment on we should switch to the full dynamics.

\begin{figure}
    \centering
    \includegraphics[width=0.7\linewidth]{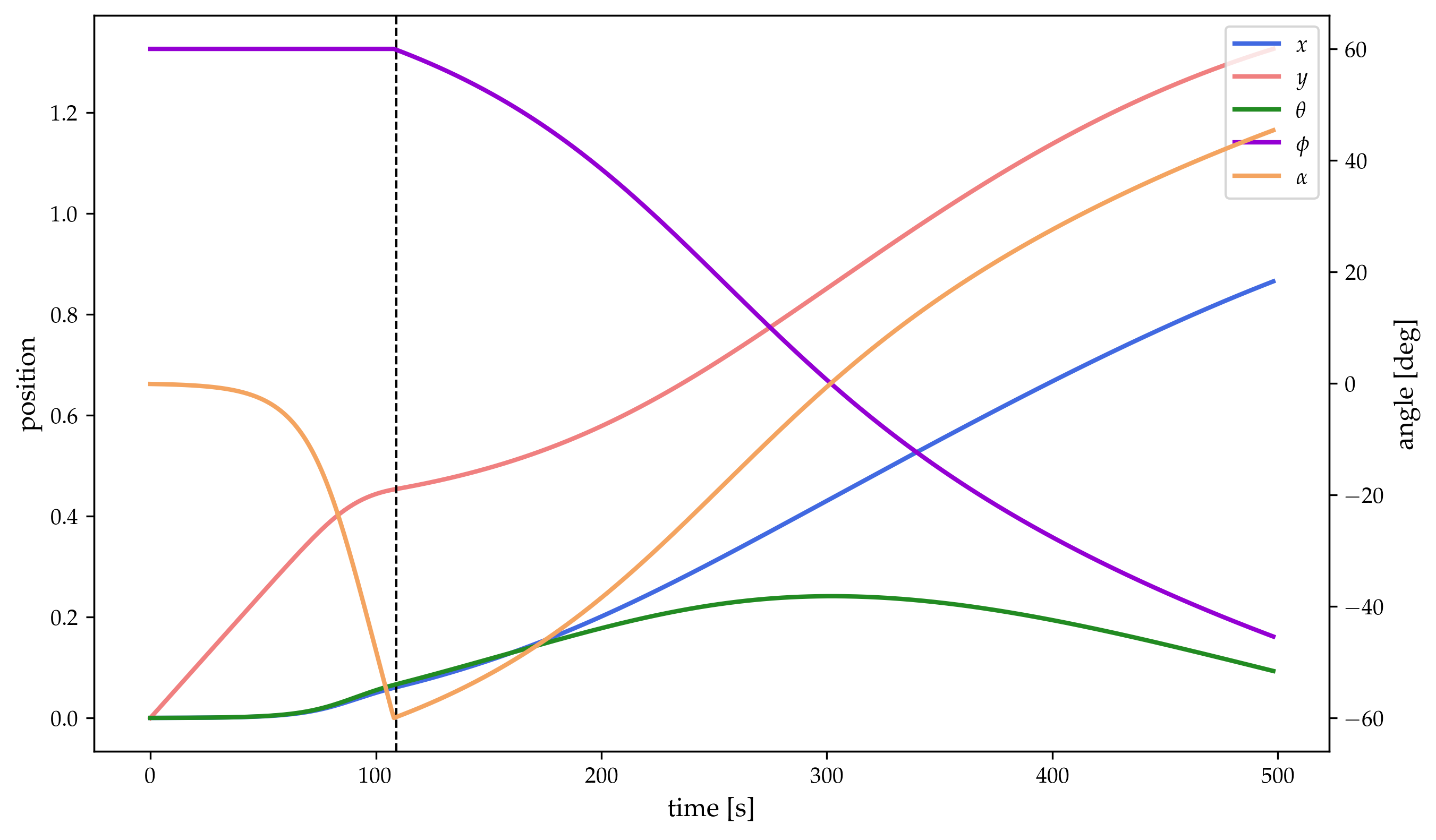}
    \caption{State variables for a forward push of a triangular slider, starting from the corner, showing transition and point contact dynamics. The dashed line indicates the transition.}
    \label{fig:push_poly_states}
\end{figure}

The forward push in Fig. \ref{fig:push_poly} starts from the corner and therefore we begin with point contact dynamics. The value of $\alpha$ gradually decreases, while $\phi$ remains constant, as seen in Fig. \ref{fig:push_poly_states}. This continues until you reach the next edge, from where we transition to the full dynamics. After the transition, we see $phi$ steadily decreasing, or in other words, we see the slider moving up the edge. The slider moves as expected; during the point contact it mainly moves up, while after the transition it moves diagonally.

\subsection{Tracking control}
Next, we can look at tracking control for rectangular sliders. As mentioned before, our control objective is to track a position signal given by $x_d$ and $y_d$. We examine tracking both static and dynamic position signals. All cases are performed using the three previously described control methods, the cascaded feedback strategy and both choices for the dynamic feedback linearization.

Additionally, all experiments are performed both in simulation with perturbed simulation models and input noise, as on a set-up. Simulated results are shown in orange, while real-world results are shown in yellow. 

The set-up includes a KUKA KR 6 R700-2 robot with a finger-like end-effector, two Intel Realsense D435 depth cameras, and a square block with ArUco markers. The system runs on ROS, with computations performed on a 64-bit Intel 8-Core i7 3.60GHz workstation with 32GB RAM. The main limitation of the current setup is that pushing is restricted to a forward direction to prevent camera obstruction. This can be resolved with an alternative perception system, or by using more cameras.

\subsubsection{Stationary point}
\begin{figure*}
    \begin{subfigure}[b]{0.33\textwidth}
        \centering
        \includegraphics[width=.9\textwidth]{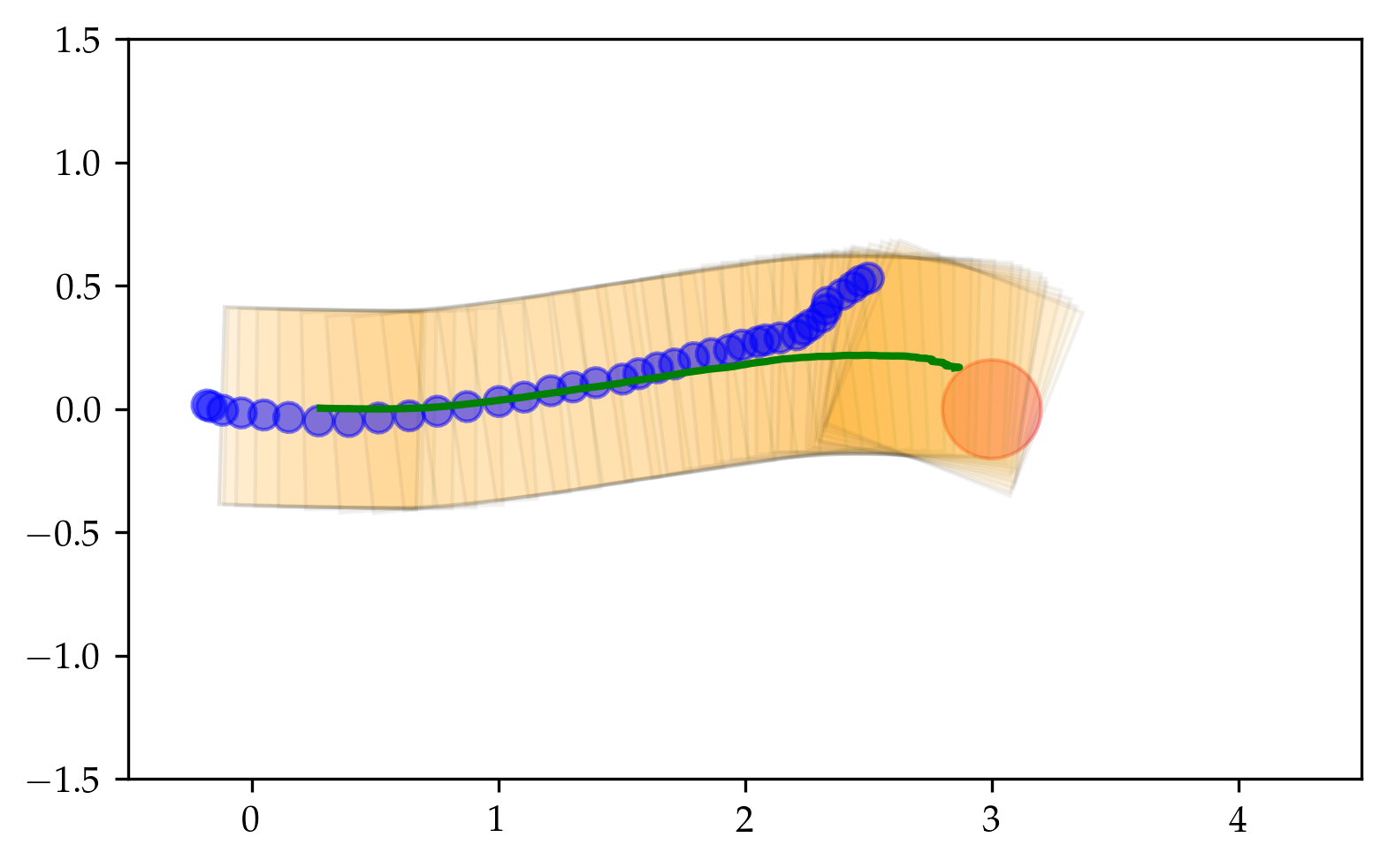}
        \caption{}
    \end{subfigure}
    \begin{subfigure}[b]{0.33\textwidth}
        \centering
        \includegraphics[width=.9\textwidth]{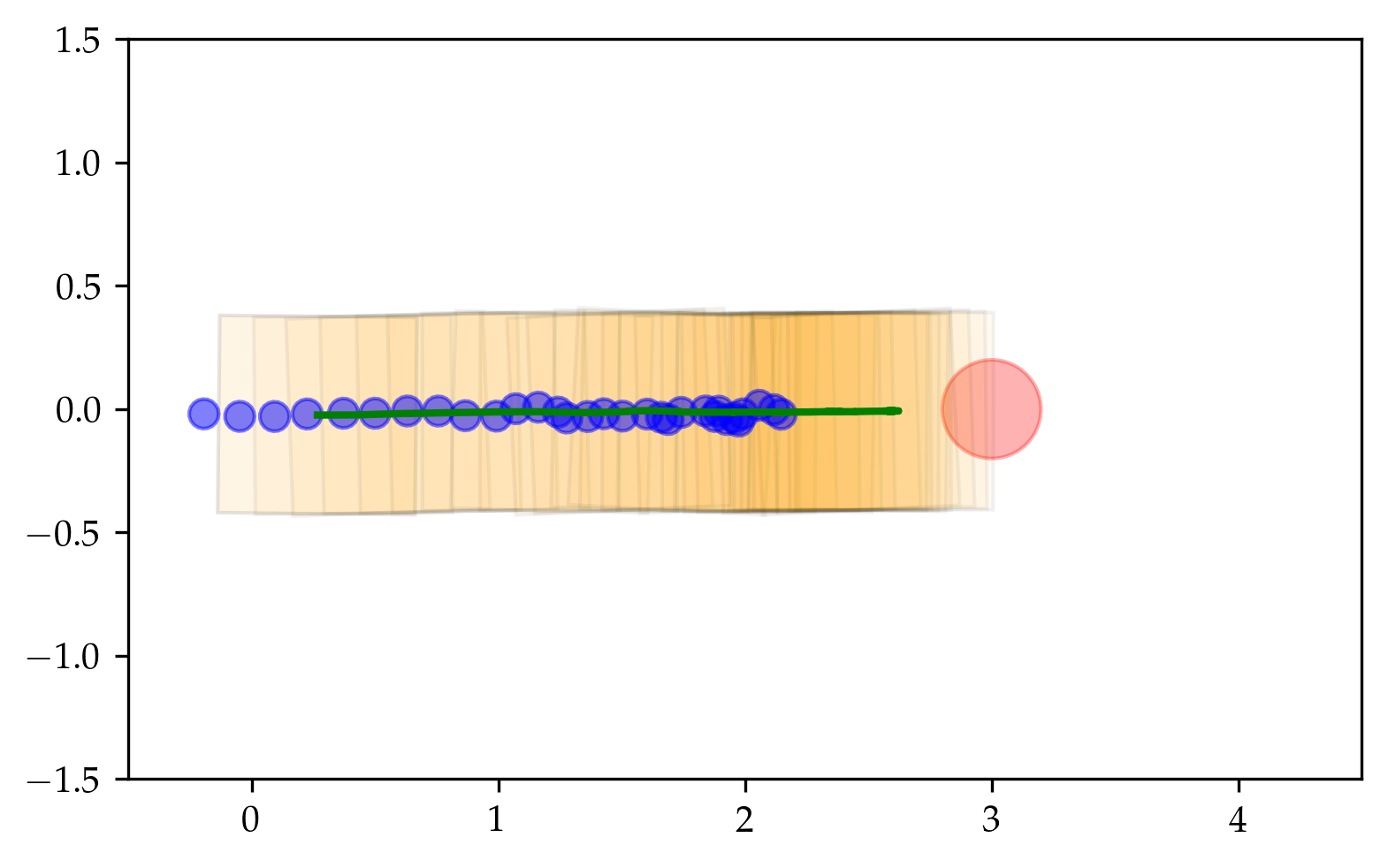}
        \caption{}
    \end{subfigure}
    \begin{subfigure}[b]{0.33\textwidth}
        \centering
        \includegraphics[width=.9\textwidth]{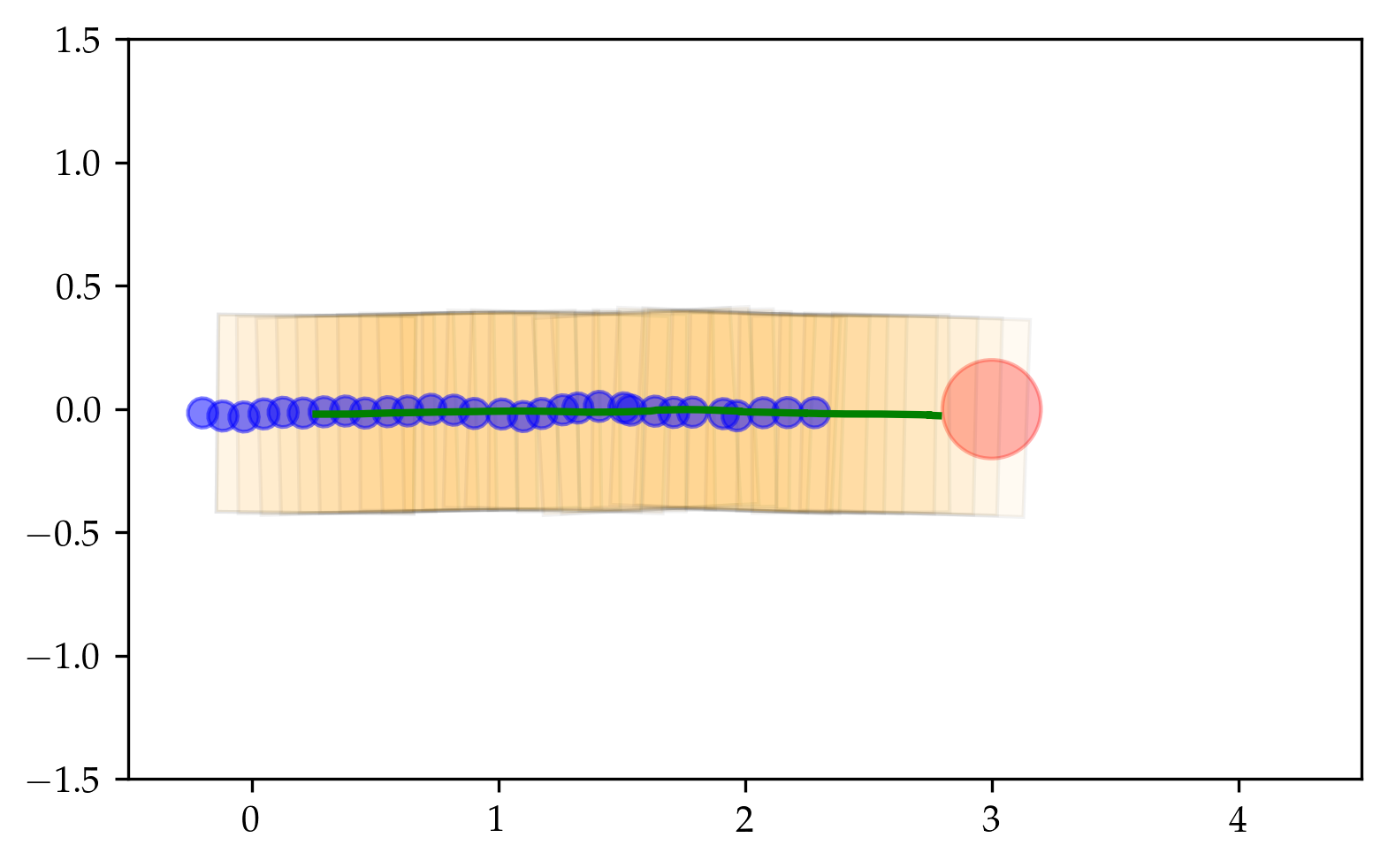}
        \caption{}
    \end{subfigure}
    \begin{subfigure}[b]{0.33\textwidth}
        \centering
        \includegraphics[width=.9\textwidth]{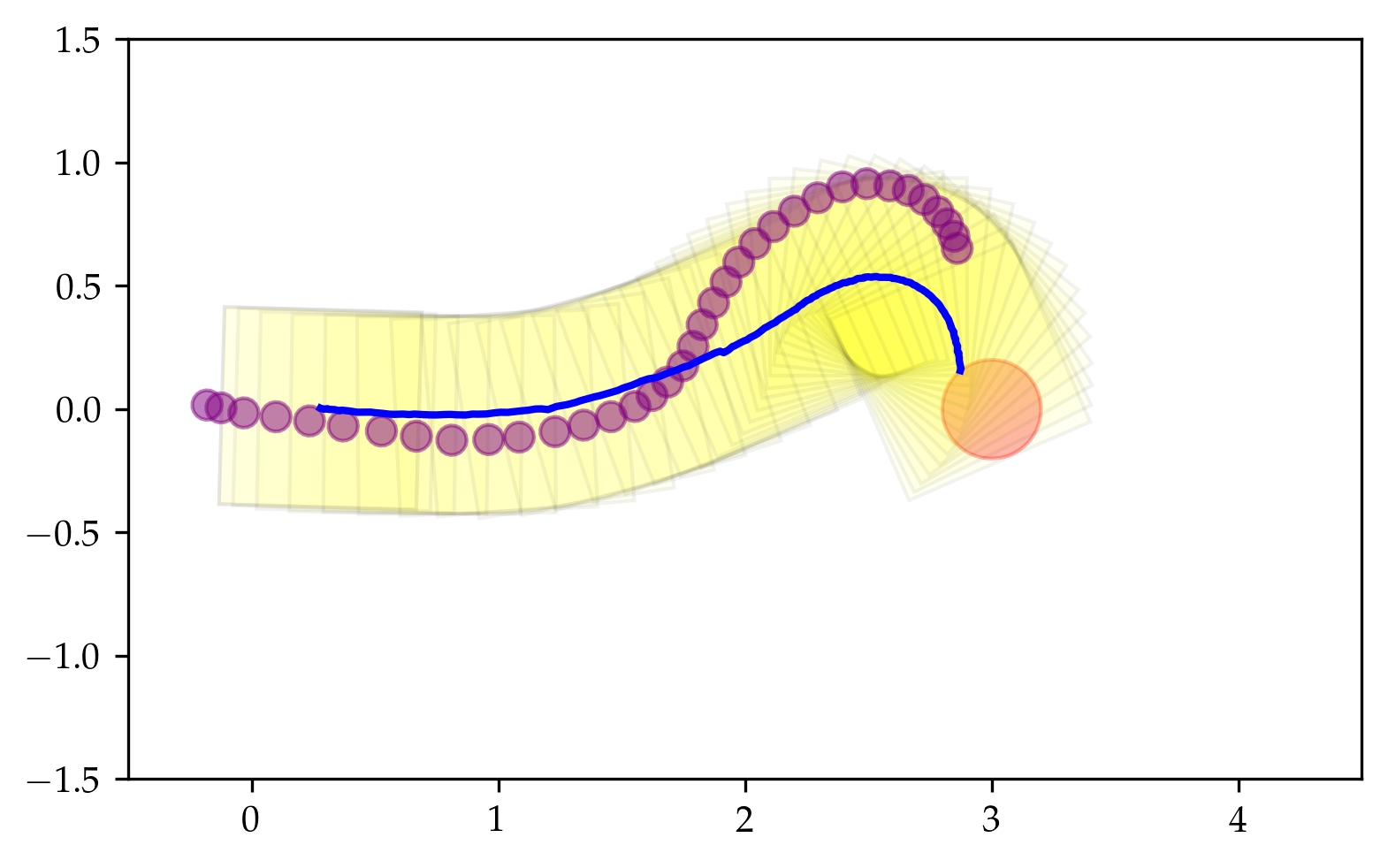}
        \caption{}
    \end{subfigure}
    \begin{subfigure}[b]{0.33\textwidth}
        \centering
        \includegraphics[width=.9\textwidth]{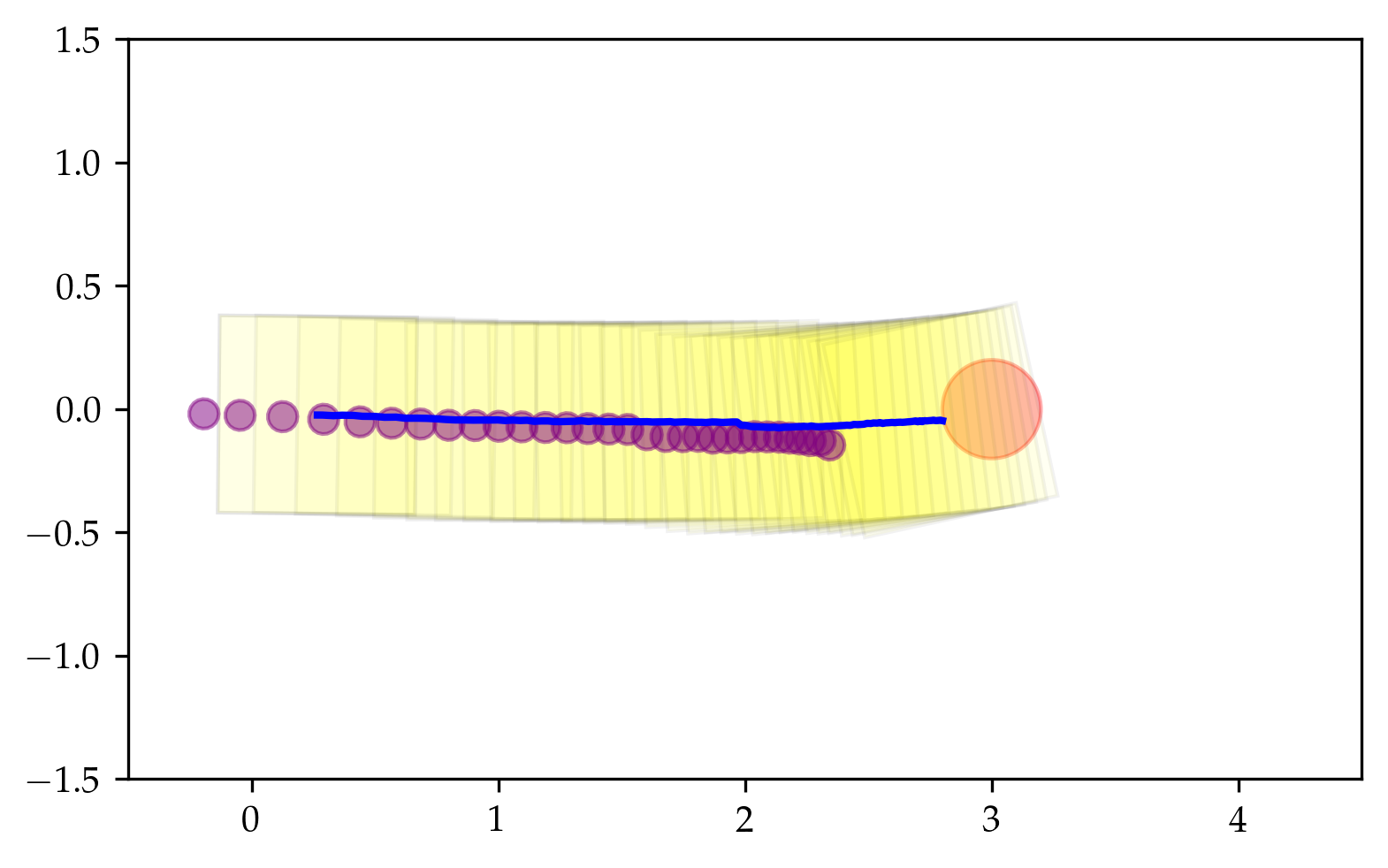}
        \caption{}
    \end{subfigure}
    \begin{subfigure}[b]{0.33\textwidth}
        \centering
        \includegraphics[width=.9\textwidth]{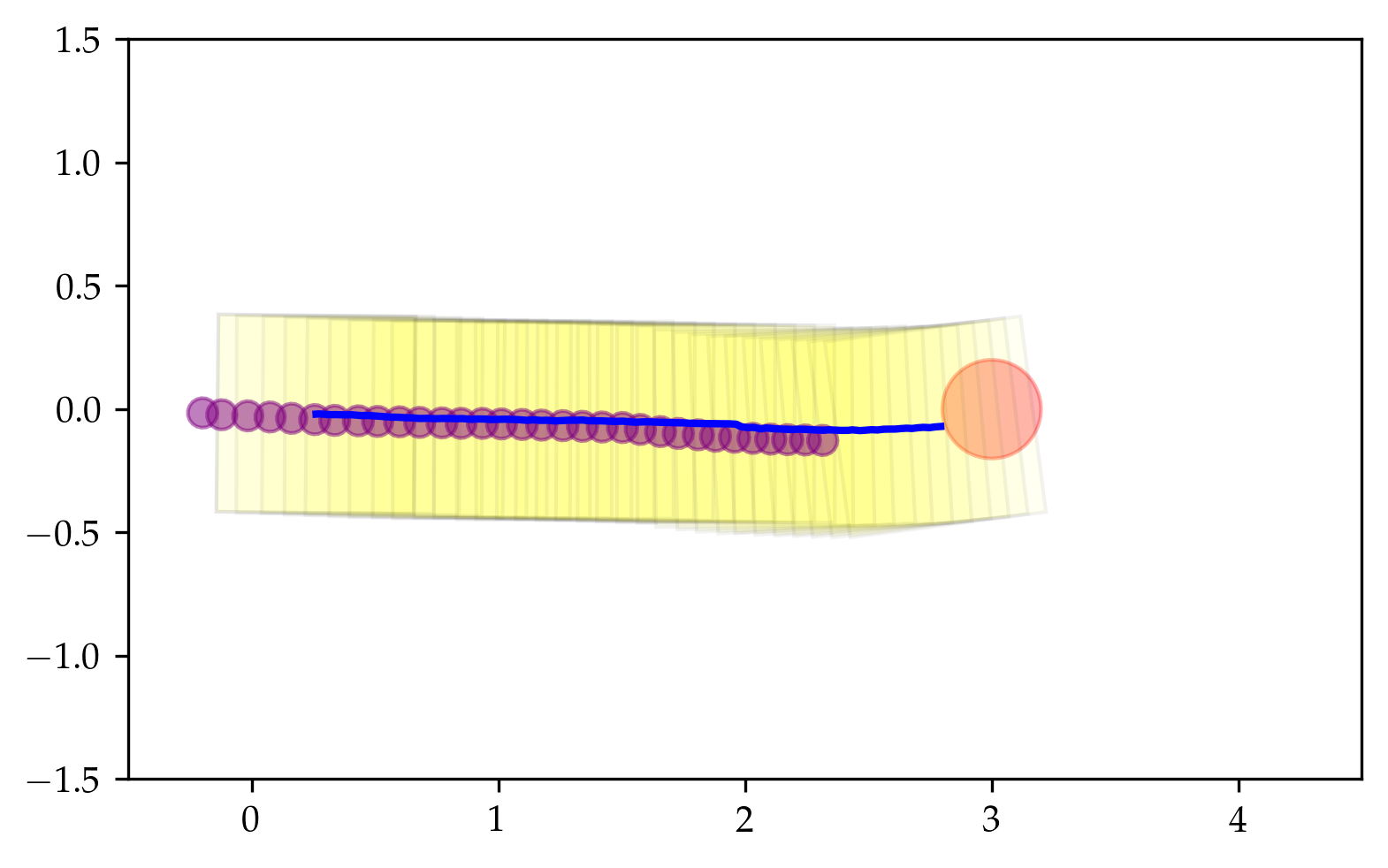}
        \caption{}
    \end{subfigure}
    \begin{subfigure}[b]{0.33\textwidth}
        \centering
        \includegraphics[width=.9\textwidth]{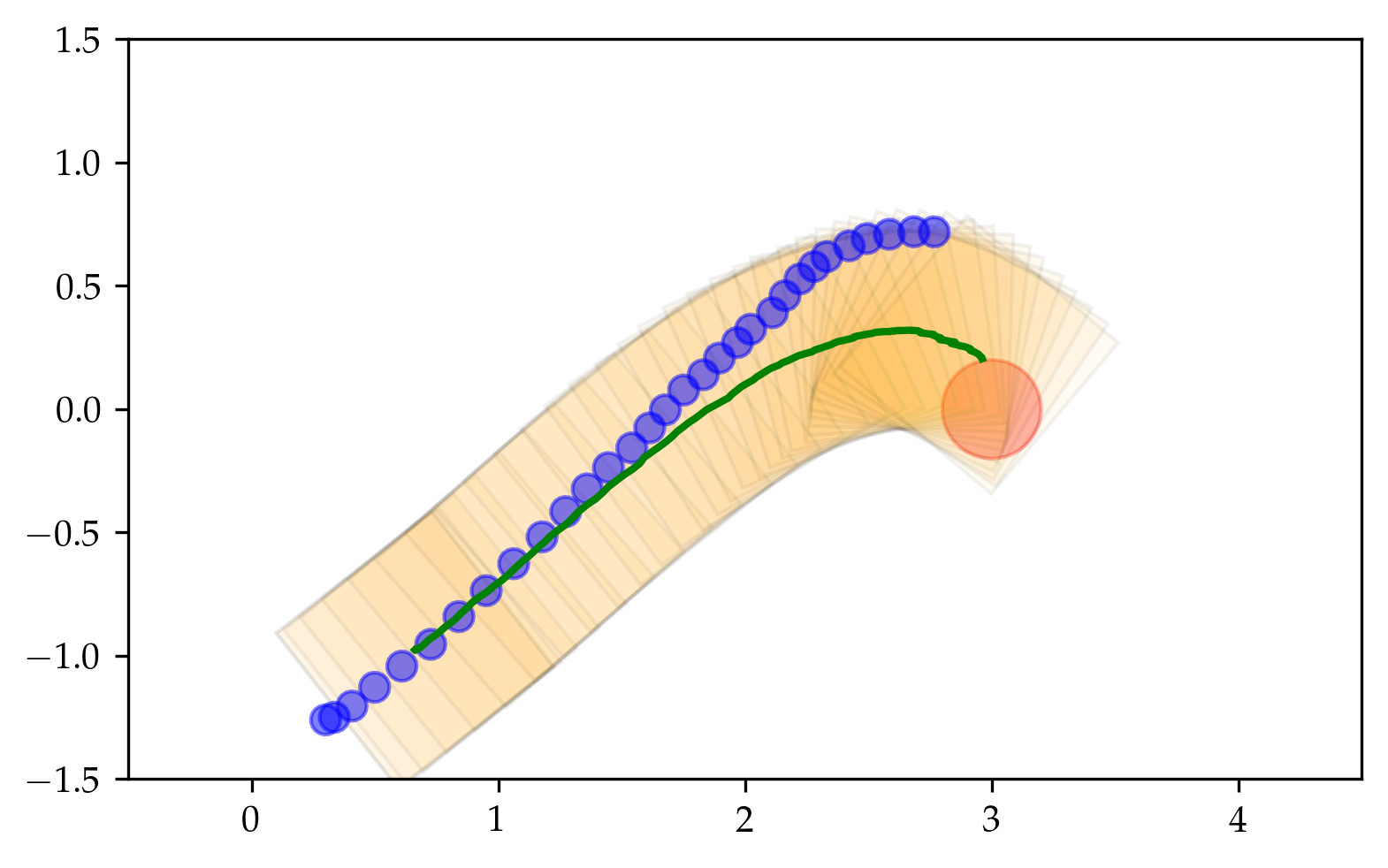}
        \caption{}
    \end{subfigure}
    \begin{subfigure}[b]{0.33\textwidth}
        \centering
        \includegraphics[width=.9\textwidth]{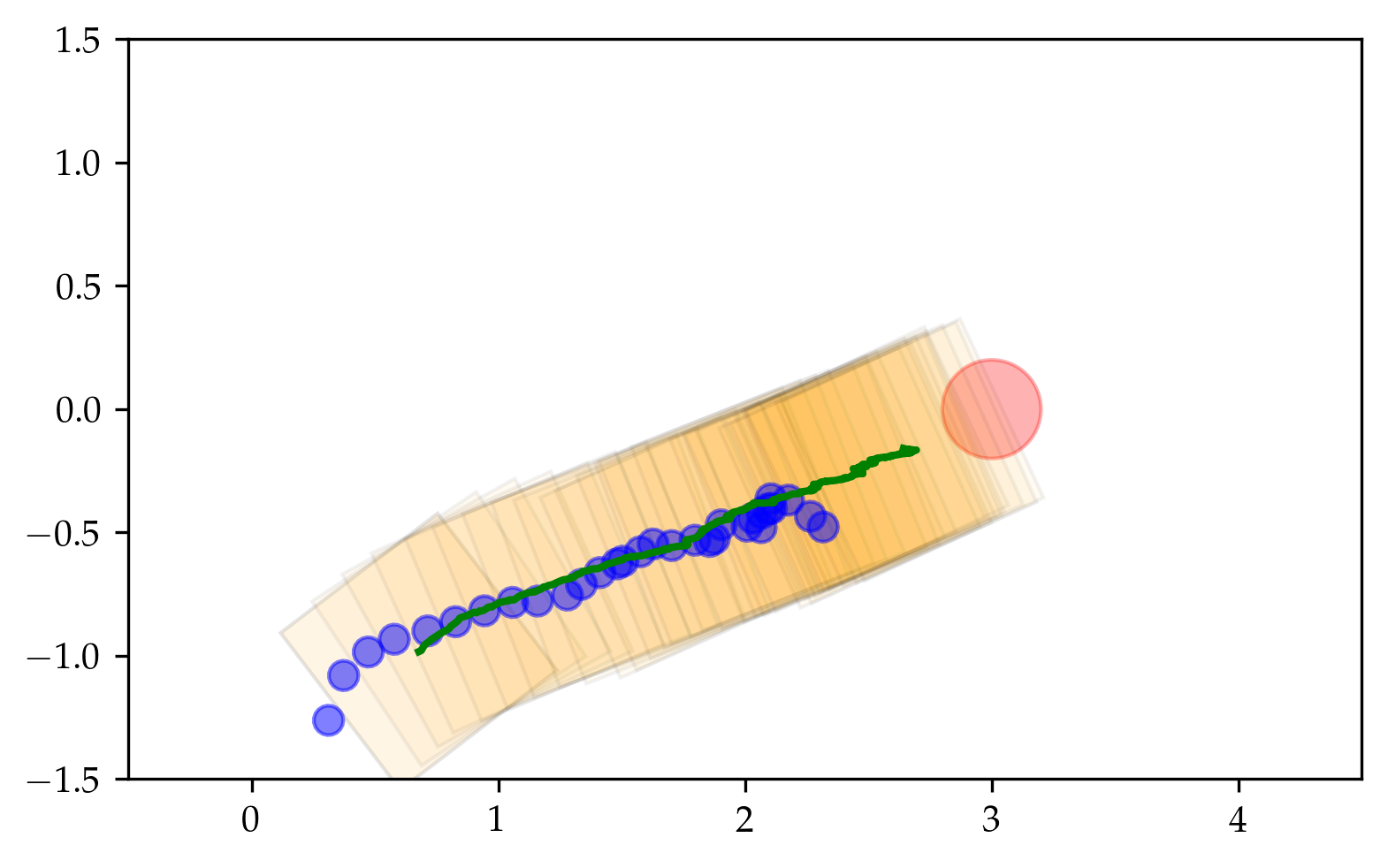}
        \caption{}
    \end{subfigure}
    \begin{subfigure}[b]{0.33\textwidth}
        \centering
        \includegraphics[width=.9\textwidth]{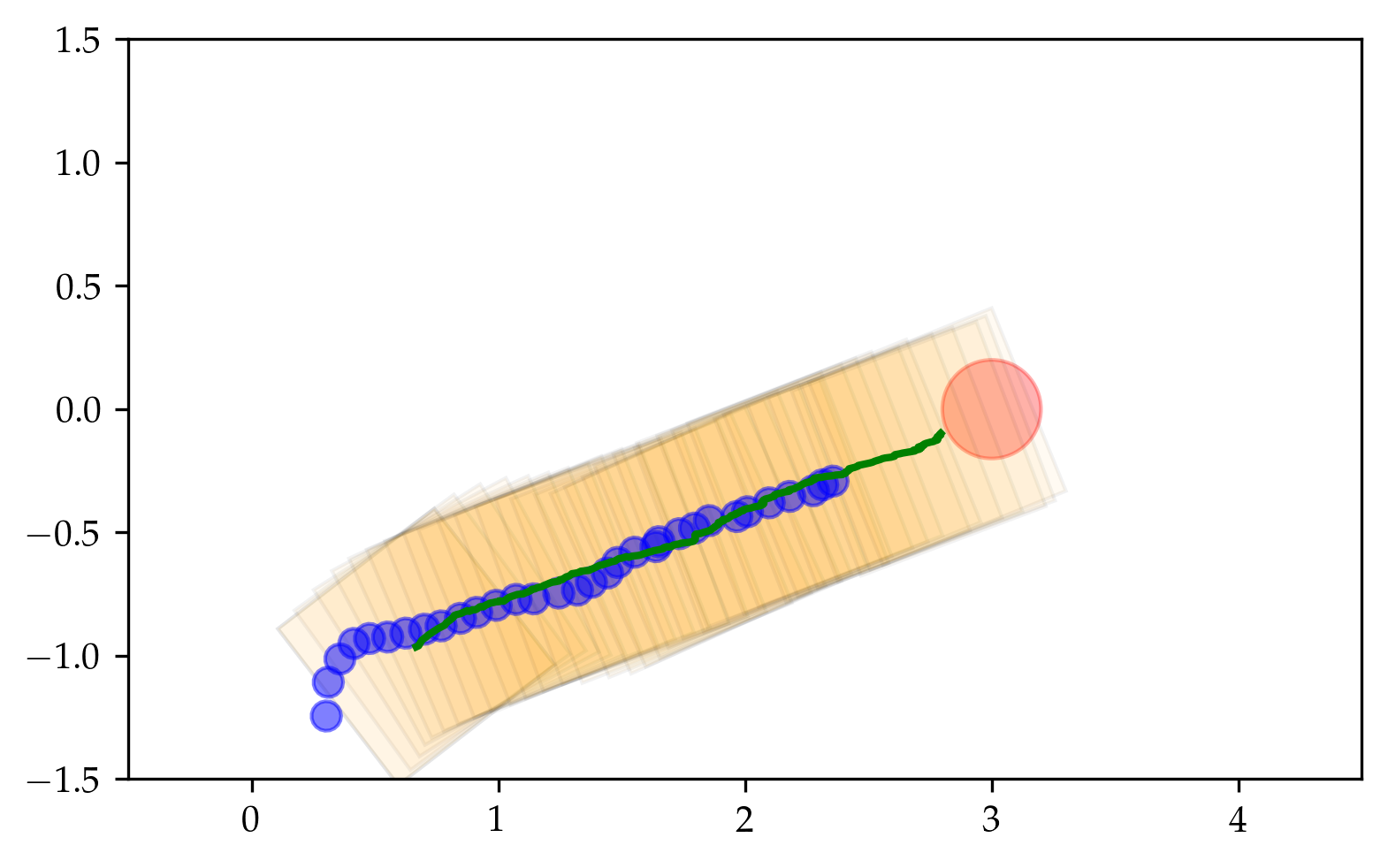}
        \caption{}
    \end{subfigure}
    \begin{subfigure}[b]{0.33\textwidth}
        \centering
        \includegraphics[width=.9\textwidth]{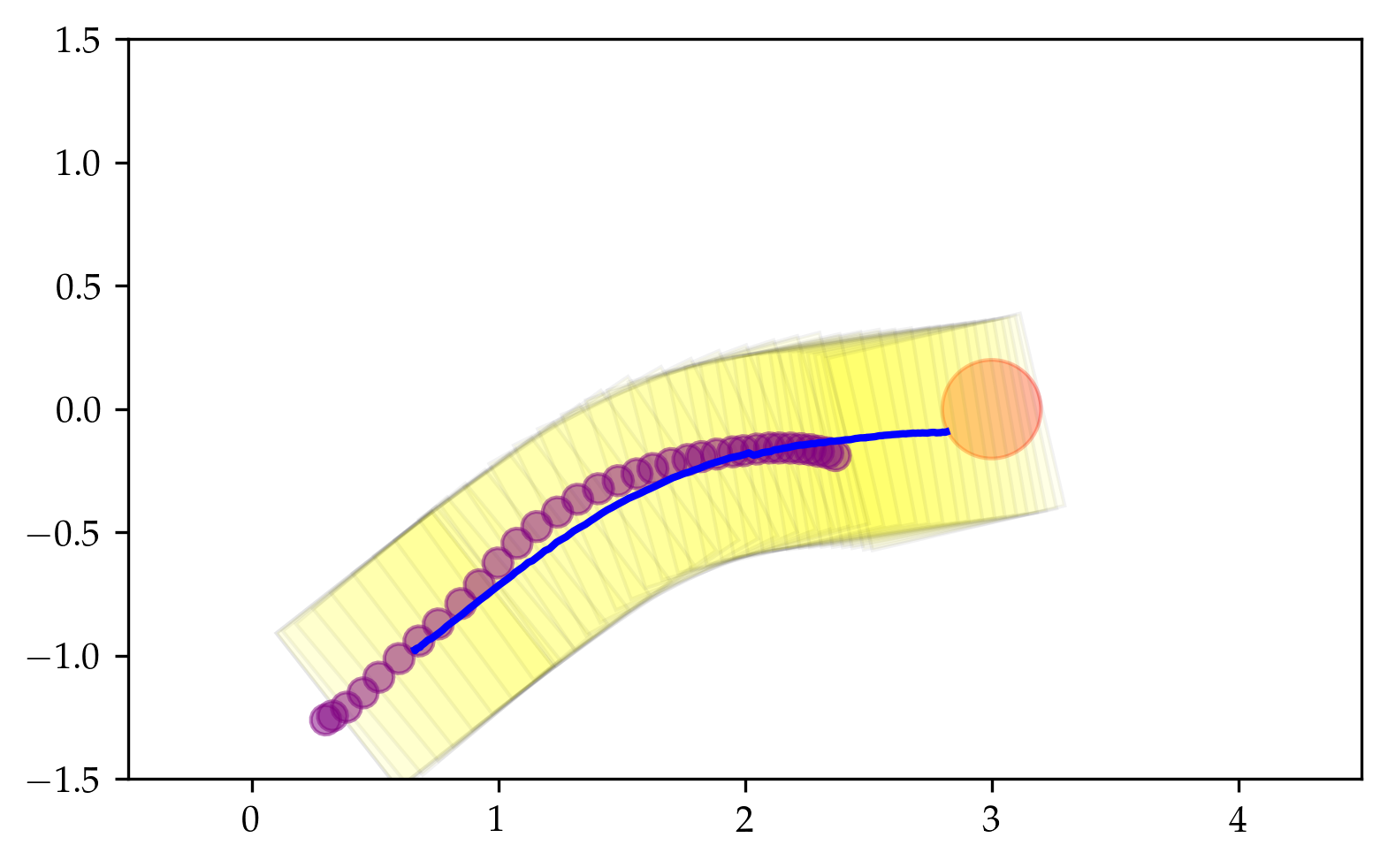}
        \caption{}
    \end{subfigure}
    \begin{subfigure}[b]{0.33\textwidth}
        \centering
        \includegraphics[width=.9\textwidth]{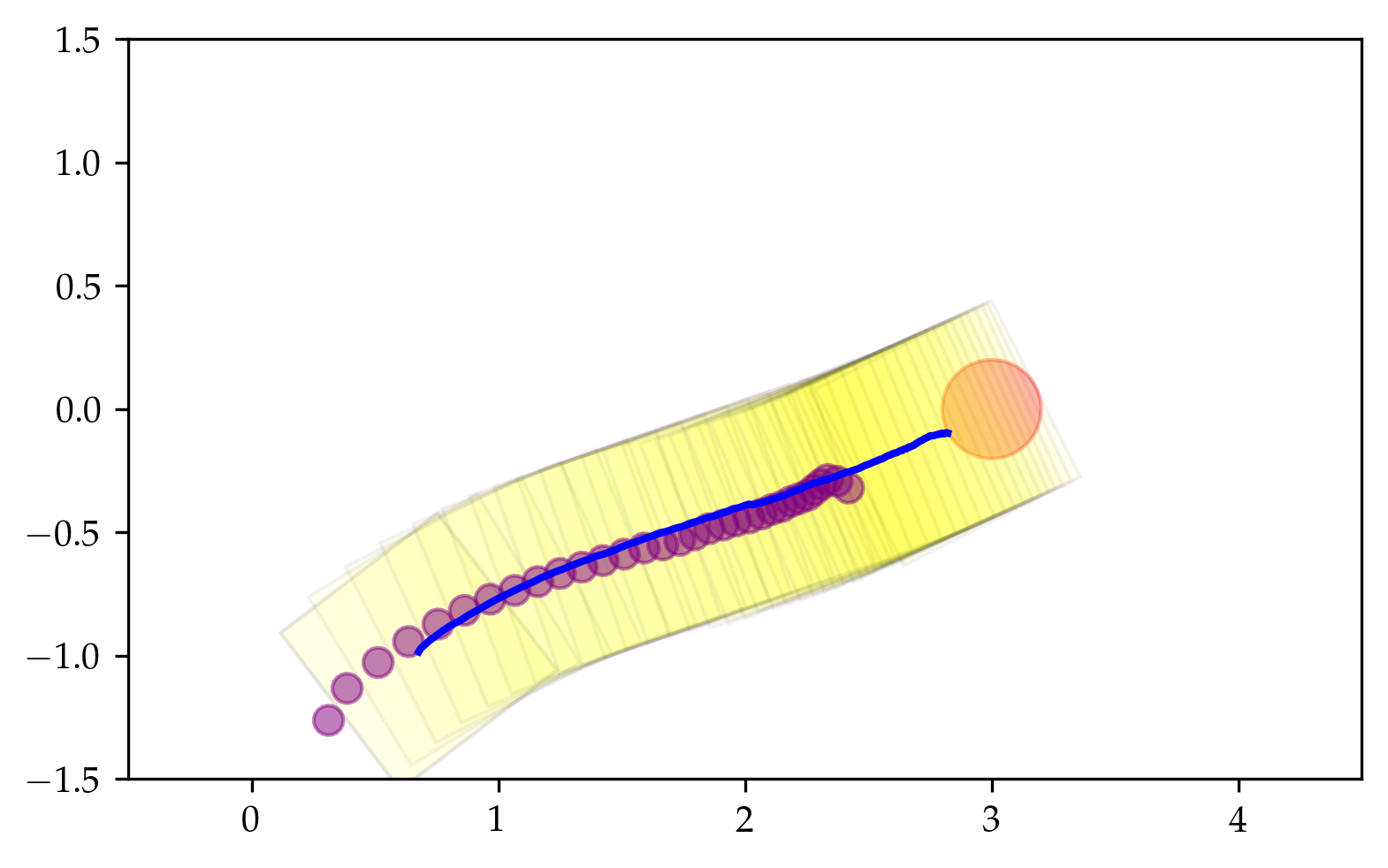}
        \caption{}
    \end{subfigure}
    \begin{subfigure}[b]{0.33\textwidth}
        \centering
        \includegraphics[width=.9\textwidth]{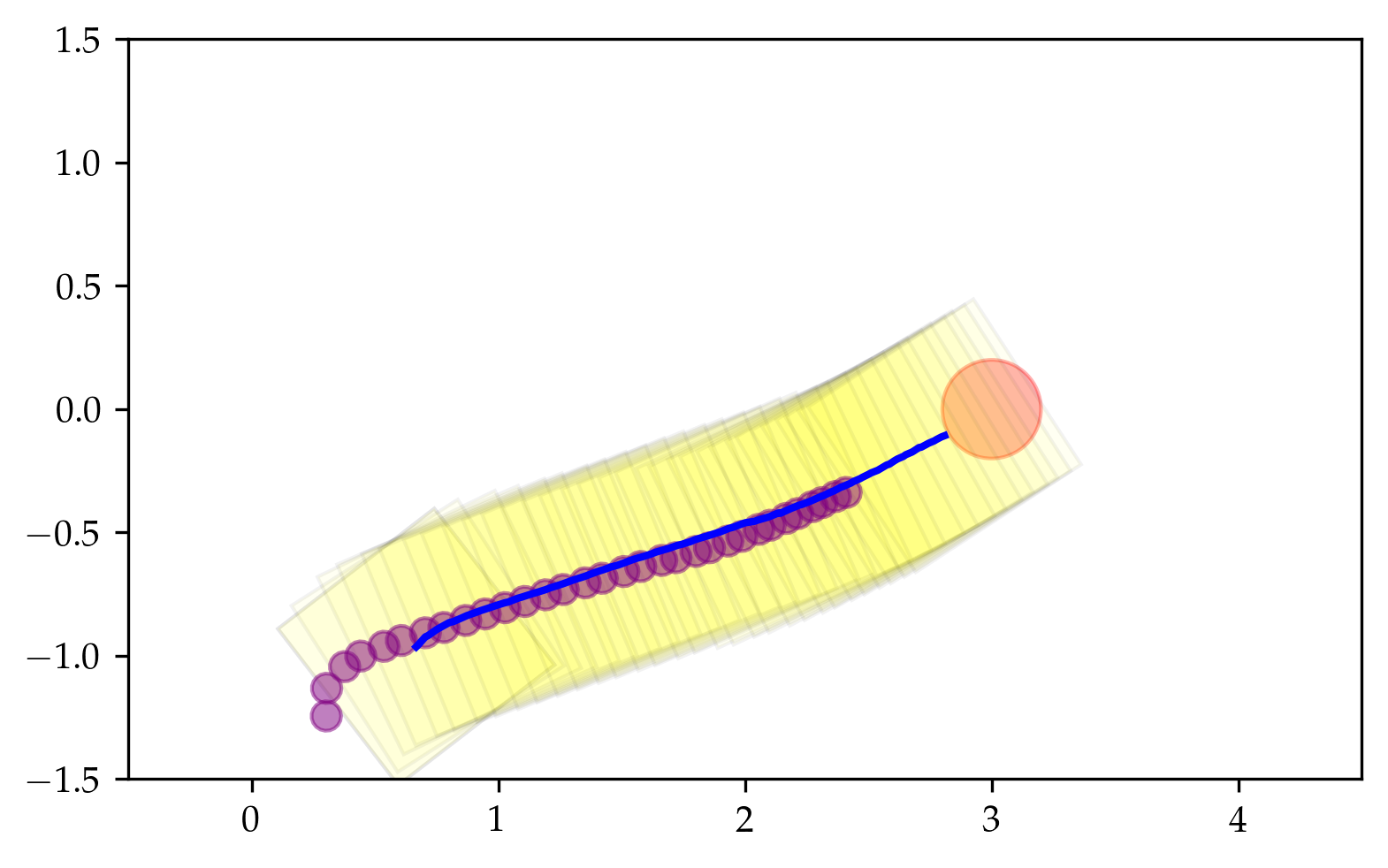}
        \caption{}
    \end{subfigure}
    \caption{Stationary point tracking, to an accuracy of $0.2$ cm indicated with a red circle. Three control methods, a cascaded (quasi-)static feedback approach and dynamic feedback linearization with two choices of $\vectorstyle{\gamma}$, are shown both in simulation (orange) and on the set-up (yellow), respectively in the first, second and third column.}
    \label{fig:point}
\end{figure*}

Stationary point tracking is tested for two starting positions, which are indicated on the set-up with two black lines as seen in Fig. \ref{fig:screengrabs}, and the same goal position. The results are shown in Fig. \ref{fig:point}.

For all three controllers, we used manually tuned control parameters. The control parameters were tuned manually until reaching satisfactory behaviour but were not optimized to maximize some performance metrics. Optimizing the gains is out of scope for this paper. For the cascaded controller, the following time scales are used: $\vectorstyle{\tau} = [\tau_x,\tau_y,\tau_{\theta}, \tau_d]=[2.0,1.6,0.6,0.5]$. The tuning of the time scales in a cascaded manner becomes apparent, to make sure that all four loops interact as desired.

The control gains for feedback linearization are obtained by solving the algebraic Riccati equation with $Q=\operatorname{diag}(0.01, 10, 20)$ and $R=20$. Higher-order derivatives are penalized more heavily to promote smoother motion in the system. Both in simulation and on the set-up the same time step is taken, $\Delta t=0.1$ s.

Since the control gains are not tuned with respect to some particular performance metric it is hard to make bold statements about which controller performs best. All three controllers manage to easily get to the desired goal, with quickly tuned parameters.

\subsubsection{Dynamic trajectory}
\begin{figure*}
    \begin{subfigure}[b]{0.33\textwidth}
        \centering
        \includegraphics[width=.9\textwidth]{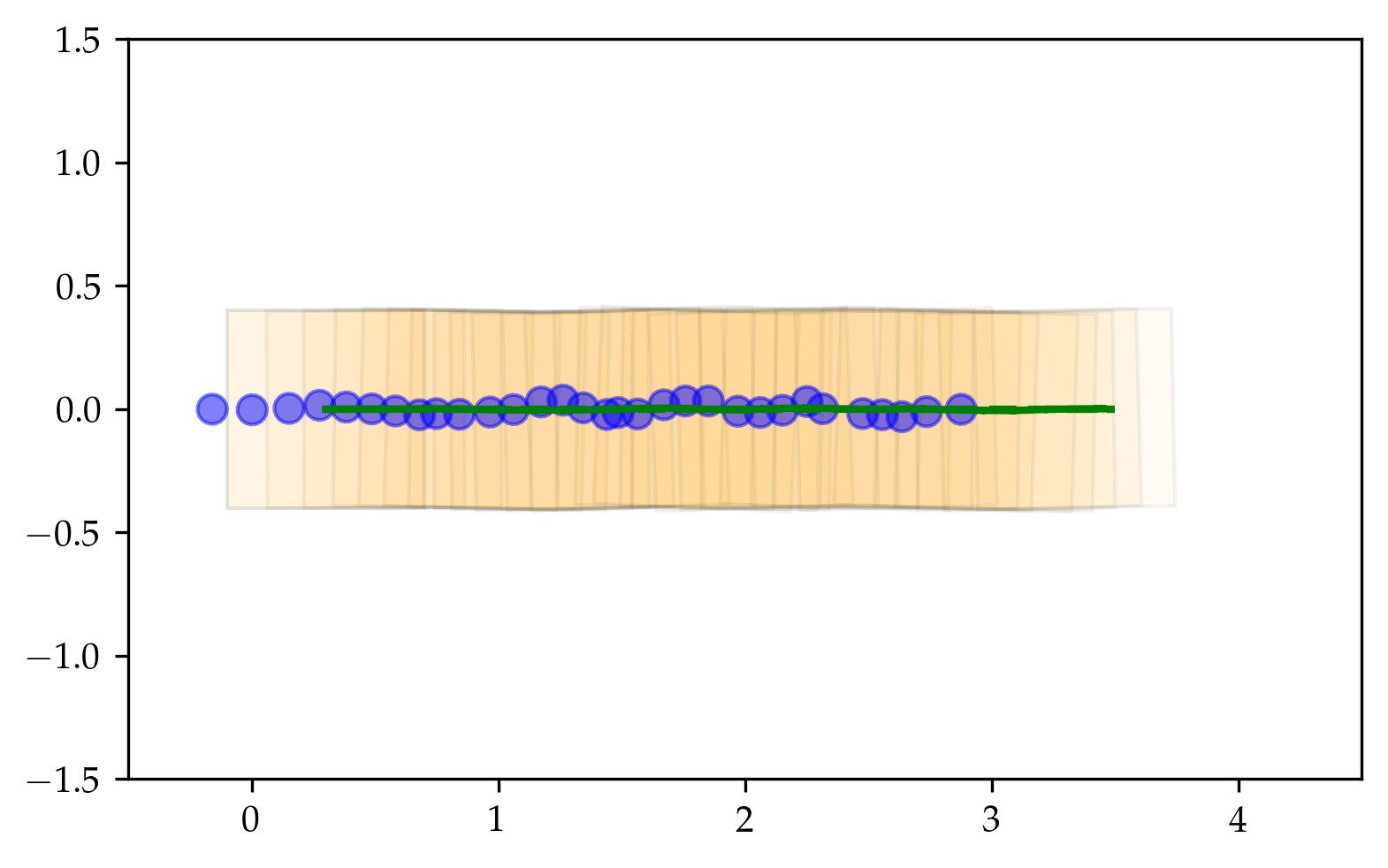}
        \caption{}
    \end{subfigure}
    \begin{subfigure}[b]{0.33\textwidth}
        \centering
        \includegraphics[width=.9\textwidth]{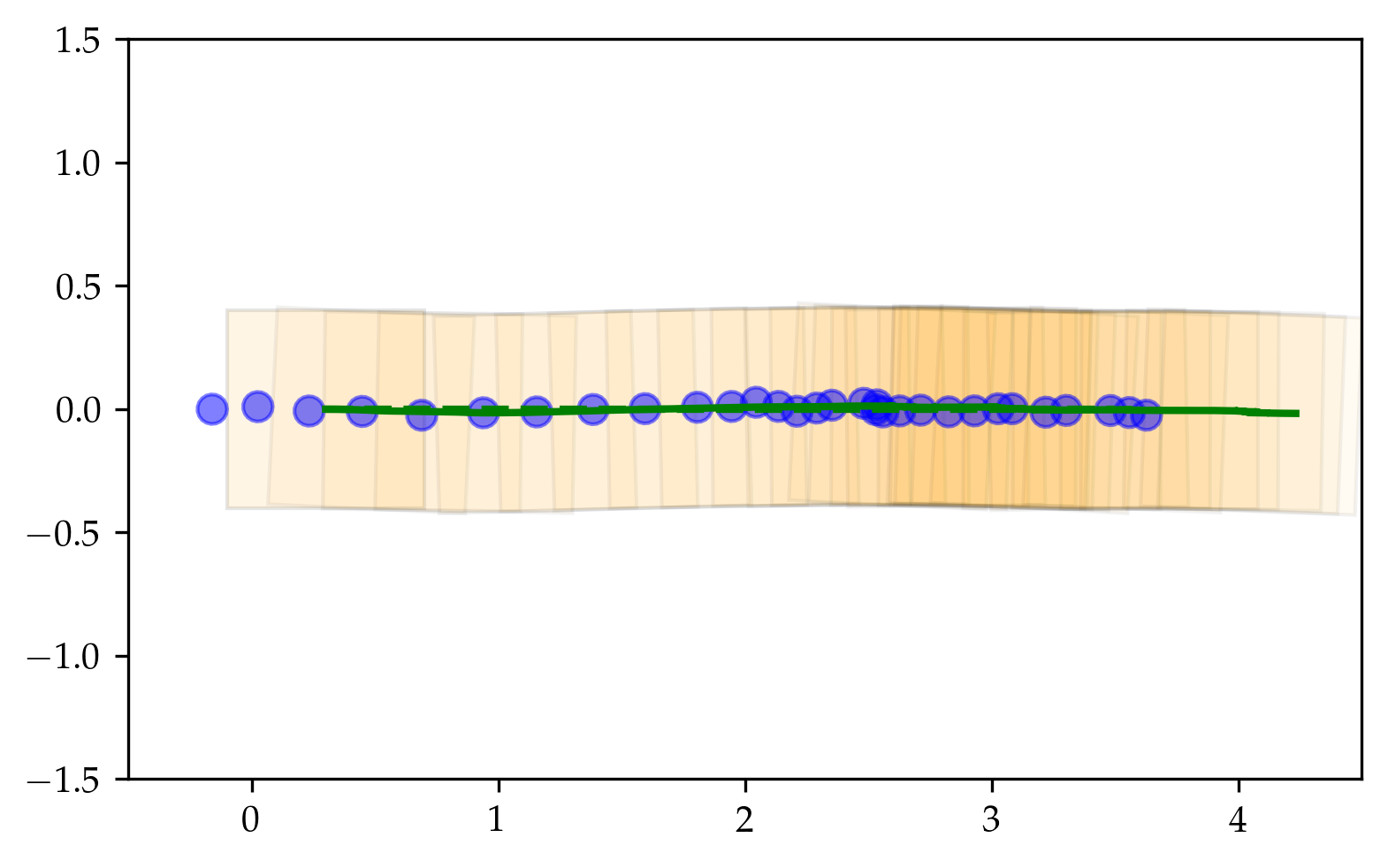}
        \caption{}
    \end{subfigure}
        \begin{subfigure}[b]{0.33\textwidth}
        \centering
        \includegraphics[width=.9\textwidth]{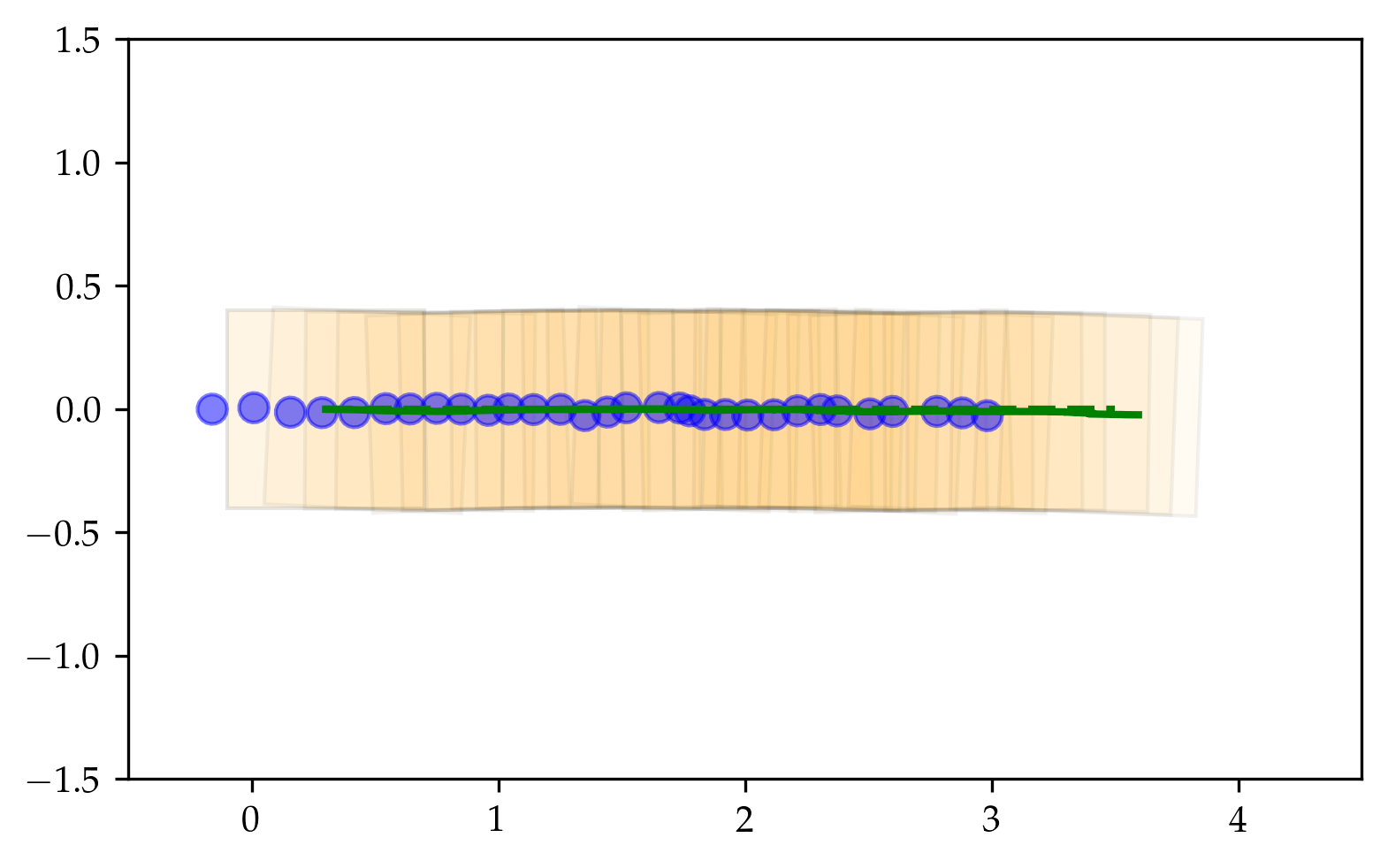}
        \caption{}
    \end{subfigure}
    \begin{subfigure}[b]{0.33\textwidth}
        \centering
        \includegraphics[width=.9\textwidth]{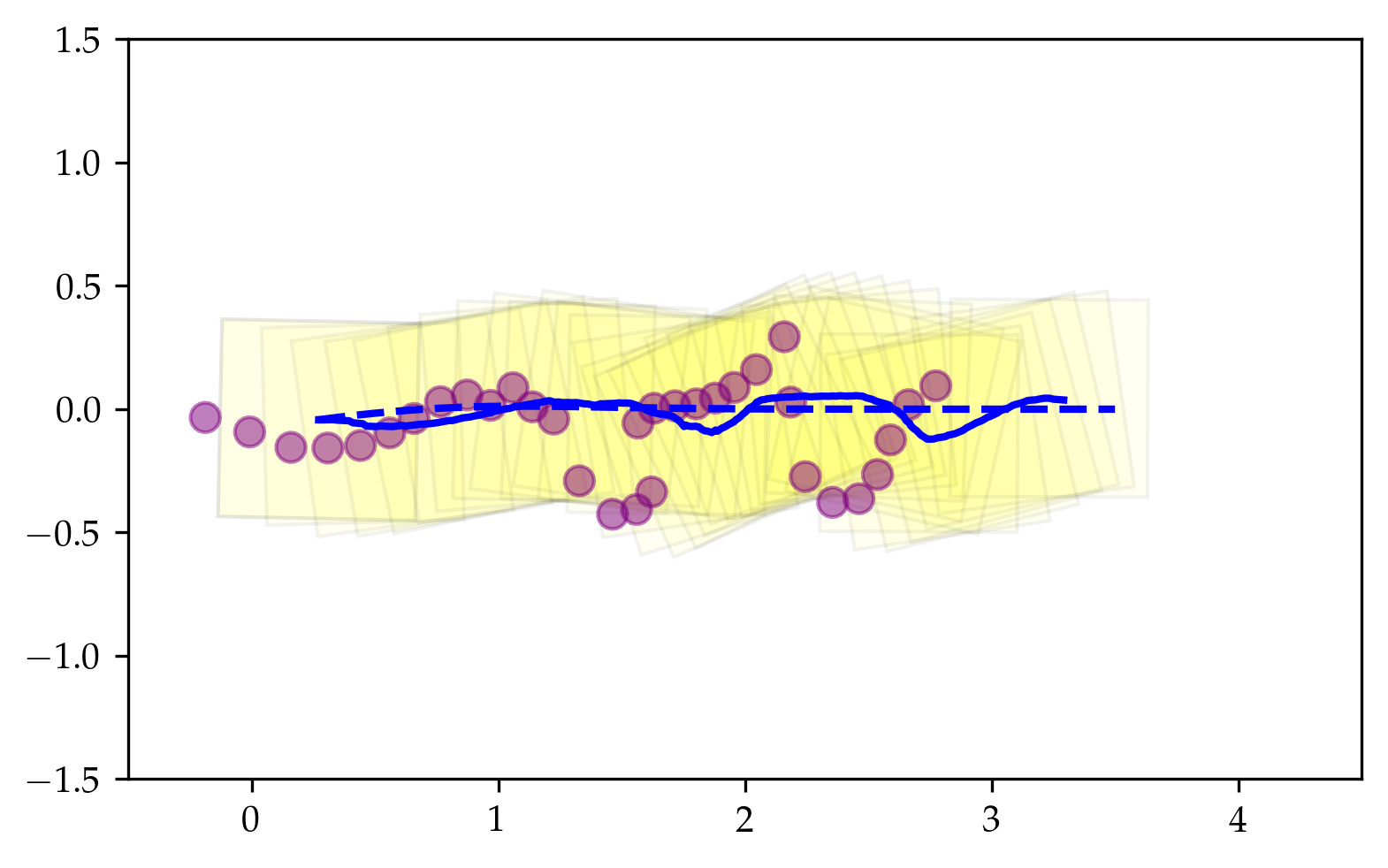}
        \caption{}
    \end{subfigure} 
    \begin{subfigure}[b]{0.33\textwidth}
        \centering
        \includegraphics[width=.9\textwidth]{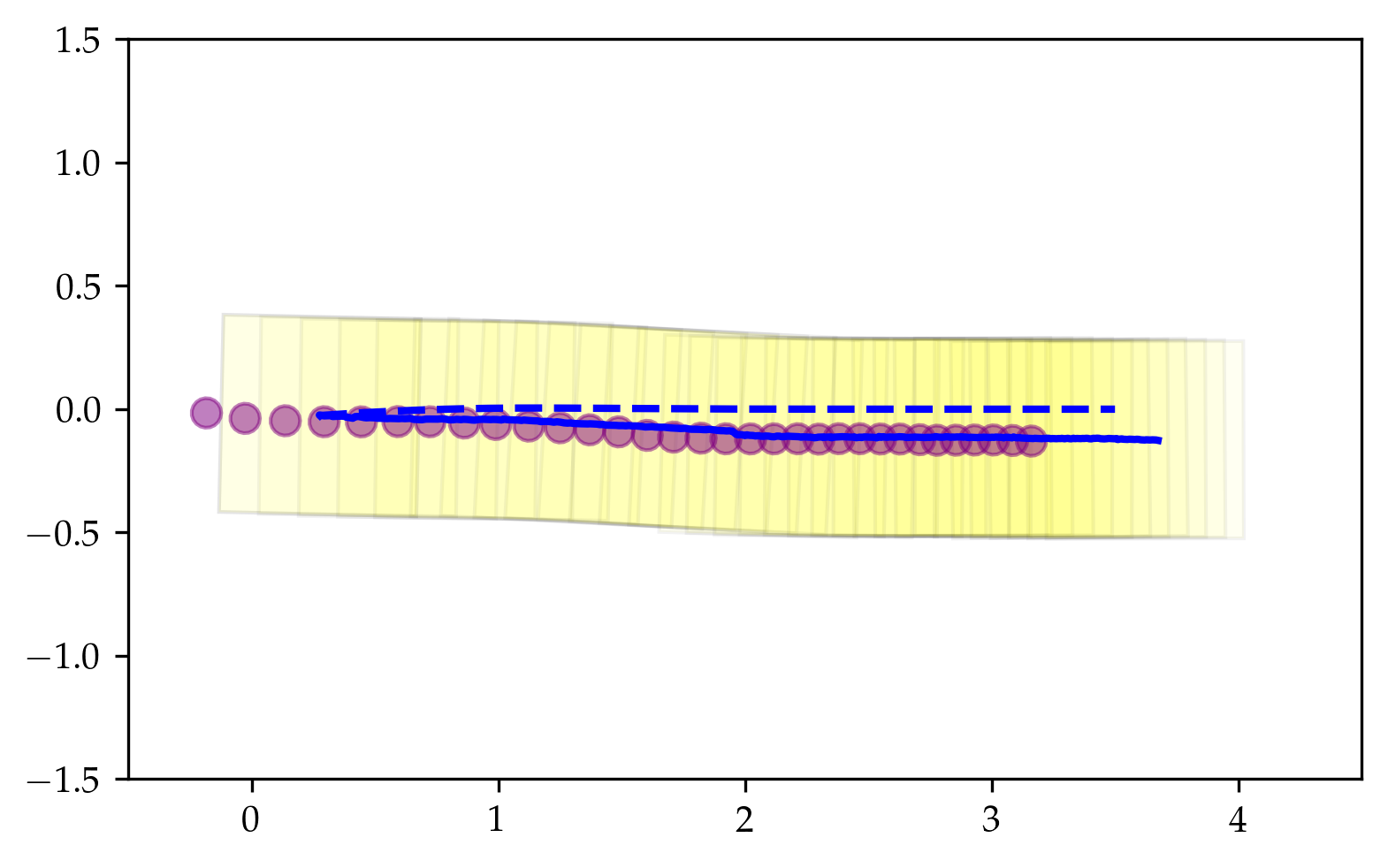}
        \caption{}
    \end{subfigure}
        \begin{subfigure}[b]{0.33\textwidth}
        \centering
        \includegraphics[width=.9\textwidth]{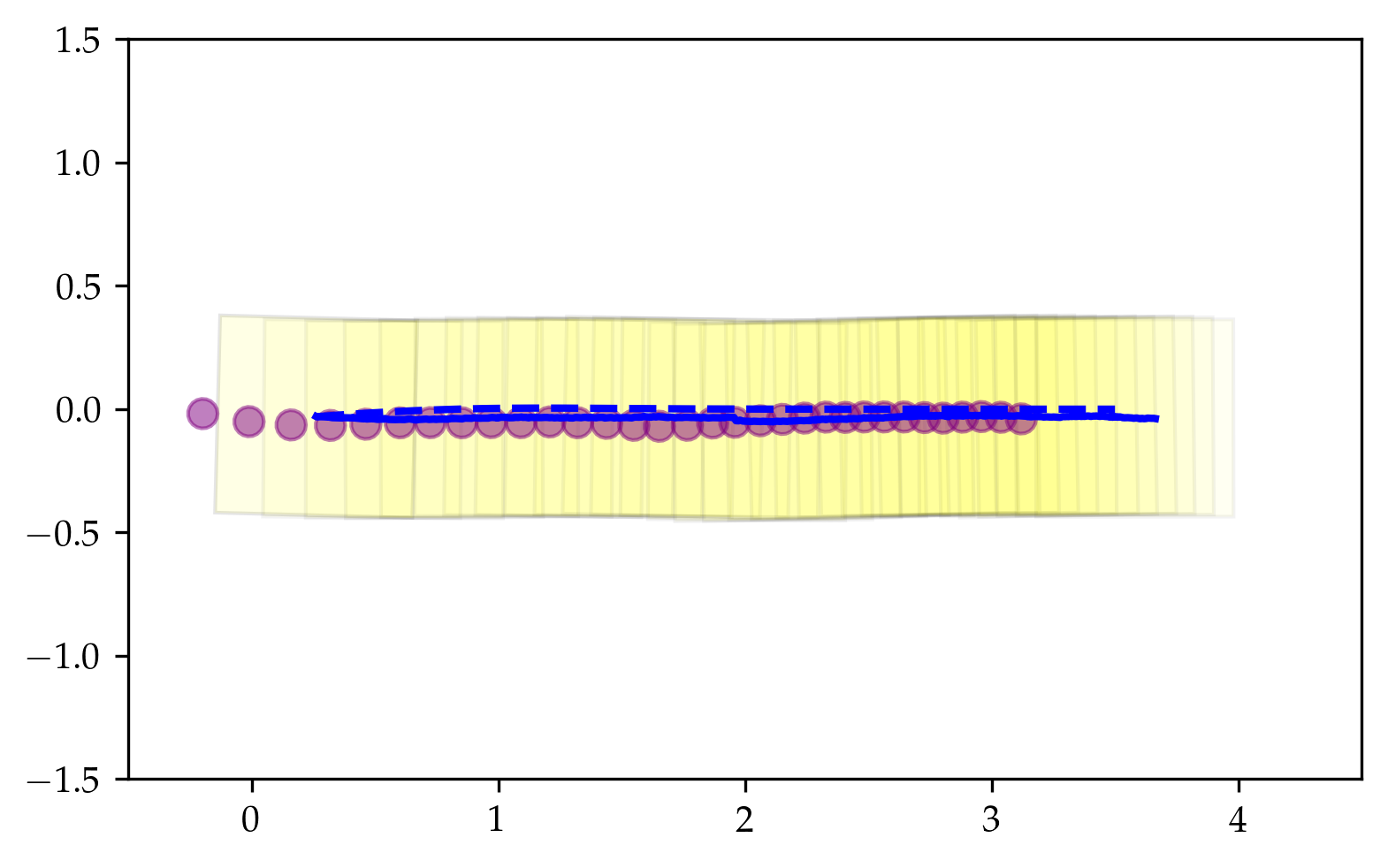}
        \caption{}
    \end{subfigure}
    \begin{subfigure}[b]{0.33\textwidth}
        \centering
        \includegraphics[width=.9\textwidth]{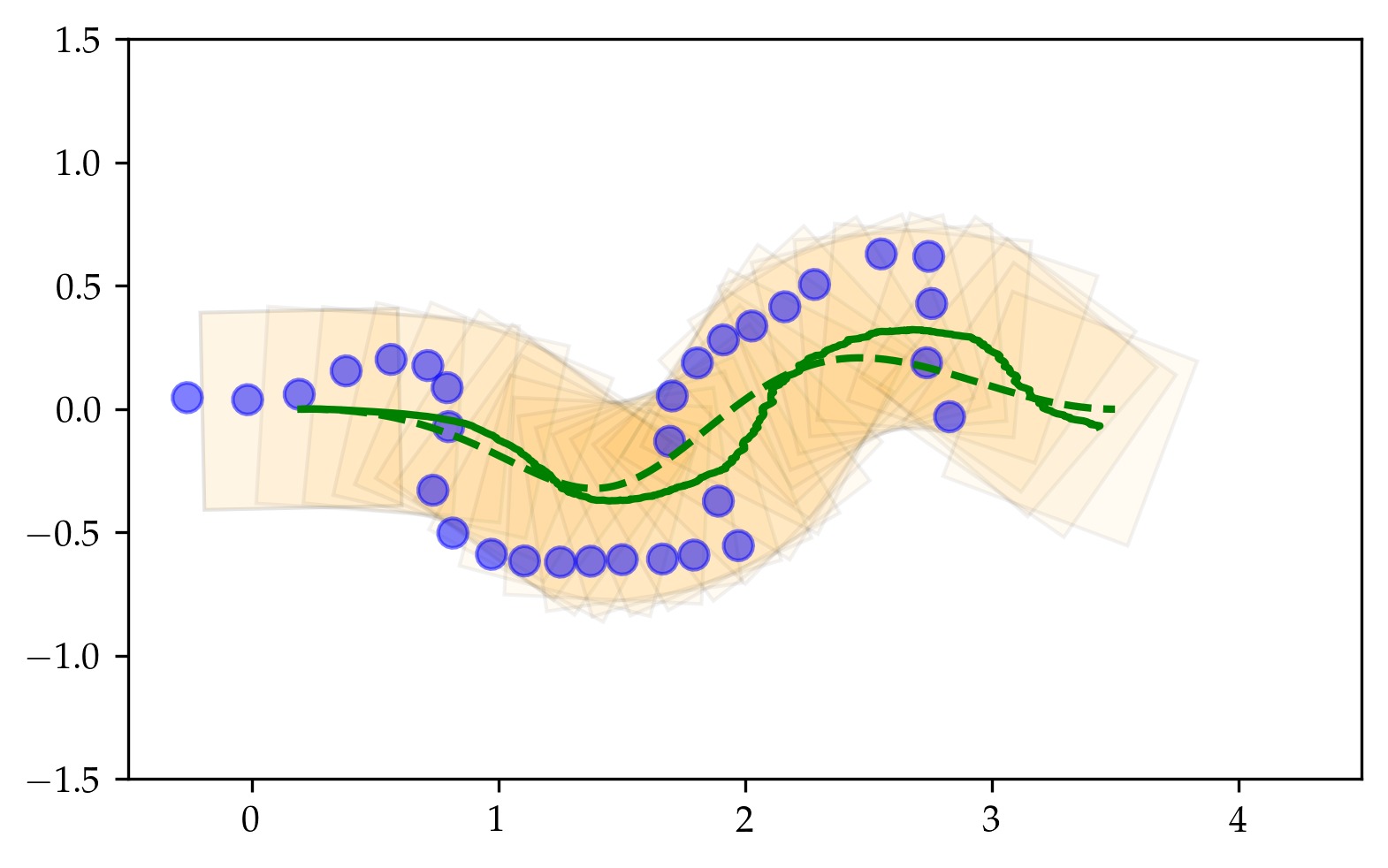}
        \caption{}
    \end{subfigure}
    \begin{subfigure}[b]{0.33\textwidth}
        \centering
        \includegraphics[width=.9\textwidth]{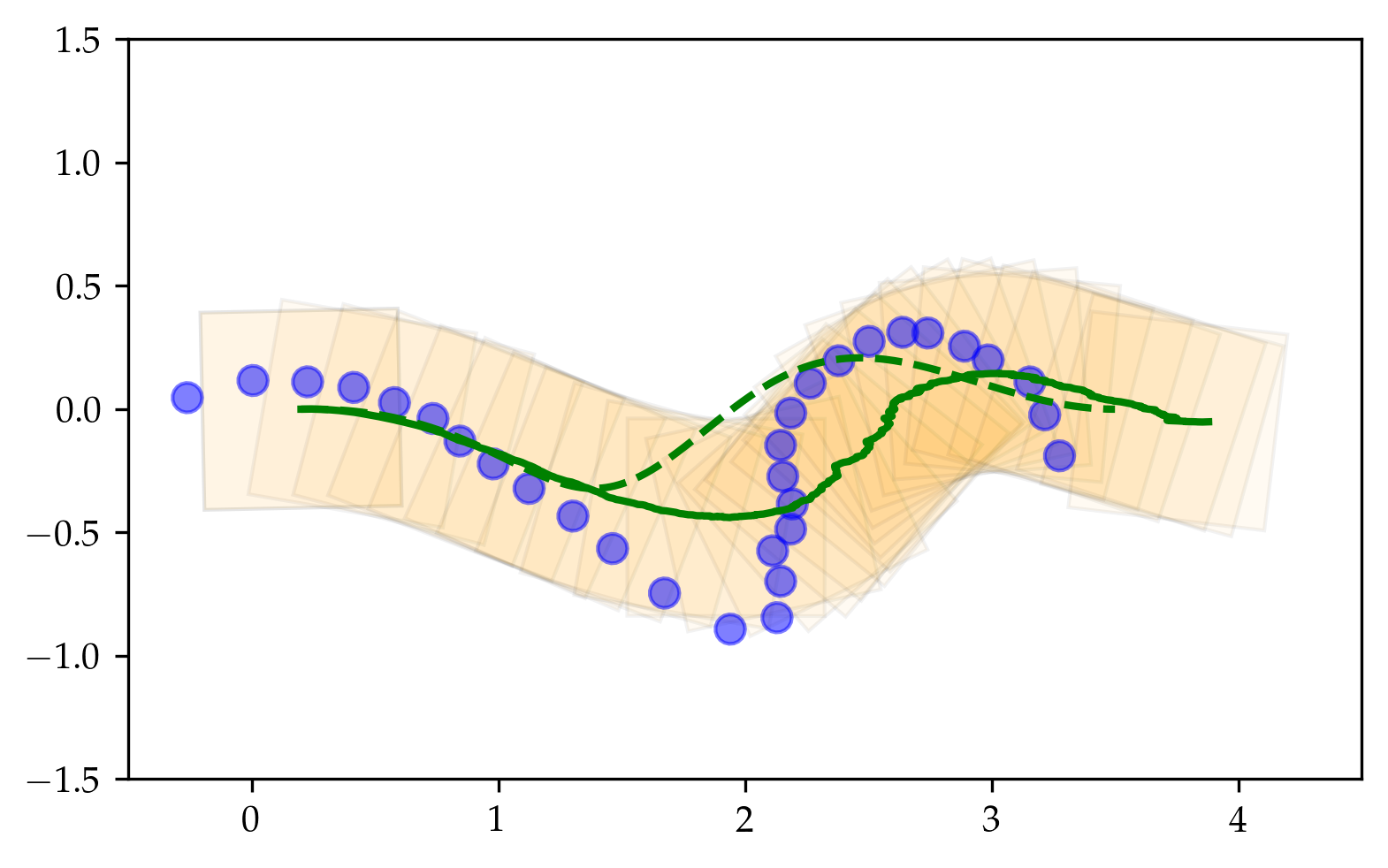}
        \caption{}
    \end{subfigure}
    \begin{subfigure}[b]{0.33\textwidth}
        \centering
        \includegraphics[width=.9\textwidth]{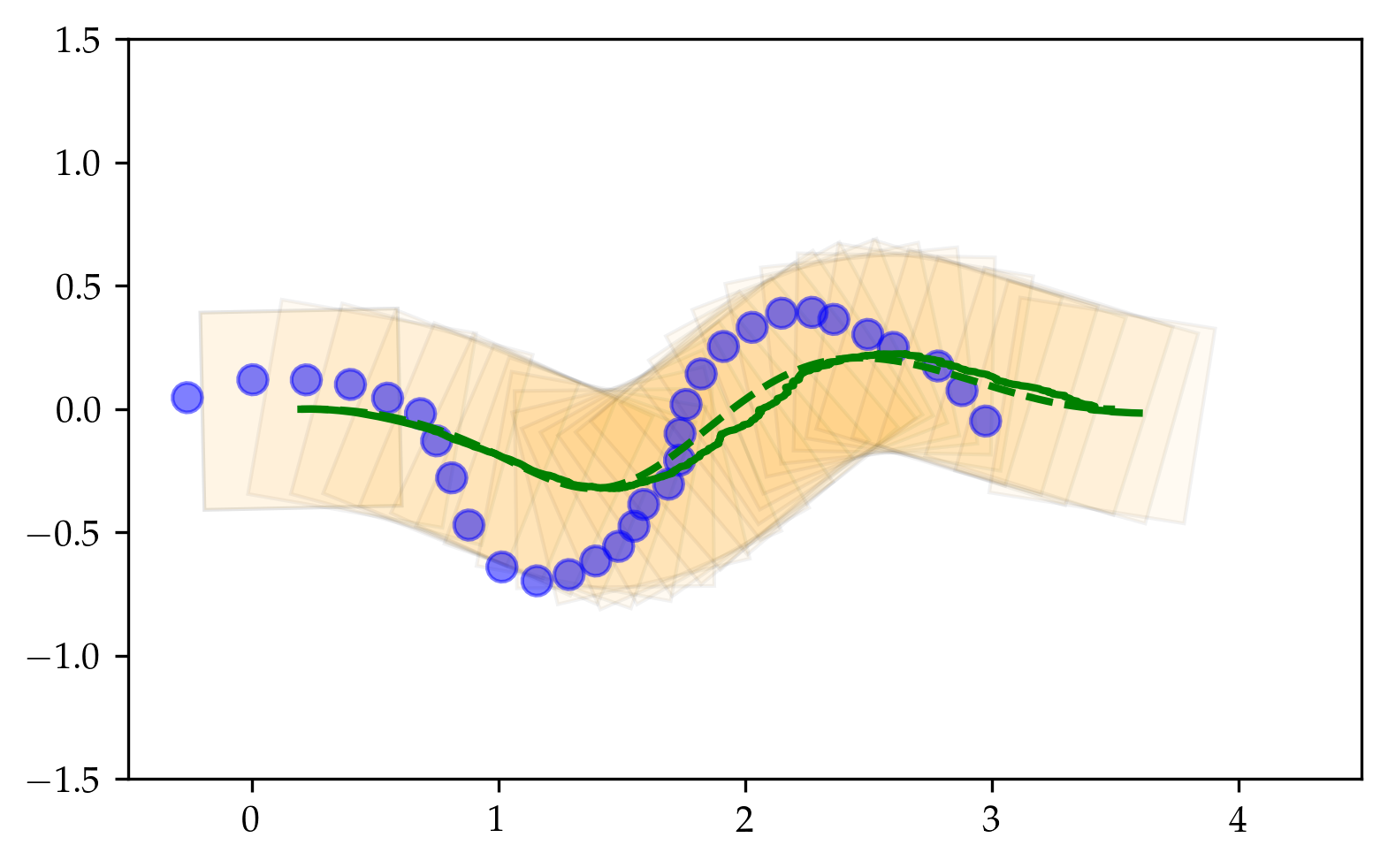}
        \caption{}
    \end{subfigure}
    \begin{subfigure}[b]{0.33\textwidth}
        \centering
        \includegraphics[width=.9\textwidth]{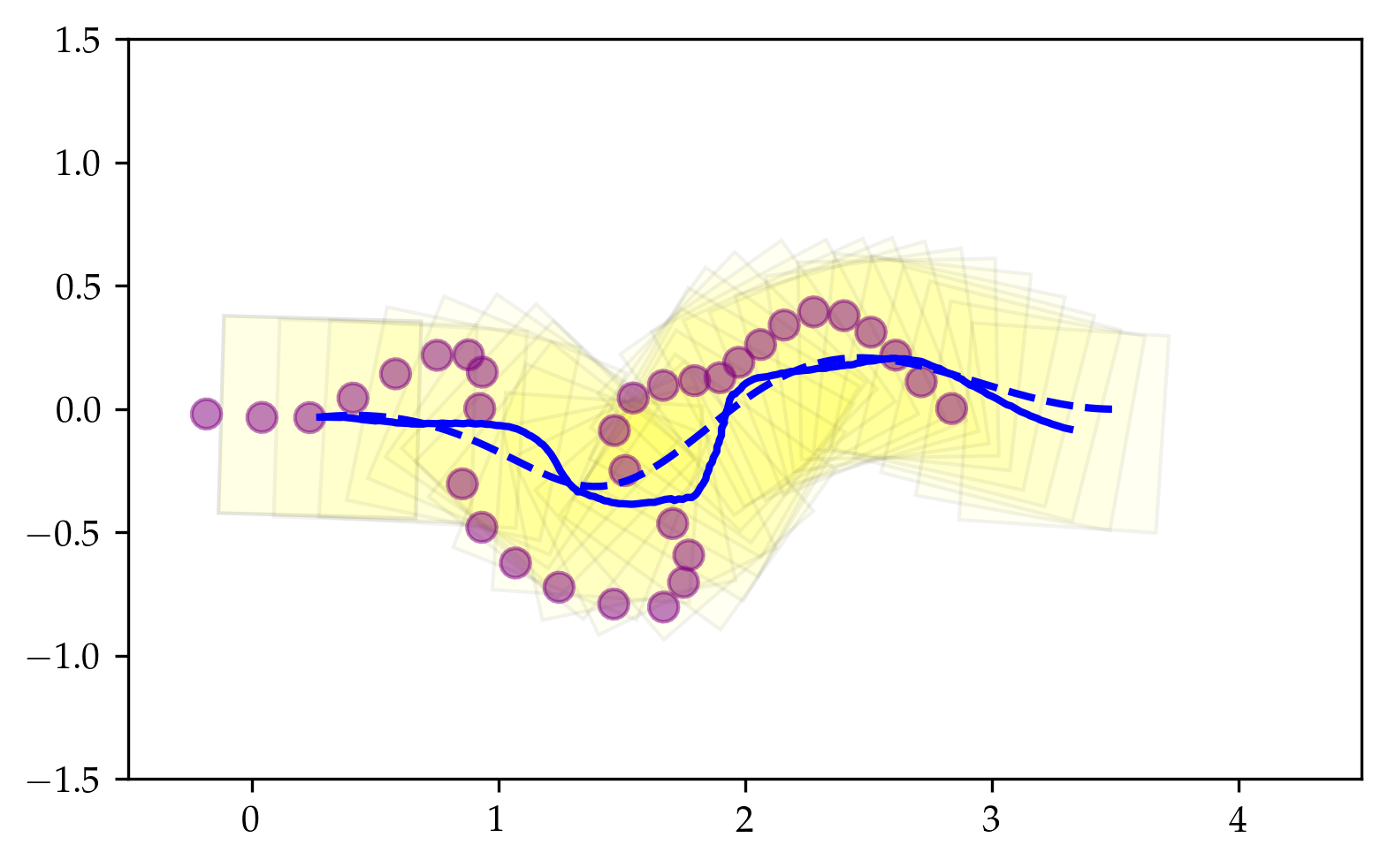}
        \caption{}
    \end{subfigure}
    \begin{subfigure}[b]{0.33\textwidth}
        \centering
        \includegraphics[width=.9\textwidth]{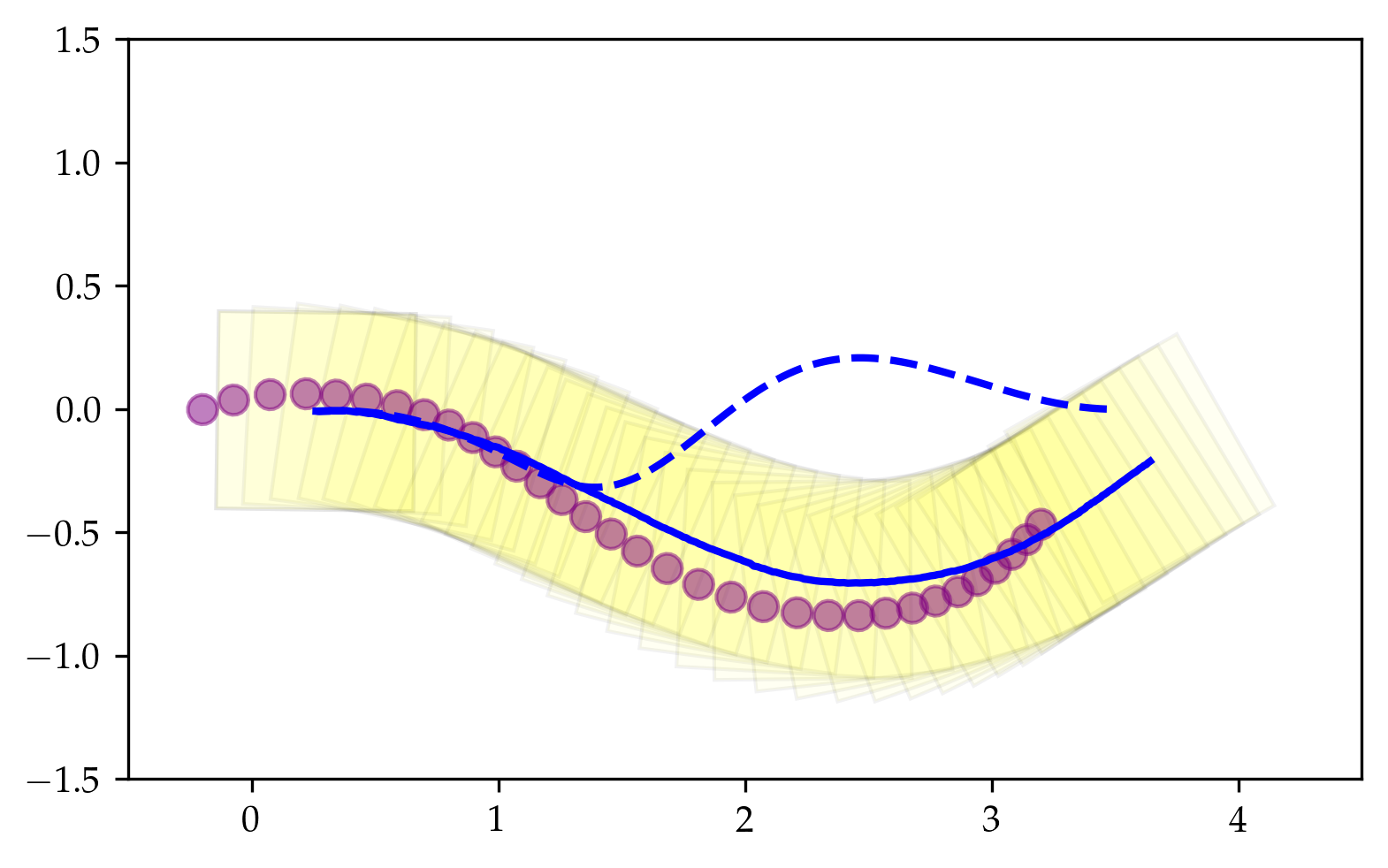}
        \caption{}
    \end{subfigure}
    \begin{subfigure}[b]{0.33\textwidth}
        \centering
        \includegraphics[width=.9\textwidth]{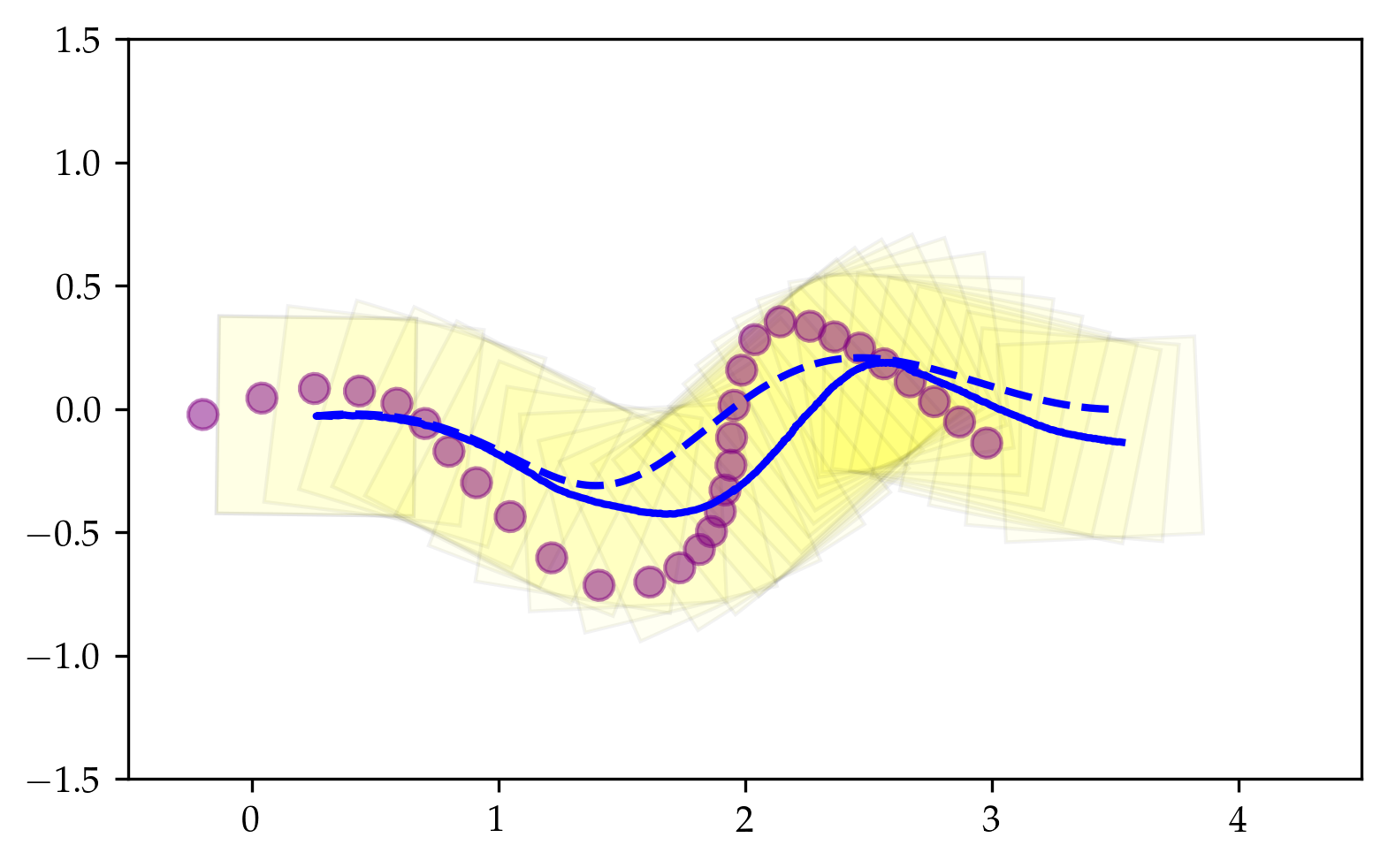}
        \caption{}
    \end{subfigure}
    \caption{Reference trajectory tracking for two different trajectories, with the reference indicated with dashed lines and the performed trajectory with a full line. Three control methods, a cascaded (quasi-)static feedback approach and dynamic feedback linearization with two choices of $\vectorstyle{\gamma}$, are shown both in simulation (orange) and on the set-up (yellow), respectively in the first, second and third column.}
    \label{fig:path}
\end{figure*}

Fig. \ref{fig:path} shows the tracking performance for the three tracking strategies. The same two reference trajectories are tested both in simulation and on the set-up, the first being a straight line and the second resembling a tilde. The trajectories should be covered in the same time, $T=40$ s, and the control frequency is the same as for the stationary point tracking, $\Delta t=0.1$ s. We have access to all time derivatives of the position signal. The controller gains for the dynamic feedback linearization are taken the same, while these for the cascaded controllers are slightly adapted, $\vectorstyle{\tau} = [2.5,2.0,0.75,0.4]$.

\begin{figure}
    \begin{subfigure}[b]{0.49\linewidth}
        \centering
        \includegraphics[width=.99\textwidth]{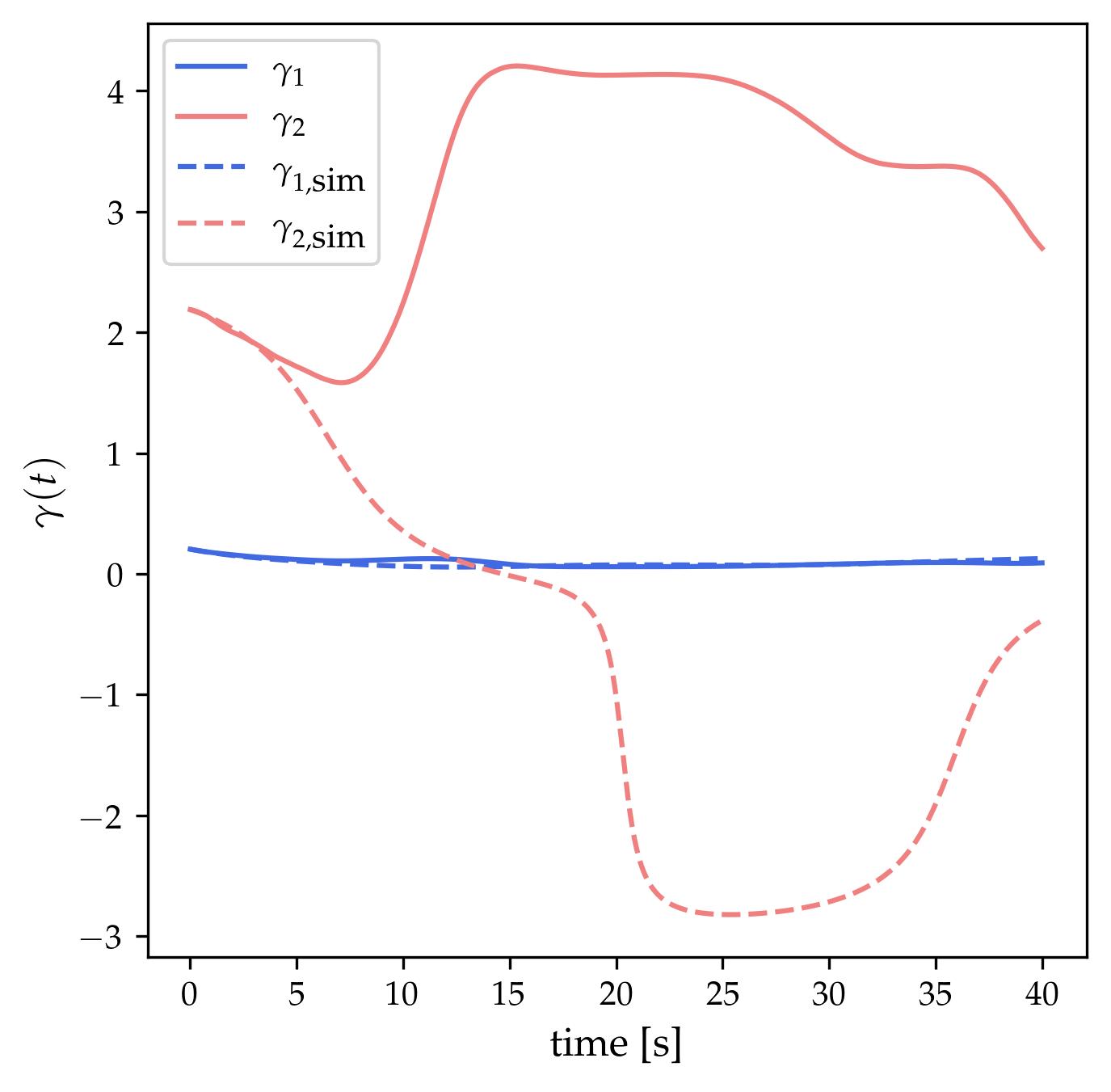}
        \caption{$\vectorstyle{\gamma}(t)$ for method 1}
    \end{subfigure}
    \begin{subfigure}[b]{0.49\linewidth}
        \centering
        \includegraphics[width=.99\textwidth]{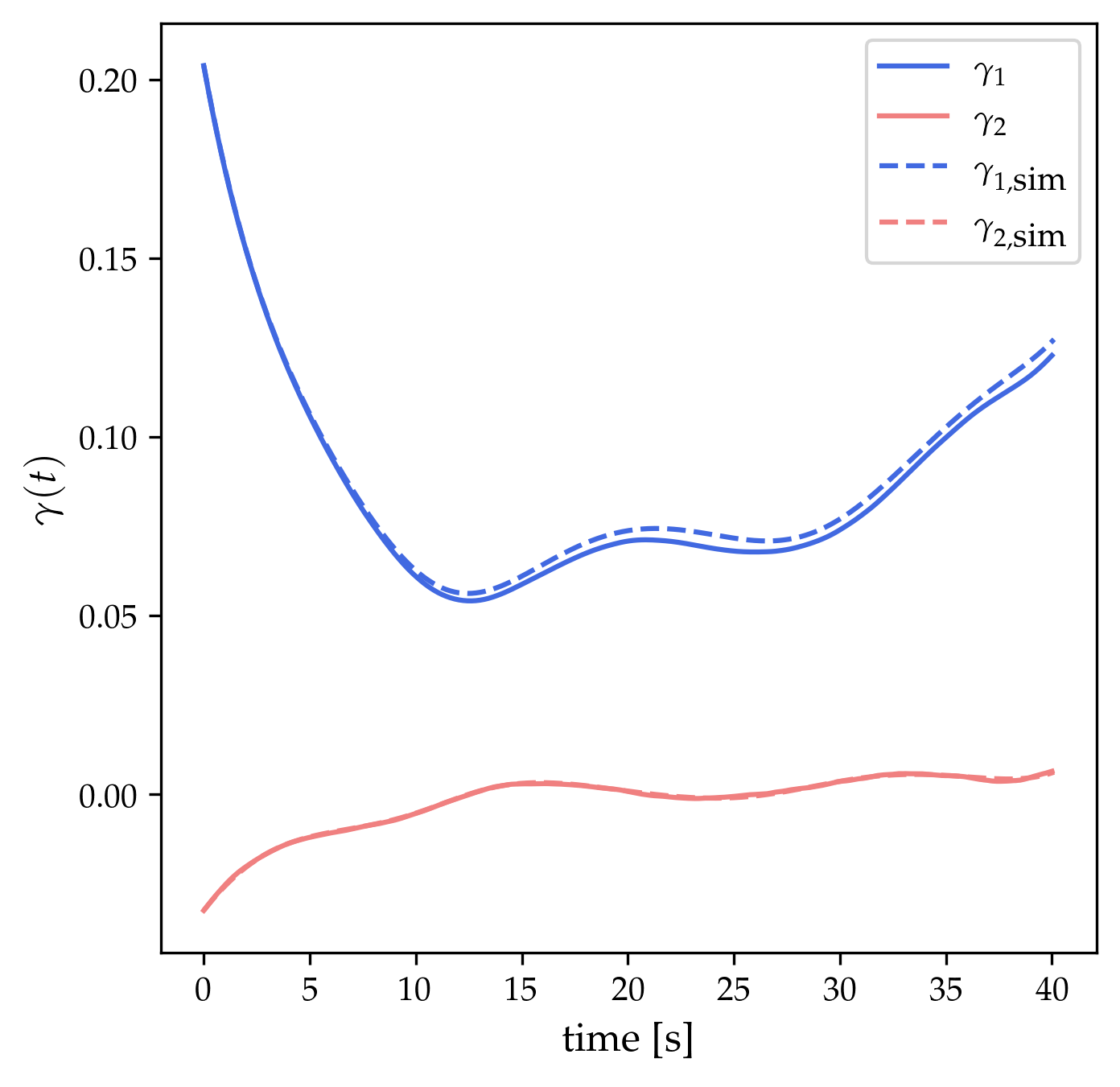}
        \caption{$\vectorstyle{\gamma}(t)$ for method 2}
    \end{subfigure}
    \begin{subfigure}[b]{0.49\linewidth}
        \centering
        \includegraphics[width=.99\textwidth]{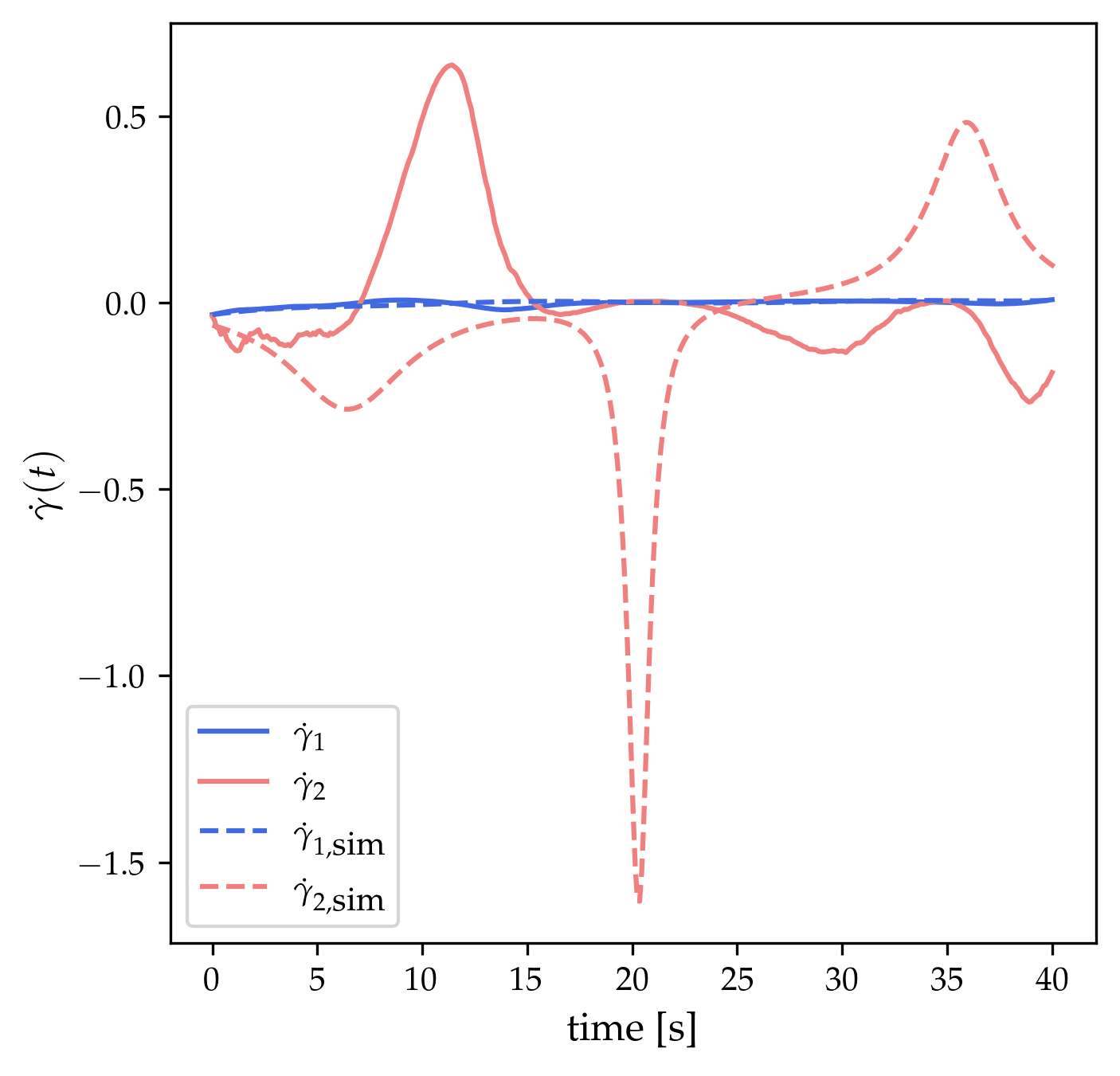}
        \caption{$\dot{\vectorstyle{\gamma}}(t)$ for method 1}
    \end{subfigure}
    \begin{subfigure}[b]{0.49\linewidth}
        \centering
        \includegraphics[width=.99\textwidth]{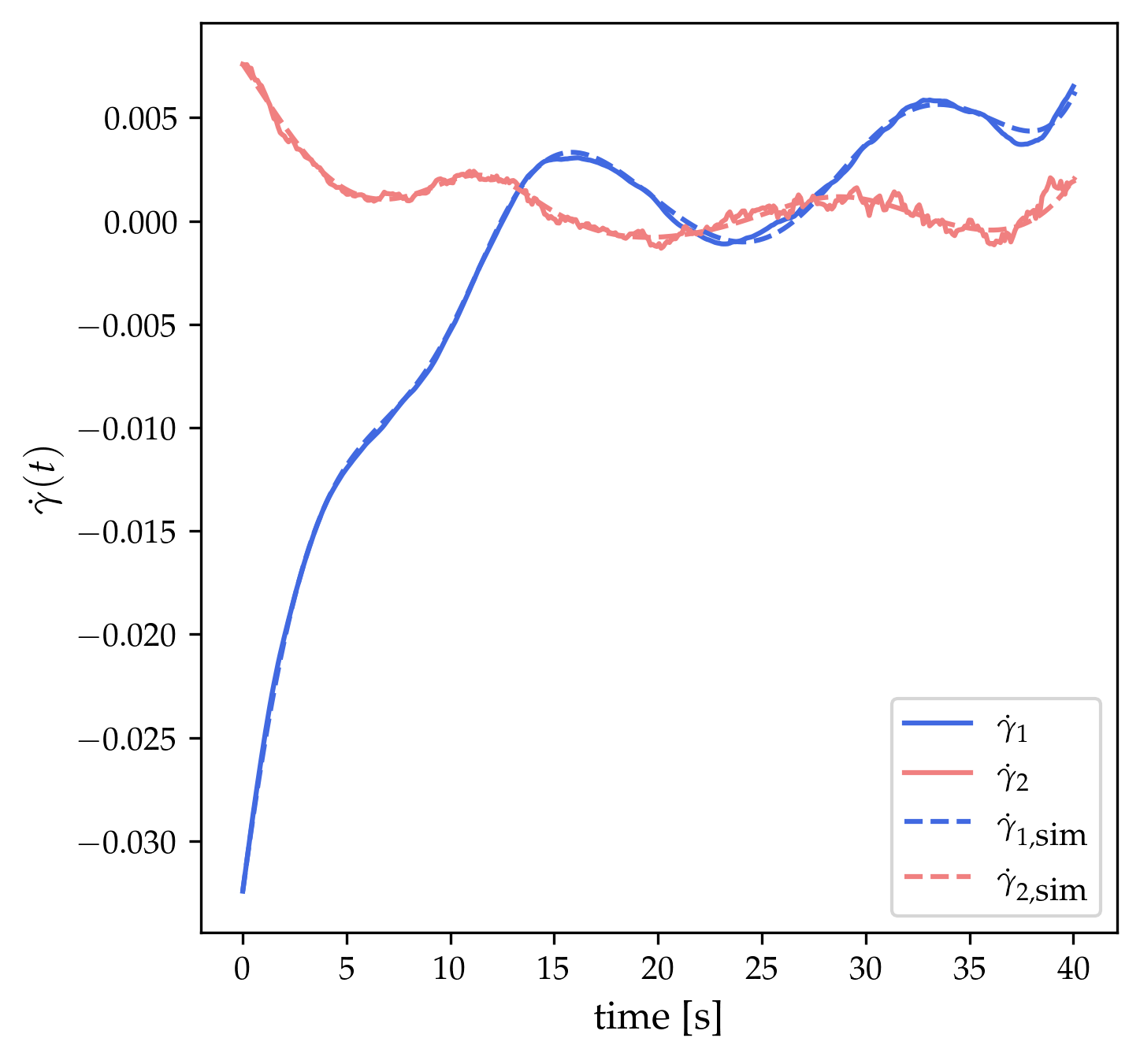}
        \caption{$\dot{\vectorstyle{\gamma}}(t)$ for method 2}
    \end{subfigure}
    \caption{Comparing $\vectorstyle{\gamma}(t)$ and $\dot{\vectorstyle{\gamma}}(t)$ for the first reference trajectory and the respective simulated trajectory for both choices of $\vectorstyle{\gamma}$ for the dynamic feedback linearization.}
    \label{fig:path_gamma}
\end{figure}

All three methods manage to track the straight line to a certain degree, with the first choice of $\gamma$ for feedback linearization performing the worst. A similar observation can be made for the tilde-like trajectory, again the first method of the linearization performs worst. The cascaded controller manages to track both trajectories accurately but with a non-smooth pusher trajectory. The latter could be because of badly tuned control gains.

If we take a closer look at the simulated and desired $\vectorstyle{\gamma}$ and $\dot{\vectorstyle{\gamma}}$ for the first trajectory and both methods, shown in Fig. \ref{fig:path_gamma}. We see that the first method does not reach the desired values, while the second accurately matches the desired values. This might be accounted to the singularity, as we can see a spike in the desired $\dot{\vectorstyle{\gamma}}_2$ around 20 seconds.

All three implemented methods demonstrate effective trajectory tracking, confirming the validity of the simplified model for control purposes. The cascaded quasi-static feedback controller, while effective, is more challenging to implement compared to the dynamic feedback linearization (DFL) approaches. Among the DFL methods, the singularity-free variant proves to be more reliable, highlighting the importance of avoiding singularities in DFL design. While the second DFL method appears to perform best in our experiments, the overall performance of all controllers depends on gain optimization, making direct comparisons dependent on tuning choices. Finally, the consistency between simulation and real-world results allows us to tune the controllers in simulation and apply them directly to the physical setup, demonstrating the practical applicability of the proposed model.

\section{Conclusion}
In this article, we analyzed the differential kinematics of slider-pusher systems with arbitrary slider geometries and circular pushers. Our approach relies on two key assumptions: (i) inertial forces are negligible compared to friction with the supporting surface by limiting to slow motion regimes, and (ii) friction at the slider-pusher contact is negligible compared to friction with the supporting surface. While the first assumption imposes a speed limitation, the second is a practical consideration influenced by material properties or enforced through design choices, such as using a bearing in the pusher, or could just be a modelling assumption that benefits the control design.

Our analysis led to general expressions for the system's differential kinematics and revealed that differential flatness is exhibited primarily by polygonal sliders with flat contact. This property simplifies trajectory planning and control synthesis, making feedback control more tractable. Based on this finding, we implemented two trajectory-tracking control strategies: a cascaded quasi-static feedback controller and a dynamic feedback linearization approach.

Experimental validation demonstrated that the same control parameters used in simulation could be applied directly to the physical setup. This would simplify automatic tuning of the control gains significantly, which is left for future work. Additionally, this result confirms that, despite its simplifying assumptions, the proposed model is sufficiently accurate for practical tracking control, reinforcing its applicability in real-world robotic manipulation tasks.

\appendix
\section*{Derivation of equation (\ref{eq:r})}\label{sec:derivation-of-equation-refeqr}
Restarting from equation (\ref{eq:flat}) 
\begin{equation}
\label{eq:A}
r(\phi)^2+2r’(\phi)^2- r’(\phi)'r(\phi) = 0
\end{equation}

Introducing the function $v_r=r’(\phi)$ as a function of $r$. It follows that
\begin{equation}
r’(\phi)' = \frac{\text{d}r’(\phi)}{\text{d}\phi} = \frac{\text{d}r’(\phi)}{\text{d}r}\frac{\text{d}r(\phi)}{\text{d}\phi} = v_r'v_r
\end{equation}

Hence
\begin{equation}
r^2 + 2 v_r^2 - r v_r' v_r = 0
\end{equation}
where we can substitute $w_r=\frac{1}{2}v_r^2$ to obtain
\begin{equation}
r^2 + 4 w_r - r w_r' = 0
\end{equation}

The former equation can be solved for $w$ as a function of $r$
\begin{equation}
w_r = C r^4 - \frac{1}{2}r^2
\end{equation}
which is substituted in $v$
\begin{equation}
v_r = \sqrt{C r^4 - \frac{1}{2}r^2}
\end{equation}

Recalling that $v_r=r’(\phi)$ we eventually find that
\begin{equation}
r(\phi) = A\frac{1}{c_{\phi-B}}
\end{equation}

\section{References}

\bibliographystyle{IEEEtran}
\bibliography{IEEEabrv,./references}

\end{document}